\documentclass[11pt, notitlepage, tightenlines,aps, prd, reprint,superscriptaddress,nofootinbib]{revtex4-1}

\usepackage[utf8]{inputenc}
\usepackage{placeins}
\usepackage{mathtools}
\usepackage{amsfonts}
\usepackage{amssymb,amsmath}
\usepackage{color}
\usepackage{xcolor}
\usepackage[breaklinks,colorlinks,urlcolor=blue,citecolor=citcolor,linkcolor=lcolor]{hyperref}
\usepackage{graphicx}
\usepackage{subfigure}
\usepackage{multirow}
\usepackage{braket}
\usepackage{bm}
\usepackage{txfonts}
\usepackage{mathrsfs} 
\usepackage{slashed}
\usepackage{cleveref}
\usepackage{hepunits}
\usepackage{overpic}

\definecolor{lcolor}{rgb}{0.5,0,0}
\definecolor{citcolor}{rgb}{0,0.3,0.0}

\newcommand*\diff{\mathop{}\!\mathrm{d}}
\graphicspath{ {image/} }

\DeclarePairedDelimiter{\floor}{\lfloor}{\rfloor}
\DeclareMathOperator{\diag}{diag}
\DeclareMathOperator{\Li}{Li}

\newcommand{\GeV}{{{\,}\textrm{GeV}}}

\begin{document}

\bibliographystyle{apsrev4-1}

\title{Scattering and gluon emission of physical quarks in a SU(3) colored field}
\author{Meijian Li}
\email{meijian.li@usc.es}
\affiliation{Instituto Galego de Fisica de Altas Enerxias (IGFAE), Universidade de Santiago de Compostela, E-15782 Galicia, Spain}

\author{Tuomas Lappi}
\email{tuomas.v.v.lappi@jyu.fi}
\affiliation{Department of Physics, P.O. Box 35, FI-40014 University of Jyv\"{a}skyl\"{a},
Finland}
\affiliation{
Helsinki Institute of Physics, P.O. Box 64, FI-00014 University of Helsinki,
Finland
}

\author{Xingbo Zhao}
\email{xbzhao@impcas.ac.cn}
\affiliation{Institute of Modern Physics, Chinese Academy of Sciences, Lanzhou 730000, China}
\affiliation{University of Chinese Academy of Sciences, Beijing 100049, China}

\author{Carlos A. Salgado}
\email{carlos.salgado@usc.es}
\affiliation{Instituto Galego de Fisica de Altas Enerxias (IGFAE), Universidade de Santiago de Compostela, E-15782 Galicia, Spain}

\begin{abstract}
We study the scattering of the gluon-dressed physical quarks, defined as the eigenstates of the vacuum QCD Hamiltonian, off a colored medium. 
 We solve the wavefunction of the physical quark state by diagonalizing the QCD Hamiltonian in vacuum in a $\ket{q}+\ket{qg}$ Fock space, with implementing the sector-dependent mass renormalization scheme.
  We then perform numerical simulations of the real-time quantum state evolution of the initially dressed quark state at various medium densities. The results are compared with those of an initially bare or off-shell quark states.
  With the obtained light-front wavefunction of the evolved state, we extract the quark jet transverse momentum broadening, the quenching parameter, the cross section, the gluon emission rate, and the evolution of the invariant mass. 
  The scenario considered is relevant for high energy scattering processes, where the quark originates from far outside the color field describing the scattering target.
  This investigation on dressed quarks complements our earlier studies of the single quark scattering in the $\ket{q}$ Fock space, and of the bare quark scattering in the $\ket{q}+\ket{qg}$ Fock space, providing a novel systematic description of quark scattering process using a non-perturbative formalism.
\end{abstract}

\maketitle
\tableofcontents

\section{Introduction}
Quark jets serve as effective probes of QCD matter in various high-energy processes~\cite{Busza:2018rrf}, including electron-ion, proton-proton, proton-ion, and ion-ion collisions, at the BNL Relativistic Heavy-Ion Collider (RHIC), CERN Large Hadron Collider (LHC), and the upcoming Electron-Ion Collider (EIC). The scattering of the quark jet off the matter field provides one of the most direct ways to study properties of the strong interaction and the role of gluons.
In high-energy collisions, a jet is observed as a collimated beam of particles from the splitting of a common ancestor, a quark or gluon, and is intrinsically a quantum state.
In the presence of a nuclear medium, the real-time evolution of the jet becomes highly complex, involving both its intrinsic dynamics and interactions with the external field, along with the interplay between them.

The light-front Hamiltonian formalism provides powerful tools for investigating a range of quantum properties, from the internal structure of self-bound states to real-time scattering dynamics. 
Basis Light-Front Quantization (BLFQ)~\cite{1stBLFQ} has been introduced as a non-perturbative approach for solving bound states by diagonalizing the Hamiltonian in a suitable basis function representation, allowing for an  efficient numerical computation. 
 This approach has been effectively applied to systems in both QED~\cite{Honkanen:2010rc, Zhao:2014xaa, Wiecki:2014ola, Chakrabarti:2014cwa, Hu:2020arv, Nair:2022evk, Nair:2023lir} and QCD~\cite{Jia:2018ary, Lan:2019vui, Lan:2019rba, Adhikari:2021jrh, Lan:2021wok, Mondal:2021czk, Li:2015zda, Li:2017mlw, Li:2018uif, Lan:2019img, Tang:2018myz, Tang:2019gvn, Mondal:2019jdg, Qian:2020utg, Xu:2021wwj, Liu:2022fvl, Hu:2022ctr, Peng:2022lte,Kaur:2023lun, Zhu:2023nhl,Zhang:2023xfe, Liu:2024umn}, with recent developments advancing beyond the valence Fock sector to include gluon dynamics~\cite{Kaur:2024iwn,Xu:2023nqv, Lin:2023ezw,Zhu:2023lst, Yu:2024mxo}.
 
The time-dependent Basis Light-Front Quantization (tBLFQ) approach\footnote{Its Quantum Mechanics counterpart--the time-dependent Basis Function (tBF) approach has been developed to address nuclear structure and scattering~\cite{Du:2018tce, Yin:2022zii}.} extends the BLFQ approach to tackle time-dependent problems in the presence of an external background field.  Initially applied in QED~\cite{Zhao:2013cma, Chen:2017uuq, Hu:2019hjx, Lei:2022nsk}, it has been further developed for QCD, where we have developed a framework to simulate the real-time evolution of a quark state traversing a SU(3) colored medium~\cite{Li:2020uhl, Li:2021zaw, Li:2023jeh}.

In this framework, the calculation of the evolution process is done on the quantum amplitude level with the full Hamiltonian, which enables us to explore physics effects in the regimes that can be difficult for other approaches.
In the tBLFQ, the eikonal limit (e.g., Ref.~\cite{Dumitru:2002qt}) can be overcome  straightforwardly by assigning a finite longitudinal momentum to the jet in the Hamiltonian.
Unlike in the pQCD-based approaches, where the number of gluon emissions is truncated by the order of the coupling expansion (e.g., Refs.~\cite{Baier:1996kr,Baier:1996sk,Zakharov:1997uu}), in tBLFQ, gluon emission and its conjugate process, gluon absorption, occur continuously throughout the evolution.
The interaction with the medium also occurs continuously in tBLFQ, different from the implementation of having a finite number of scattering centers in an opacity expansion~\cite{Gyulassy:1999zd,Gyulassy:2000fs,Wiedemann:2000za}.

\begin{figure*}[t]
  \centering 
  \subfigure[Bare quark evolution\label{fig:setup_incoherent}]{
    \includegraphics[width=0.44\textwidth]{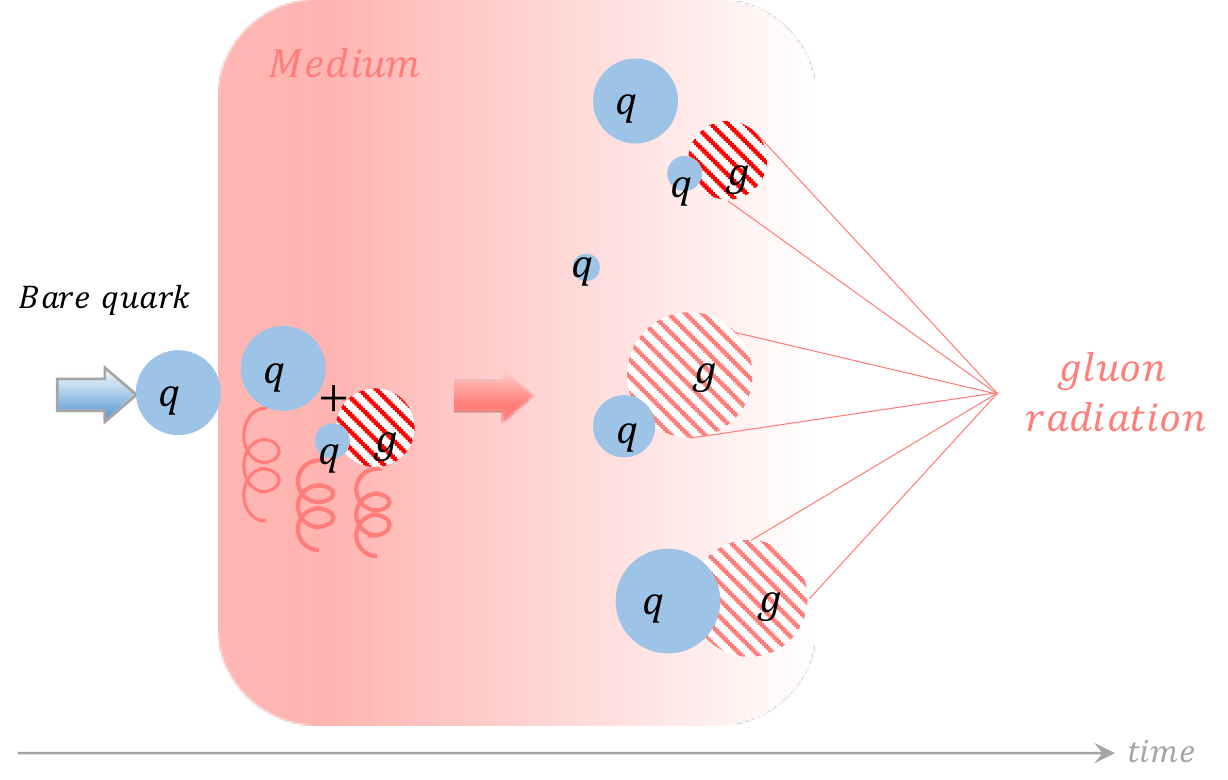} 
  }  
  \quad
  \subfigure[Dressed quark evolution \label{fig:setup_coherent}]{
    \includegraphics[width=0.44\textwidth]{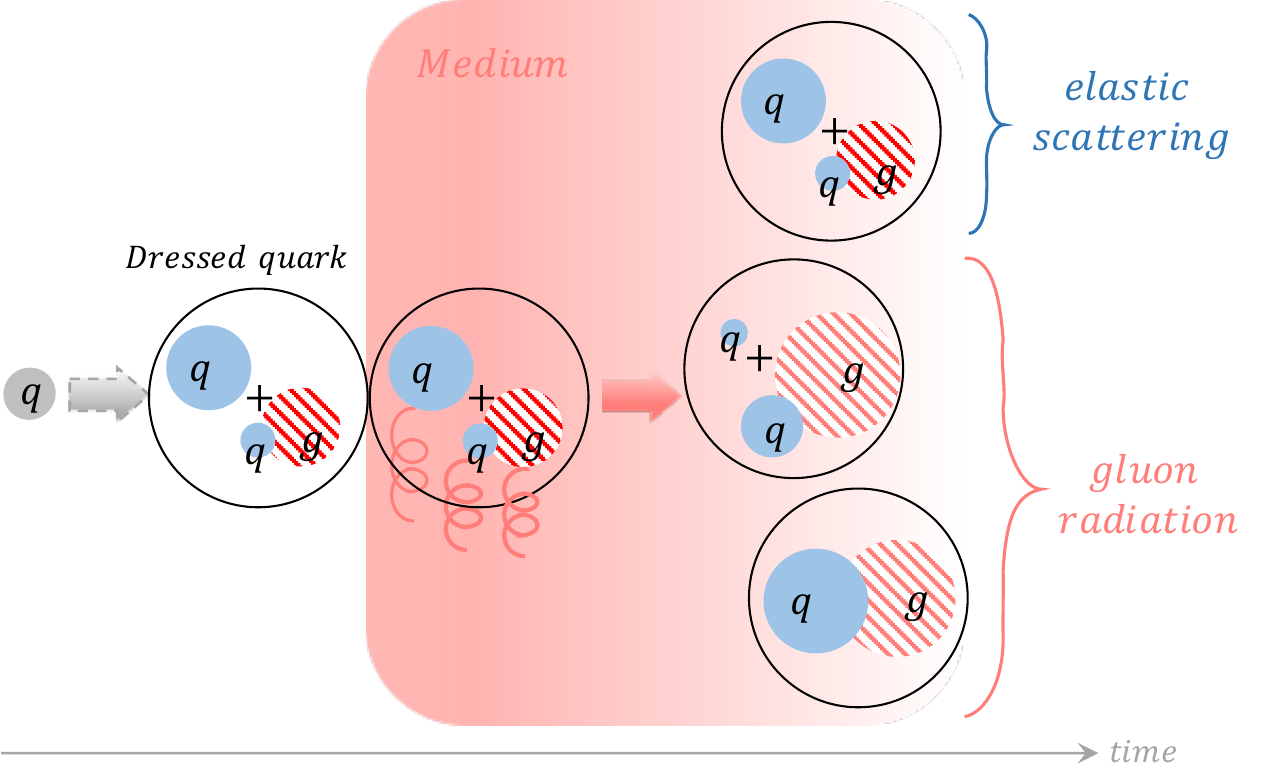} 
  }
  \caption{
  Schematic illustration of in-medium quark evolution in the $\ket{q}+\ket{qg}$ Fock space.  
  (a) The initial state is a bare quark state, and the emergent $\ket{qg}$ components can be interpreted as radiated gluon.
  (b) A quark originated far outside the medium develops into a dressed quark state, which then traverses the medium.
  The transition to different coherent states can be interpreted as gluon radiation.
  }
  \label{fig:setup}
\end{figure*}

In our preceding work~\cite{Li:2023jeh}, we formulated the jet state in the $\ket{q}+\ket{qg}$ Fock space. We used a single bare $\ket{q}$ state with a definite momentum as the initial state to investigate the jet's in-medium momentum broadening, a process illustrated in Fig.~\ref{fig:setup_incoherent}. 
In this work, we study a different scenario where the quark originates far outside the medium. In this case the initial state traversing the medium is a dressed quark, corresponding to the eigenstate of the QCD Hamiltonian in the $\ket{q}+\ket{qg}$ Fock space and characterized in terms of its light-front wavefunction. 
Such a process is illustrated in Fig.~\ref{fig:setup_coherent}.

The layout of this paper is as follows. 
In Sec.~\ref{sec:formalism}, we introduce the formalism of in-medium quark jet evolution using the light-front Hamiltonian approach, first for solving the physical quark states and studying their various features in Sec.~\ref{sec:phys_q}, then for simulating the time evolution process and analyzing the evolved states in Sec.~\ref{sec:time_evolution}. 
In Sec.~\ref{sec:results}, we provide and discuss the results of the in-medium quark jet simulation. We compare the results using four different initial states, the bare quark, the onshell dressed quark, timelike and spacelike quarks. 
We conclude the work in Sec.~\ref{sec:summary}.

\section{Formalism}\label{sec:formalism}
Our method is based on the light-front Hamiltonian formalism, in which the field is quantized on the equal light-front time surface $x^+=0$. The Hamiltonian is obtained through the standard Legendre transformation of the QCD Lagrangian in the light-cone gauge $A^+=0$. 
In this work, we study the physical quark and its in-medium evolution by truncating its Fock space expansion to $\ket{q}+\ket{qg}$. 

The QCD Hamiltonian in $\ket{q}+\ket{qg}$ can be written as \cite{Li:2021zaw}, 
\begin{align}\label{eq:Pmin_QCD}
  P^-_{QCD}=P^-_{KE}+V_{qg}\;.
\end{align}
Note that the instantaneous quark/gluon interactions are not included, as their contributions will be canceled by the mass counterterm through renormalization. Further details can be found in Appendix~\ref{app:inst}.

In the presence of a background gluon field, the full Hamiltonian contains an additional interaction term,
\begin{align}\label{eq:P_full}
  P^-(x^+)=P_{QCD}^- + V_{\mathcal{A}}(x^+)\;.
\end{align}

Here, we are interested in how a physical quark evolves due to interactions with a background field. To this purpose, we first solve for the light-front wavefunction of the physical quark as an eigenstate of the QCD Hamiltonian $P^-_{QCD}$, we then simulate the time evolution process with the full Hamiltonian $P^-(x^+)$.

\subsection{Physical quark}\label{sec:phys_q}
In the light-front Hamiltonian formalism, the physical quark state is described as the eigenstate of the QCD Hamiltonian. 
The mass spectrum and the light-front wavefunctions (LFWFs) are obtained by solving the time-independent eigenvalue equation,
\begin{align}\label{eq:TISE_Pmn}
  P^-_{QCD}\ket{\phi}=P^-_{\phi}\ket{\phi}\;,
\end{align}
in which $\ket{\phi}$ is the physical quark eigenstate, and the eigenvalue $ P^-_{\phi}$ its light-front energy.
It is often intuitive to write the above equation by defining a light-cone Hamiltonian as the operator $\mathrm H_{\text{LC}} = P^\mu P_\mu = P^+ P^-_{QCD} -\vec P_\perp^2$, in which $P^+$ is the longitudinal momentum operator, and $\vec P_\perp$ the transverse momentum operator.
In such a way, its eigenvalues correspond to the invariant mass spectrum of the theory,
\begin{align}\label{eq:TISE_HLC}
  \mathrm H_{\text{LC}}\ket{\phi}=M^2\ket{\phi}\;.
\end{align}
Here, $M^2= P^+_{\phi} P^-_{\phi} -\vec P_{\phi,\perp}^2$ gives the invariant mass of the eigenstate, in which $P^+_{\phi}$ and $\vec P_{\phi,\perp}$ are the momenta of the state.

Schematically, we can write the LFWF $\{\psi_Q, \psi_{qg}\}$\footnote{
  Here and in the following, we use the subscripts ``$Q$" and ``$q$" to distinguish the quark in the $\ket{q}$ sector and that in the $\ket{qg}$ sector of the Fock space.
} of a physical quark in the truncated Fock space as 
\begin{align}
    \ket{q_{\text{phy}}} = \psi_Q\ket{q}+\psi_{qg}\ket{qg}\;.
\end{align}
In a chosen basis representation, Eq.~\eqref{eq:TISE_Pmn} or~\eqref{eq:TISE_HLC} is solved by diagonalizing the Hamiltonian matrix, and the basis coefficients of the obtained eigenstates give the LFWFs. 

\subsubsection{Mass renormalization}
We renormalize the quark mass in a procedure guided by the sector-dependent renormalization approach~\cite{PhysRevD.77.085028,PhysRevD.86.085006}. 
In this approach, the mass counterterm is introduced to compensate for the mass correction due to the quantum fluctuations to higher Fock sectors. 
Accordingly, in the Fock space $\ket{q}+\ket{qg}$, the mass counterterm $\delta m$ is only applied to the quark in the $\ket{q}$ sector, which compensates for the fluctuation to $\ket{qg}$, such that the renormalized mass is $m_Q= m_q+\delta m$, with $m_q$ being the physical quark mass. For the quark in the $\ket{qg}$ sector, on the other hand, there is no higher sector (e.g., $\ket{qgg}$) for it to fluctuate into, therefore it does not receive any mass correction.

The value of the mass counterterm $\delta m$ can be determined numerically through iterative diagonalization until the resulting mass in terms of the ground state eigenvalue (for each  $P_\perp$ and $P^+$ separately) matches the physical quark mass,
  \begin{align}\label{eq:HLC_dm}
   \mathrm H_{\text{LC}}(\delta m)
    \ket{\phi_0}
    = m_q^2 \ket{\phi_0}\;.
  \end{align}
  The ground state $\ket{\phi_0}$, being a dressed quark state, is the physical quark state on its mass shell.
In addition to the ground state, the spectrum has several different types of excited states. Many of them, which we call ``uncoupled'' $\ket{qg}$ states, have quantum numbers (color, angular momentum, \dots) that do not correspond to a single quark. There are also excited states with the quantum numbers of a single quark. We call these dressed quark-gluon states, or ``coupled'' $\ket{qg}$ states, since the quantum numbers allow them to mix with the single quark state.  In terms of the asymptotic physical particle content (as an ``in'' or ``out'' state in a scattering problem) they correspond to a quark-gluon system, i.e. the decay products of a time-like off-shell particle. These different excited states are discussed in more detail in Sec.~\ref{subsec:spectrum}.

To obtain an off-shell quark state, one could find the mass counterterm $\delta m'$ by equating the state (not necessarily the ground state) eigenvalue to the intended virtuality $Q^2$ instead. Here, 
$Q^2>0$ corresponds to a time-like quark, while 
$Q^2<0$ corresponds to a space-like quark.  
More details are addressed in Sec.~\ref{subsec:offshell}.

For the Hamiltonian we are dealing with here, the mass counterterm can alternatively be determined analytically, without the iterative algorithm with a  matrix diagonalization at each step, as we will illustrate in Sec.~\ref{sec:analytical_solution}.

\subsubsection{Symmetries and basis}\label{sec:symmetry}
In this investigation we use the same discrete basis representation constructed with momentum eigenstates in our preceding works~\cite{Li:2021zaw,Li:2023jeh}. 
We will further develop it by exploiting symmetries in the problem Hamiltonian.  
In doing so, we can both develop a physically meaningful classification of the eigenstates, and reduce the computational complexity of matrix diagonalization.

In the basis representation that we use here, each single particle state carries five quantum numbers, $\beta_l=\{k^+_l, k^x_l, k^y_l, \lambda_l, c_l \}$ ($l=q \text{ or } g$), the three momentum components $k^+_l, k^x_l, k^y_l$, the light-front helicity $\lambda_l$, and the color $c_l$. The two-particle basis states are direct products of single particle states, $\ket{\beta_{qg}}=\ket{\beta_q}\otimes\ket{\beta_g}$. 
A physical state $\ket{\phi}$ is expanded as a linear combination of these basis states
\begin{align}
  \ket{\phi}=\sum_\beta c_{\phi,\beta} \ket{\beta}\;.
\end{align}
The LFWF is the column vector $\bm{c}_\phi$ consisting of the basis coefficients $c_{\phi,\beta}=\braket{\beta|\phi}$.
The Hamiltonian operator $P^-_{QCD}$ is in the matrix form, denoted by $\mathcal{P}$, and its matrix elements encodes the transition amplitude between two basis states, $\mathcal P_{\beta\beta'}=\bra{\beta}P^-_{QCD}\ket{\beta'}$.
The Hamiltonian has several symmetries, which enable us to classify its eigenstates. Let us go through these symmetries and how to use them to decompose our Fock space into the corresponding parts. 
 
\begin{itemize}
\item[i.] \textbf{Longitudinal boost invariance}
The longitudinal momentum operator commutes with the Hamiltonian, i.e., $[P^-_{QCD},P^+]=0$. As a result, each $P^-_{QCD}$ eigenstate has a definite total longitudinal momentum $P^+$, and it can only contain single-quark basis states with $p^+_Q=P^+$ and quark-gluon basis states with $p^+_q+p^+_g=P^+$.

In constructing the basis space, we discretize the longitudinal momentum by imposing an (anti)periodic boundary condition for the bosons (fermions) within the length $ x^-\in [-L,L]$. The single-particle longitudinal momentum is therefore parameterized as $ p_l^+= k^+_l 2\pi/L$, with $ k^+_q=1/2,3/2,\ldots $ and $k^+_g=1,2,\ldots$. 
For each basis state, quark or quark-gluon, the total longitudinal momentum is a sum of the individual partons, $K=\sum_l k^+_l$, so the total longitudinal momentum is $P^+=K\times 2\pi/L $ with possible values $K= 1/2, 2/2, \dots$. Because the total $P^+$ is conserved, we keep $K$ fixed to one value. Then, in the $\ket{qg}$ sector, the value of $\floor*{K} (=K-0.5)$ indicates the number of different ways this total $P^+$ can be split between the quark and the gluon. In other words, $1/\floor*{K}$ is the resolution in the momentum fraction $k^+_g/P^+$.

For a fixed $K$, the dependence of the $P^-_{QCD}$ matrix on the value of $P^+$ is solely an overall factor $1/P^+$. 
Consequently, the $P^-_{QCD}$ eigenstates at different $P^+$ are the same, and their eigenvalues scale by $1/P^+$.
Physically, it means that the intrinsic structure of a physical quark is invariant under a longitudinal boost.
\footnote{
Note that boosting the physical quark states to a different $P^+$ is not the same as allowing multiple total $P^+$ states simultaneously. The latter would be necessary for a $x^-$-dependent external potential, which could change the $P^+$ of the partonic state during the interaction.
}
So in practice, we can diagonalize the $P^-_{QCD}$ matrix once, and specify the value of $P^+$ afterwards. When $K$ is fixed, the value of $P^+$ in physical units specifies the value of $L$ in physical units, or vice versa.

The dimension of the $P^-_{CM}$ matrix, $\mathcal{P}$, summing over all the quark and quark-gluon basis states, is 
  \begin{equation}  
  \begin{split}
    \dim_{tot} = &N_c \times 2 \times \dim_\perp\\
    &+ \floor*{K}\times N_c(N_c^2-1)\times 4 \times \dim_\perp^2\;,
  \end{split}
  \end{equation}
  in which $N_c=3$ is the color dimension, the $2$ and $4$ count the different helicity states, and 
  $\dim_\perp$ is the number of degrees of freedom in the transverse dimension.
In our basis, the transverse space is discretized on a periodic lattice that spans $[-L_\perp, L_\perp]$ with number of sites $2 N_\perp$, so $\dim_\perp=(2 N_\perp)^2$.

\item[ii.] \textbf{Translational and Galilean transverse boost invariance}
The transverse momentum operator $\vec P_\perp$ also commutes with the Hamiltonian, i.e., $[P^-_{QCD},\vec P_\perp]=0$. As a result, each physical quark eigenstate has a definite $ \vec P_\perp$ value.
Light front dynamics also has a Galilean 2-dimensional transverse boost invariance. These two invariances together allow us to split the renormalized light-front Hamiltonian into two terms, $P^-_{QCD}= P^-_{CM}+ P^-_{rel}$, the first part depending on the center-of-mass (CM) and the second depending only on the relative ($k^+$-weighted) momenta.

 To see this explicitly , we first write the kinetic energy term as in Eq.~\eqref{eq:Pmin_QCD} for a bare quark as
  \begin{align}\label{eq:Pmn_Q}
    \begin{split}
      & p^-_Q=\frac{\vec p_{\perp,Q}^2+m_Q^2}{p^+_Q}
            =\frac{\vec P_{\perp}^2+m_q^2}{P^+} +\delta P^-
      \;,
    \end{split}
  \end{align}
  in which we define $\delta P^-$ corresponding to the mass counterterm, such that $\delta m=m_Q-m_q= \sqrt{m_q^2+\delta P^- P^+}-m_q$.
 Note that for a single-quark state, the momentu of the quark is the total momentum of the state. 
 The kinetic energy of a $\ket{qg}$ state can be written as 
  \begin{align}\label{eq:pmn_qg}
   \begin{split}
    p^-_{qg}
    =\frac{\vec p_{\perp,q}^2+m_q^2}{p^+_q}&+\frac{\vec p_{\perp,g}^2}{p^+_g}\\
    = & \frac{\vec P_{\perp}^2+m_q^2}{P^+}
    + \underbrace{\frac{\Delta_m^2 + z^2 m_q^2}{z(1-z)P^+}}_{P^-_{rel,qg}} 
    \;.
     \end{split}
  \end{align}
  In the second equation, we switch from the single-particle coordinate to the relative coordinate, by recognizing the CM momentum $\{\vec P_\perp, P^+ \}=\{\vec p_{\perp,q} + \vec p_{\perp,g}, p^+_q+p^+_g \}$, the relative momentum $ \vec \Delta_m =-z \vec p_{\perp,q}+(1-z)\vec p_{\perp,g} $, and the gluon longitudinal momentum fraction $z=p^+_g/P^+$. On a discrete lattice with periodic boundary conditions, such basis transformation is non-trivial, for which we elaborate the detailed implementation in Appendix~\ref{app:basis}.

  Secondly, the interaction term $V_{qg}$ as in Eq.~\eqref{eq:Pmin_QCD} preserves the CM momenta, and it only depends on the relative motion $\{\vec \Delta_m,z \}$, but not $\vec P_\perp$. 

  Thus, overall the Hamiltonian $P^-_{QCD}$ is a sum of
    \begin{align}\label{eq:Pmn_CM_rel}
    P^-_{CM} \equiv \frac{\vec P_{\perp}^2+m_q^2}{P^+}\,,\quad
    P^-_{rel} \equiv  \delta P^-  + P^-_{rel,qg} + V_{qg} \;.
  \end{align}
In this way, the $P^-_{CM}$ matrix is completely diagonal, and the $P^-_{rel} $ matrix block-diagonal in a basis where states have a definite $\vec P_\perp$, as illustrated in Fig.~\ref{fig:matrix_P}, 
  \begin{align}
  \begin{split}
  & \mathcal{P}^-_{CM}=\bigoplus_{i=1}^{\dim_\perp} P^-_{CM (i)} \mathbb{I}_{\dim \mathbb{V}_{rel}} \;,\\
  & \mathcal{P}^-_{rel}=\bigoplus_{i=1}^{\dim_\perp} B_\perp \;.
  \end{split}
  \end{align}
Here, $i$ is the index of $\vec P_\perp$ in the basis, and $P^-_{CM (i)}$ is the value of $P^-_{CM}$ calculated using the $i$-th $\vec P_\perp$. 
The number of basis states with a same $\vec P_\perp$ is 
\begin{equation}  
  \begin{split}
\dim \mathbb{V}_{rel} =& N_c  \times 2 \\
&+ \floor*{K} \times N_c(N_c^2-1) \times 4  \times\dim_\perp \;,
  \end{split}
\end{equation}
in which one can think of $ \floor*{K}\times \dim_\perp $ as the number of different ${z,\vec \Delta_m}$ states.
Note that $\dim_{tot} =\dim_\perp \dim \mathbb{V}_{rel}$.
The $P^-_{rel} $ matrix in each $\vec P_\perp$ subspace is the same, and we denote it by $B_\perp$.
\begin{figure}[t]
\centering 
\includegraphics[width=0.49\textwidth]{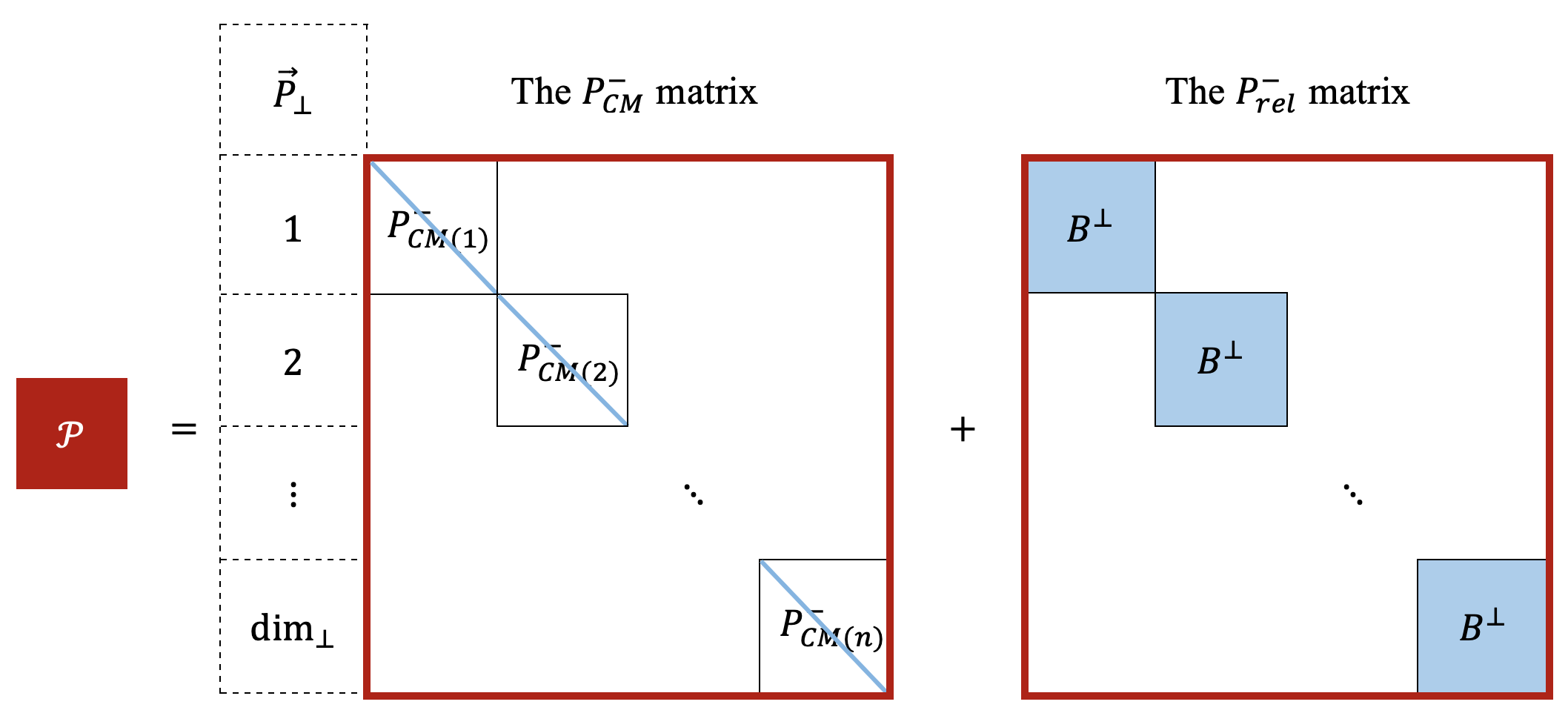}
\caption{The matrix representation of $P^-_{CM}$, $\mathcal{P} $, is written as a summation of the diagonal $P^-_{CM}$ matrix and the block-diagonal $P^-_{rel} $ matrix.}
\label{fig:matrix_P}
\end{figure}

Accordingly, we can write the LFWFs of the eigenstates as 
\begin{align}
  \ket{\phi}=\ket{\phi}_{CM}\otimes \ket{\phi}_{rel}\;,
\end{align}
and the eigenvalue equation then decouples into a CM and a relative parts,
\begin{subequations}
    \begin{align}
        & P^-_{CM}\ket{\phi}_{CM}= P^-_{\phi,CM}\ket{\phi}_{CM} \;,\\
        \label{eq:eigen_rel}
        & P^-_{rel}\ket{\phi}_{rel}= P^-_{\phi,rel} \ket{\phi}_{rel}\;.
      \end{align}
\end{subequations}
Since $P^-_{CM}$ is diagonal in the discrete momentum basis space, we find $\ket{\phi}_{CM}$ immediately as the basis states. We only need to solve Eq.~\eqref{eq:eigen_rel} for the relative wavefunction. Our problem reduces to the diagonalization of $B^\perp$, which is of the dimension $\dim \mathbb{V}_{rel}$.

\item[ii.] \textbf{Color rotation invariance}
The Hamiltonian $P^-_{QCD}$ is SU(3) gauge invariant. Its eigenstates have a definite color index, and are degenerate in each irreducible representation. 
The physical quark states must be in the same color representation as  the bare quark states, $N_c=3$.
The color dimension of the $\ket{qg}$ sectors is $N_c\times(N_c^2-1)=24$, and it can be decomposed into irreducible representations as  $3\oplus \bar 6 \oplus 15$. Since only the triplet quark-gluon states can couple to the quark states through the $V_{qg}$ interaction, we can decouple the Hamiltonian blocks of different color subspaces.

Accordingly, as shown in Fig.~\ref{fig:matrix_Bperp_Bc}, we write $B^\perp$ as, 
\begin{align}
& B^\perp = B^c_3 \oplus B^c_{qg}\,,
\quad  B^c_3 = \bigoplus^3 B^\perp_{q+qg}\;,
\end{align}  
in which $B^c_3$ is the Hamiltonian matrix in the triplet color subspaces. The block $B^\perp_{q+qg} $ is the same for each of the three colors, and its dimension counts the number of quark and quark-gluon states with the same color,
  \begin{align}
    \dim B^\perp_{q+qg} = 2 + \floor*{K} \times 4 \times\dim_\perp \;.
  \end{align}
The block $B^c_{qg}$ is in the $ \bar 6 \oplus 15$ color space, which contains only quark-gluon states, with a dimension 
  \begin{align}
    \dim B^c_{qg} = \floor*{K}\times 21 \times 4\times \dim_\perp\;.
  \end{align}
The color-excited quark-gluon basis states are already $P^-_{QCD}$ eigenstates, and what remains to be diagonalized is $ B^\perp_{q+qg}$.

\begin{figure}[t]
  \centering 
  \includegraphics[width=0.35\textwidth]{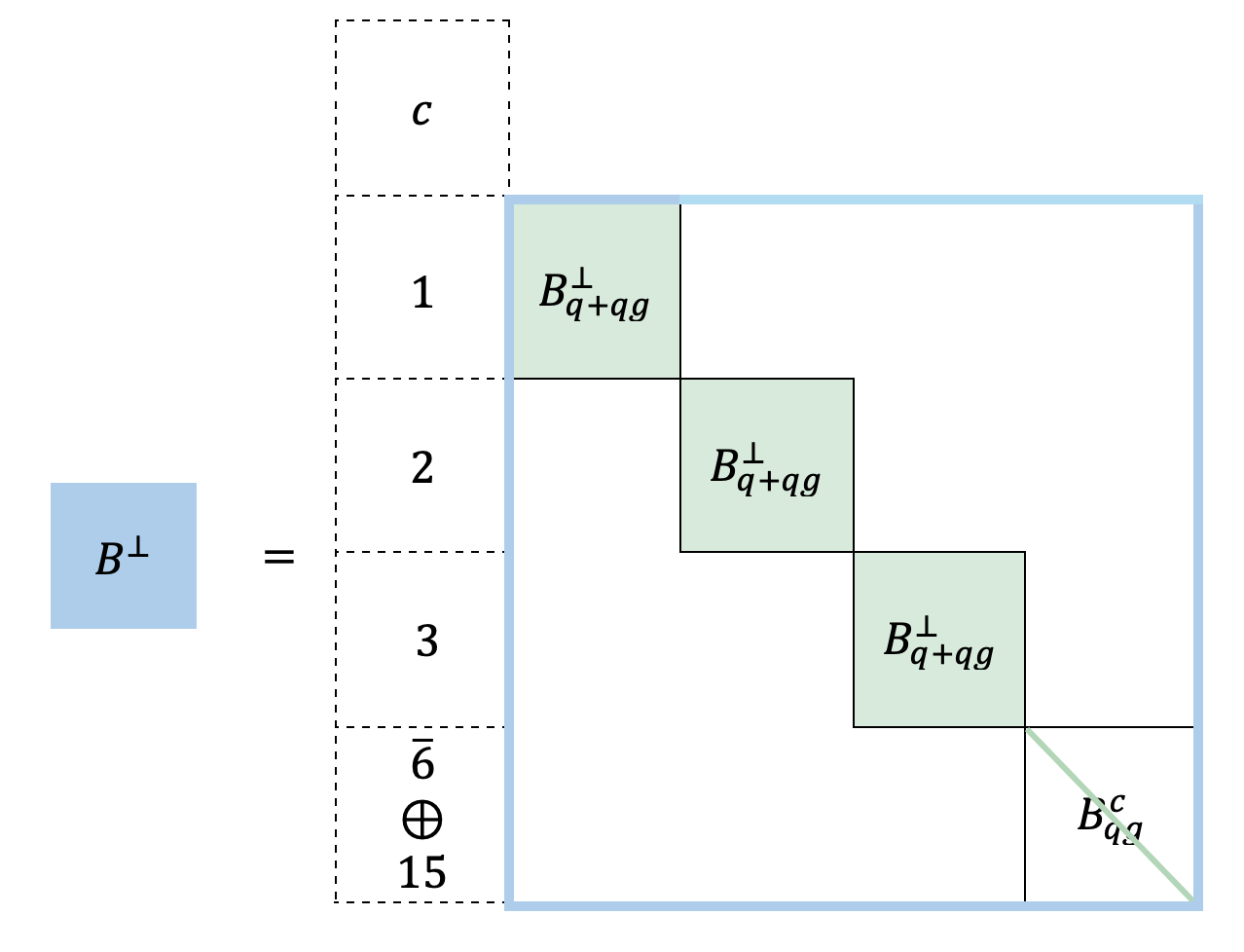}
  \caption{The matrix representation of the Hamiltonian, $B^\perp$, is block-diagonal in the color space.}
  \label{fig:matrix_Bperp_Bc}
\end{figure}

\item[iii.] \textbf{Spin rotation symmetry}
As QCD is rotational invariant, each eigenstate of the Hamiltonian can be classified by its spin projection along the z-axis, the light-front helicity.
The physical quark states can have the same helicity as a bare quark, $\lambda_Q=\pm 1/2$.
Each bare quark helicity state selects the quark-gluon helicity states that can couple to it through the $V_{qg}$ interaction. 
To see the `selection rule` explicitly, let us write out the Hamiltonian matrix $B_{q+qg}^\perp$.

The structure of $B_{q+qg}^\perp$ is represented in Fig.~\ref{fig:Bperp_i_q_qg_new}.
In the top panel, the basis space, which we call the $\beta$ basis, is formed by first having the two bare quark helicity states, \[\beta_1=\{\lambda_Q=1/2\}, \qquad \beta_2=\{\lambda_Q=-1/2\}\;,\] 
and then grouping the quark-gluon basis states according to their relative momentum, $\{z, \vec\Delta_m\}$. 
Each relative momentum group iterates over the four quark-gluon helicity states, 
\begin{align*}
   &\beta_{i(1)}=\{\lambda_q,\lambda_g=1/2,1\}\,,\\
   & \beta_{i(2)}=\{\lambda_q,\lambda_g=1/2,-1\}\;,\\
   &\beta_{i(3)}=\{\lambda_q,\lambda_g=-1/2,1\}\,,\\
   & \beta_{i(4)}=\{\lambda_q,\lambda_g=-1/2,-1\}\;,
\end{align*}
in which $i=1,2,\ldots, \floor*{K}\times \dim_\perp$ is the index of $\{z, \vec\Delta_m\}$ in the basis.
In this basis, the Hamiltonian consists of the diagonal blocks from the kinetic energies of the quark and quark-gluon states
\begin{align}\label{eq:BQ_Bqg}
  \begin{split}
     & B_Q =\delta P^- \mathbb{I}_2\,,\qquad
     B_{qg,i} =  P^-_{qg,rel(i)} \mathbb{I}_4\;,
  \end{split}
\end{align}
and the off-diagonal blocks $\Gamma_i$ and $\Gamma_i^\dagger$ from the gluon emission and absorption vertices, which in the $\beta$-basis are not diagonal: 
\begin{align}\label{eq:Gamma_i}
\begin{array}{c c} 
\begin{array}{c c c c c} 
~~~~~~~~~~~~~~&\beta_{i(1)}~& \beta_{i(2)}~ &\beta_{i(3)}~ &\beta_{i(4)}  \\ 
\end{array} \\
      \Gamma_i
=\dfrac{\mathcal{C}_\Gamma}{P^+}
  \begin{bmatrix}
   u^R~~~
    & v^L~~~
    & w~~~
    & 0\\
    0~~~
    & -w~~~
    & v^R~~~
    &u^L
  \end{bmatrix}
 &  \begin{array}{c c c c}\beta_1\\ \beta_2 \end{array}
\end{array}
\;.
\end{align}
Here, for each $\{z, \vec\Delta_m\}$, we have defined  
\begin{align*}
  \begin{split}
    &u^R \equiv \Delta_m^R/z^{3/2}/(1-z),\qquad
    v^R \equiv \Delta_m^R/z^{3/2},\\
    & w \equiv -m_q z^{1/2}/(1-z)\;.
  \end{split}
\end{align*}
We use the notations that for an arbitrary transverse vector $\vec v_\perp=\{v_x, v_y\}$, $v^R\equiv v_x + iv_y $, $v^L\equiv v_x - iv_y $, and $ v=|\vec v_\perp|=\sqrt{v^R v^L}$. 
The coefficient is constant,
\begin{align}
    \mathcal{C}_\Gamma\equiv g \sqrt{\frac{N_c^2-1}{2 N_c}}\frac{1}{2 L_\perp\sqrt{2\pi K}}\;.
\end{align}

\begin{figure}[t]
  \centering 
  \includegraphics[width=0.42\textwidth]{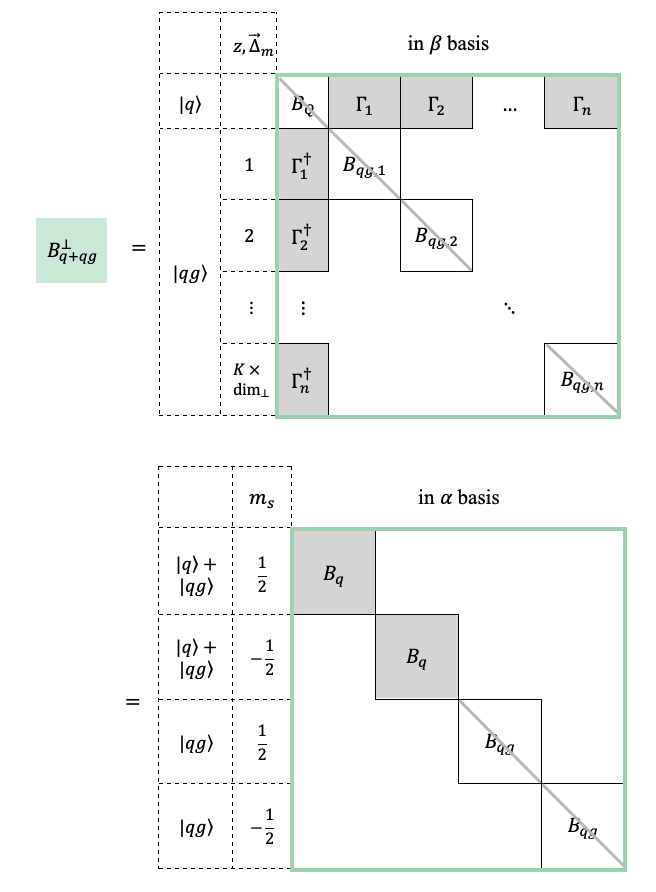}
  \caption{
  Matrix structure of the $B^\perp_{q+qg}$ block. Top panel: in the original $\beta$ basis, with the components expressed in Eqs.~\eqref{eq:BQ_Bqg} and ~\eqref{eq:Gamma_i}; bottom panel: in the new $\alpha$ basis, with the components expressed in Eqs.~\eqref{eq:VQ_scheme} and \eqref{eq:Bqg_alpha_basis}. 
  }
  \label{fig:Bperp_i_q_qg_new}
\end{figure}

In order to decouple the helicity-up and the helicity-down quarks, we make a basis transformation,
\begin{align}\label{eq:CHD_basis_transf}
  \begin{split}
    &\alpha_1=\beta_1,
    \qquad \alpha_2=\beta_2,\\  
    &\begin{pmatrix}
      \alpha_{i(1)}\\
      \alpha_{i(2)}\\
      \alpha_{i(3)}\\
      \alpha_{i(4)}
    \end{pmatrix}
    =
    \frac{1}{\sigma_i }
    \begin{pmatrix}
      u^L  & v^R  & w  & 0\\
      0 & -w  & v^L  & u^R  \\
    -w  u^L /u  & 0 & u  & -v^R  u^R /u  \\
      -v^L  u^L /u   & u   & 0 & w  u^R /u 
    \end{pmatrix}
    \begin{pmatrix}
      \beta_{i(1)}\\
      \beta_{i(2)}\\
      \beta_{i(3)}\\
      \beta_{i(4)}
    \end{pmatrix}
    \;,
  \end{split}
\end{align}
in which $\sigma_i\equiv \sqrt{u^2+v^2+w^2}$.\footnote{
For the basis transformation at $\Delta_m=0$, a direct application of Eq.~\eqref{eq:CHD_basis_transf} would run into indeterminate as the limit $\lim_{x\to 0, y\to 0}\frac{x+i y}{\sqrt{x^2+y^2}}$ is not well defined. Since our purpose of changing the basis is to separate the coupled and the uncoupled quark-gluon states, we can simply define the basis transformation at $\Delta_m=0$ as
\begin{align*}
    \begin{pmatrix}
      \alpha_{i(1)}\\
      \alpha_{i(2)}\\
      \alpha_{i(3)}\\
      \alpha_{i(4)}
    \end{pmatrix}_{\Delta_m=0 }
    =
    \begin{pmatrix}
      0 & 0 & -1 & 0\\
      0 & 1 & 0 & 0 \\
     1 & 0 & 0 & 0 \\
     0  & 0  & 0 & -1
    \end{pmatrix}
        \begin{pmatrix}
      \beta_{i(1)}\\
      \beta_{i(2)}\\
      \beta_{i(3)}\\
      \beta_{i(4)}
    \end{pmatrix}
    \;.
\end{align*}
}

In the new basis, the $\Gamma_i$ reads as
\begin{align}\label{eq:Gamma_i_new_basis}
\begin{array}{c c} 
    \begin{array}{c c c c c} 
    ~~~~~~~~~~~~~& \alpha_{i(1)}& \alpha_{i(2)} &\alpha_{i(3)}&\alpha_{i(4)}  \\ 
    \end{array} 
    & \\
    \Gamma_i
    =\dfrac{\mathcal{C}_\Gamma}{P^+}
    \begin{bmatrix}
    ~~\sigma_i~~
    & 0~~~~
    & 0~~~~
    & 0~~\\
    ~~0~~
    & \sigma_i~~~~
    & 0~~~~
    &0~~
    \end{bmatrix}
    &
    \begin{array}{c c c c} \alpha_1\\ \alpha_2 \end{array} 
\end{array}\;.
\end{align}
It indicates that only the $\alpha_{i(1)}$ states can couple to the helicity-up quark, and the $\alpha_{i(2)}$ to the helicity-down quark, whereas $\alpha_{i(3)}$ and $\alpha_{i(4)}$ states stay uncoupled.
 We will refer to $\{\alpha_i \}$ the color-helicity-degenerate (CHD) basis. One way to interpret the new basis is to consider the total spin projection.

The total angular momentum projection sums over the spin projection and the orbital angular momentum projection, namely $m_j=m_s+m_l$. Since we have already factorized out the CM momentum, the orbital angular momentum of the motion of the CM of the quark or the quark-gluon system with respect to the (arbitrary) origin of the coordinate system is no longer included here. For a single particle state, $m_j$ is its spin projection, while for a quark-gluon state, $m_j$ also includes of the orbital angular momentum projection of their relative motion denoted as $l_\Delta$,
\begin{align}\label{eq:ms}
    m_j=
    \begin{cases}
    \lambda_Q, &\ket{q}\\
    \lambda_q + \lambda_g + l_\Delta, &\ket{qg}
    \end{cases}
\end{align}
In the continuum, $l_\Delta$ is the eigenvalue of the relative orbital angular momentum operator, which in momentum space is $\hat l_\Delta=i\diff/\diff \theta_\Delta$ with $\theta_\Delta=\arg \vec\Delta_m$. As such, an $l_\Delta$ eigenstate has an angular dependence of $e^{ i l_\Delta \theta_\Delta}$ in its wavefunction.
A quark-gluon state can only couple to a bare quark if they have the same $m_j$. 
For example, a quark-gluon state with helicity $\{\lambda_q, \lambda_g\}=\{1/2,1\}$ can couple to a helicity-up quark if its $l_\Delta=-1$. 

On a discrete momentum basis space, an $l_\Delta$ eigenstate can be approximated by combining different momentum states with $e^{ i l_\Delta \theta_\Delta}$ as the coefficient. 
In forming the $\alpha_{i(1)}$ state, as in Eq.~\eqref{eq:CHD_basis_transf}, the $\beta_{i(1)}$ state has $\Delta^L= \Delta_m e^{ -i \theta_\Delta}$ in its coefficient, which we recognize as one unit of angular momentum $l_\Delta=-1$. Similarly, $\Delta^R= \Delta_m e^{ i \theta_\Delta}$ multiplying the $\beta_{i(2)}$ state adds one unit of angular momentum $l_\Delta=1$. The  $\beta_{i(3)}$ state does not have any angular dependence in its coefficient, so it can make into an angular momentum state of $l_\Delta=0$. Consequently, we can say that the $\alpha_{i(1)}$ state is a valid ingredient in making an $m_j=1/2$ state.
In the same way, one can find $\alpha_{i(2)}$ valid in making an $m_j=-1/2$ state, $\alpha_{i(3)}$ for $m_j=1/2$, and $\alpha_{i(4)}$ for $m_j=-1/2$. 
An interpretation of the $\alpha_{i(3)} $ and $\alpha_{i(4)} $ states is that despite having the correct $m_j$, they combine different helicity states in such a way that their total coupling to the bare quark vanishes.

Let us now group the basis states by their corresponding $m_s$, firstly $\{\alpha_1,\alpha_{1(1)},\alpha_{2(1)},\ldots, \alpha_{\floor*{K}\dim_\perp (1)} \}$, secondly $\{\alpha_2,\alpha_{1(2)},\alpha_{2(2)},\ldots, \alpha_{\floor*{K}\dim_\perp (2)} \}$, then the $\alpha_{i(3)} $s and the $\alpha_{i(4)} $s. 
We call such basis space the $\alpha$ basis, in which the Hamiltonian is block diagonal, as shown in the bottom panel of Fig.~\ref{fig:Bperp_i_q_qg_new}.
The two blocks in the first two $m_s$-subspaces are the same, explicitly,
\begin{align}\label{eq:VQ_scheme}
  B_q= 
  \begin{pmatrix}
    \delta P^- & V_1  & V_2  &\ldots &V_n\\
   V_1 & P^-_{rel,qg(1)} & & &\\
    V_2 &  &P^-_{rel,qg(2)}& &\\
   \vdots & & & \ddots&\\
   V_n &  & & &P^-_{rel,qg(n)}
  \end{pmatrix}
  \;,
\end{align}
in which $V_i=\mathcal{C}_\Gamma\sigma_i/P^+$ as obtained according to Eq.~\eqref{eq:Gamma_i_new_basis}, and 
  \begin{align}
    \dim B_q = 1+n= 1 + \floor*{K}\times\dim_\perp \;.
  \end{align}
The Hamiltonian in the $\alpha_{i(3)} $ and $\alpha_{i(4)} $ subspaces are the same and diagonal, 
\begin{align}\label{eq:Bqg_alpha_basis}
 B_{qg} = \diag\{ 
    P^-_{qg,rel(1)}, 
    P^-_{qg,rel(2)},
    \cdots,
    P^-_{qg,rel(K\dim_\perp)} \} \;.
\end{align}
Accordingly, those quark-gluon basis states are $P^-_{QCD}$ eigenstates, with their kinetic energy being the eigenvalues. 
What remains to be diagonalized is $ B_q $.

\item[iv.] \textbf{Discrete rotational symmetry}
Having identified the ``ingredients" to make the $l_\Delta$ and thus the $m_s$ eigenstates, we can further assemble those ``ingredients" into angular momentum states by utilizing the rotational symmetry in the discrete lattice space.

In constructing the $\alpha$ basis, the basis transformation as defined Eq.~\eqref{eq:CHD_basis_transf} has already taken into account the required $l_\Delta$ value to make the correct $m_s$ state as in Eq.~\eqref{eq:ms}, so any angular excitation introduced by combining those basis states will be an additional $l_\Delta'$. 
Therefore by angular momentum conservation, only the $l_\Delta'=0$ states can couple to a quark, and the $l_\Delta'\neq 0$ states will remain uncoupled. We will see this explicitly by constructing the $l_\Delta'$ states in the following.

There are quark-gluon states with the same $\{z,\Delta_m\}$, but with different $\vec \Delta_m$. We call those states $\Delta_m$-degenerate states. For example, the four basis states with a same $z$ but $\vec \Delta_m=\{\pm 1,\pm 2\}d_p$, where $d_p=\pi/L_\perp$ is the transverse momentum unit, are $\Delta_m$-degenerate states with degeneracy $d=4$. This  degeneracy is what remains on the lattice from the rotational symmetry in the continuum.  We can choose the degenerate states as eigenstates of a discrete angular momentum quantum number $l_\Delta'$, by taking the appropriate linear combinations of the $\Delta_m$-degenerate states
\begin{align} 
  \psi_{l_\Delta'}  = \frac{1}{\sqrt{d}}\sum_{k=1}^d e^{i l_\Delta' k \frac{2\pi}{d}}  \alpha^k\,,
  \quad
  l_\Delta'=0,\pm 1, \ldots, \pm \frac{d}{2}\;.
\end{align}
Here, $\alpha^k$ are $\Delta_m$-degenerate basis states in the $\alpha$ basis, and the index $k=1,2,\ldots,d$ runs over the degenerate subspace. Note that the $l_\Delta'=\pm d/2$ states are the same, so there are in total $d$ number of different $l_\Delta'$ states.
Most often, in the discrete lattice space, we have $d=4$ and $d=8$, and there is a limited number of angular excitations that we can access at each $\Delta_m$.\footnote{
A potential way to overcome such limitation and represent higher $\ell$($l_\Delta'$ in the continuum) states would be to combine multiple groups of $\Delta_m$-degenerate states that are close in $\Delta_m$, but we will not dive into that topic as our main focus here is the physical dressed quark states. 
}

Since the matrix elements of both $V_i$ and the $P^-_{qg,rel(i)}$ depend only on $\{z,\Delta_m\}$ but not the vector $\vec \Delta_m $, the matrix element between the bare quark and an $ l_\Delta'$ state is nonzero only if $l_\Delta'=0$. One can see this easily by applying the relation
\begin{align} 
\frac{1}{d}\sum_{k=1}^d e^{i l_\Delta' k \frac{2\pi}{d}}= \delta_{l_\Delta',0}\;.
\end{align}
As a result, the $ l_\Delta'\neq 0$ state is automatically a $ P^-_{QCD}$ eigenstate, with its kinetic energy being the eigenvalue. In other words, the coupling of an angularly excited quark-gluon states ($l_\Delta' \neq 0$) to the bare quark is forbidden by angular momentum conservation.
On the other hand, the $l_\Delta'=0$ state will mix with bare quarks in the physical quark states, which are yet to be solved by matrix diagonalization.

\end{itemize}

We have identified the symmetries of the problem Hamiltonian in the momentum, color, and helicity spaces, and have alongside reduced the dimension of the matrix to be diagonalized from $\dim_{tot}$ to $\dim B_q$. 
We made the discussions in terms of $ P^-_{QCD}$, which later will facilitate the thinking of shifting to the time evolution Hamiltonian, $ P_{QCD}^- + V_{\mathcal{A}}(x^+)$ as in Eq.~\eqref{eq:P_full}.
The procedure and conclusion, however, apply equivalently to the light-cone Hamiltonian as in Eq.~\eqref{eq:TISE_HLC}, 
\begin{align}\label{eq:HLC_mq_rel}
\mathrm H_{LC}=m_q^2
+P^+ P^-_{rel} \;,
\end{align} 
which is manifestly boost invariant.

\subsubsection{Analytical solution}\label{sec:analytical_solution}
\begin{figure*}[t!]
  \centering 
  \subfigure[
    Mass counterterm
    \label{fig:dm_mq}
  ]{ \includegraphics[width=0.4\textwidth]{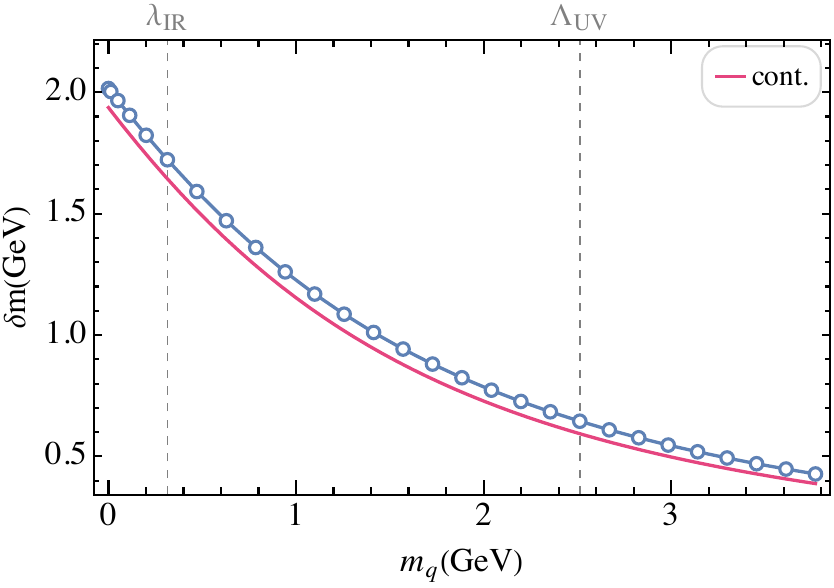} }\qquad
  \subfigure[
    Wavefunction renormalization factor
    \label{fig:Z2_mq}
  ]{ \includegraphics[width=0.4\textwidth]{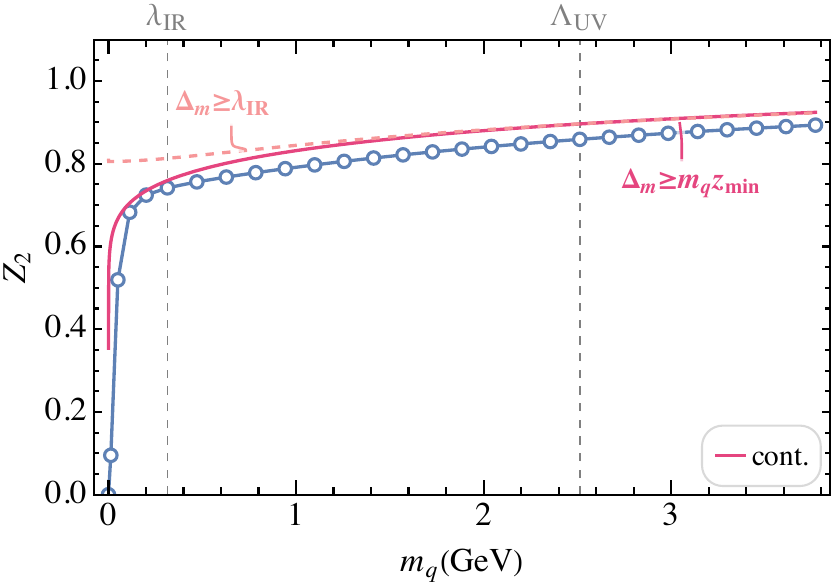} }
  \caption{Panel (a) shows the mass counterterm $\delta m$ as a function of the physical quark mass $m_q$. The red solid line is the result in the continuum according to Eq.~\eqref{eq:dm_1loop} with $\Lambda_\perp=\sqrt{2}\Lambda_{UV}$. Panel (b) shows the $Z_2$ factor at different quark mass. Both results are calculated at $N_\perp=8$, $K=8.5$, and $L_\perp=10~\GeV^{-1}$, and the basis IR and UV cutoffs are labeled with dashed lines. The red solid(dashed) line is the result in the continuum according to Eqs.~\eqref{eq:q_LFWF_qg_anal_cont} and \eqref{eq:unint_PQ_anal}, taking $\Lambda_\perp=\sqrt{2}\Lambda_{UV}$ and $\lambda_\perp=m_q z_{min}(\lambda_{IR})$ as indicated in the figure.}
  \label{fig:MC_Z2_mq}
\end{figure*}
\begin{figure*}[htp!]
  \centering 
  \subfigure[ Transverse basis dependence \label{fig:dmT_Nperp}]{
   \includegraphics[width=0.4\textwidth]{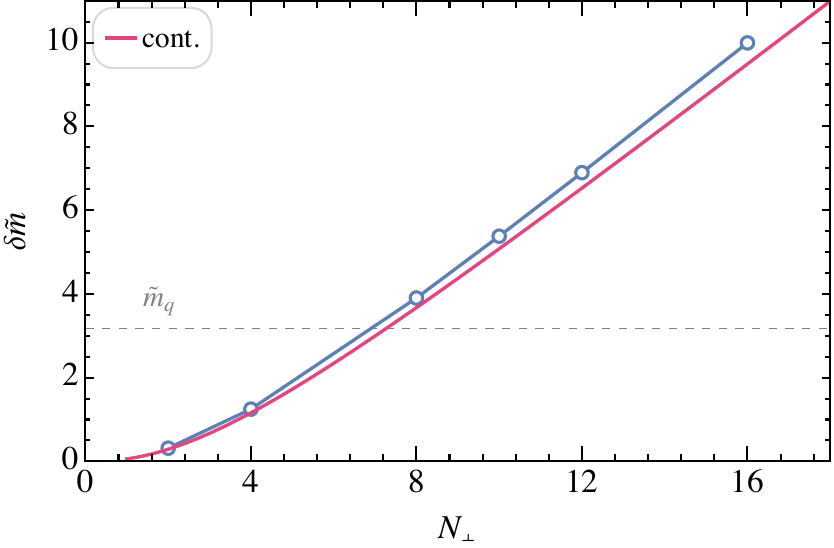} \qquad
   \includegraphics[width=0.4\textwidth]{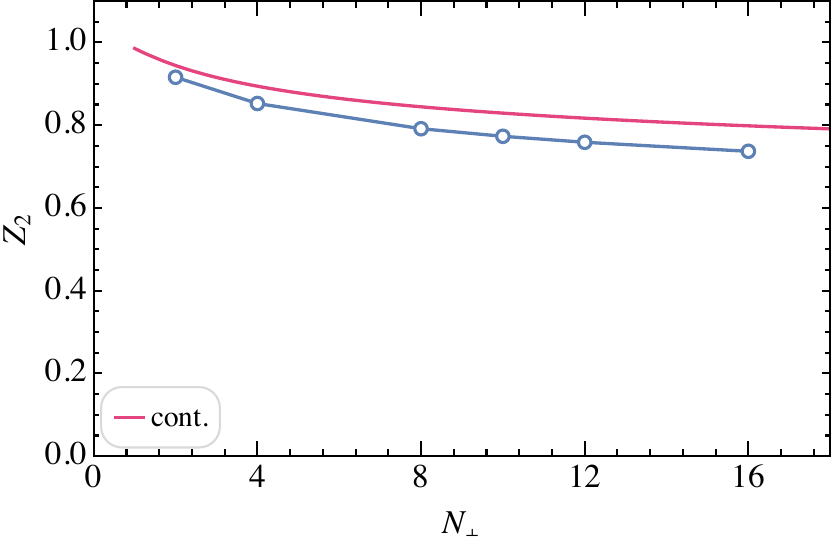} 
   }
   \subfigure[ Longitudinal basis dependence \label{fig:dmT_K}]{
    \includegraphics[width=0.4\textwidth]{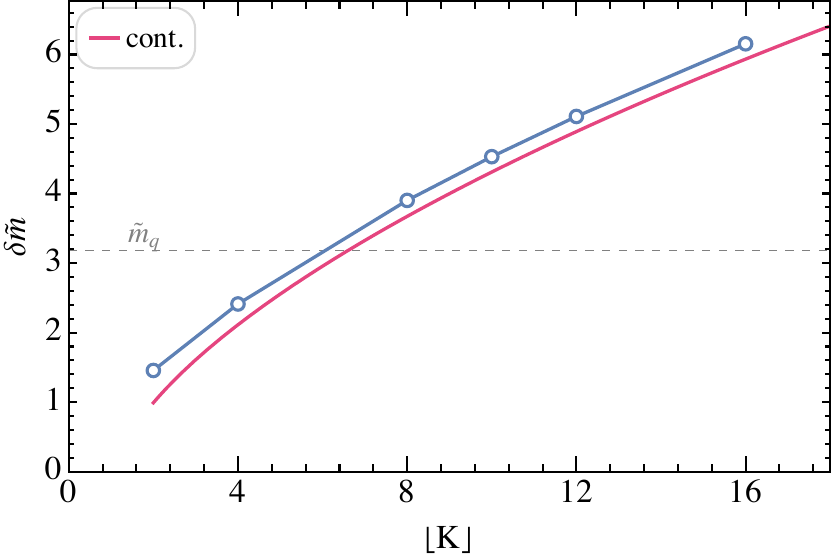} \qquad
    \includegraphics[width=0.4\textwidth]{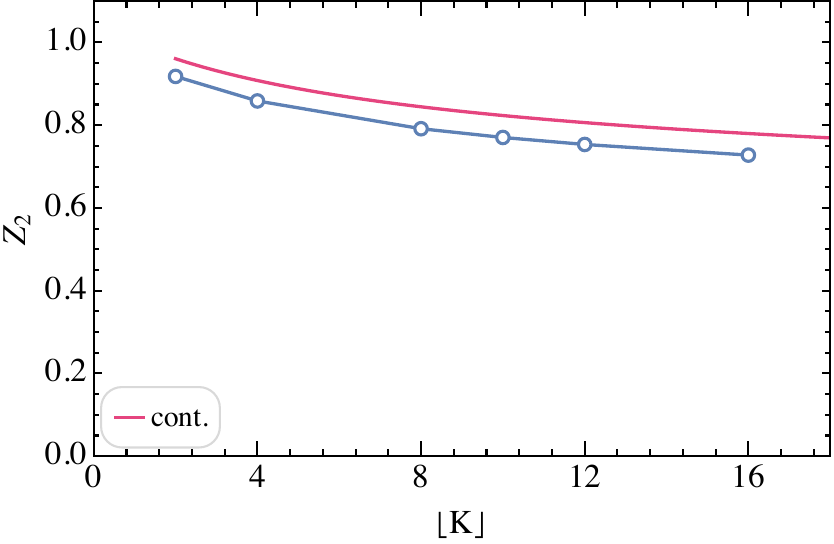} 
    }
  \caption{ 
  Dependence of the mass counterterm $\delta m$ and the $Z_2$ factor on the basis parameters: (a) the transverse lattice sites $N_\perp$ (at $K=8.5$) and (b) the longitudinal resolution $\floor{K}$ (at $N_\perp=8$). The red solid lines are the result in the continuum, $ \delta \tilde m$ calculated according to Eq.~\eqref{eq:dm_1loop} with $\Lambda_\perp=\sqrt{2}\Lambda_{UV}$, and  $ Z_2$ calculated according Eqs.~\eqref{eq:q_LFWF_qg_anal_cont} and \eqref{eq:unint_PQ_anal}, taking $\Lambda_\perp=\sqrt{2}\Lambda_{UV}$ and $\lambda_\perp=\lambda_{IR}$. All results are calculated at $\tilde m_q=3.18$, as labeled with dashed lines in the left panels.
  }
  \label{fig:MC_Z2_Nperp_K}
\end{figure*}
Now that we have understood the symmetries of the discretized QCD Hamiltonian, let us come
 back to the eigenvalue equation of the physical quarks, Eq.~\eqref{eq:HLC_dm}. We shall diagonalize the relative Hamiltonian $\mathrm  H_{rel}= P^+ P^-_{rel} $ and determine the mass counterterm $\delta m$ such that the ground state eigenvalue is $0$. 
In a discrete momentum space, it is helpful to think of physical quantities in terms of the transverse momentum unit $d_p=\pi/L_\perp$.
We define the dimensionless light-cone Hamiltonian as $\tilde{\mathrm  H}_{LC}=\mathrm H_{LC} /d_p^2$. 
Similarly, we have $\tilde{\mathrm  H}_{rel}=\mathrm H_{rel} /d_p^2$, $\tilde\Delta_m=\Delta_m/d_p $, $\tilde m_q=m_q/d_p $, and $ \delta \tilde m= \delta  m/d_p $. 

It is straightforward to obtain the part of the Hamiltonian corresponding to the relative motion and gluon emission and absorption $\tilde{\mathrm  H}_{rel}$ from the $B_q$ matrix as in Eq.~\eqref{eq:VQ_scheme},
\begin{align}
  \tilde{\mathrm  H}_{rel}= 
  \begin{pmatrix}
    \delta  \tilde{\mathrm  H} &  \tilde{V}_1  & \tilde{V}_2  &\ldots &\tilde{V}_n\\
   \tilde{V}_1 & \tilde{D}_1 & & &\\
    \tilde{V}_2 &  &\tilde{D}_2& &\\
   \vdots & & & \ddots&\\
   \tilde{V}_n &  & & &\tilde{D}_n
  \end{pmatrix}
  \;,
\end{align}
in which, with the $i$-th $\{z, \vec\Delta_m\}$,  
\begin{align}
  \begin{split}
  &\tilde{D}_i=\frac{\tilde\Delta_m^2 + z^2 \tilde m_q^2}{z(1-z)}\;,\\
  &\tilde{V}_i= g \sqrt{\frac{N_c^2-1}{2 N_c}}
  \frac{
    \sqrt{ [1+(1-z)^2]\tilde\Delta_m^2 + z^4 \tilde m_q^2 }
  }{(2\pi)^{3/2}\sqrt{ K}  z^{3/2}(1-z)} \;.
  \end{split}
  \end{align}
The mass counterterm is related to $\delta \tilde{\mathrm  H} $ by
\begin{align}
\delta \tilde{m}=\sqrt{ \delta \tilde{\mathrm  H} + \tilde{m}_q^2 }-\tilde{m}_q\;.
\end{align}

Parameterizing the eigenvector as
\begin{align}\label{eq:phi_lambda}
 \ket{\phi_\lambda}=[a, b_1, b_2,\ldots, b_n]^T
\end{align}
 for an eigenstate with eigenvalue $\lambda$, we can write the eigenvalue equation as
\begin{align}
  \begin{split}
    &\delta \tilde{\mathrm  H}~ a+\sum_{i=1}^n \tilde{V}_i b_i =\lambda a \;,\\
    & \tilde{V}_i a +\tilde{D}_i b_i=\lambda b_i\;.
  \end{split}
\end{align}
The state is normalized, such that
\begin{align}
  |a|^2+\sum_{i=1}^n |b_i|^2=1\;.
\end{align}
To carry out the mass renormalization, we already know the eigenvalue, $\lambda=0 $ (or a non-zero value determined by virtuality) for an on(off)-shell physical quark, so we can write out the eigenstate and the counterterm directly,
\begin{align}\label{eq:bi_dPmn}
  \begin{split}
    &b_i= \frac{\tilde{V}_i}{\lambda-\tilde{D}_i} a \;,\\
    &\delta \tilde{\mathrm  H}=-\sum_{i=1}^n  \frac{ |\tilde{V}_i|^2}{\lambda-\tilde{D}_i} +\lambda \;.
  \end{split}
\end{align}

The unnormalized LFWF of the quark-gluon component is therefore
\begin{align}\label{eq:q_LFWF_qg_anal}
  \begin{split}
  \phi_{qg}&(z,\vec \Delta_m) =\frac{b_i}{a}\\
  = & g \sqrt{\frac{N_c^2-1}{2 N_c}} \frac{1}{(2\pi)^{3/2}\sqrt{ zK} 
  }
  \frac{
    \sqrt{ [1+(1-z)^2]\tilde\Delta_m^2 + z^4 \tilde m_q^2 }
  }{
\lambda z(1-z)
-\tilde\Delta_m^2 - z^2 \tilde m_q^2
  }\;.
\end{split}
\end{align} 

The normalization condition gives the probability of the quark component, $P_{\ket{q}}=|a|^2$, which is also the state renormalization coefficient $Z_2$,
\begin{align}\label{eq:unint_PQ_anal}
  \begin{split}
  Z_2 
    =\bigg\{
       1+ \sum_{i=1}^n \left|\phi_{qg}(z,\vec \Delta_m) \right|^2
       \bigg\}^{-1}
       \;.
  \end{split}
\end{align}
Qualitatively, the quark component increases as the physical quark mass increases. 
The above result agrees with the dressed quark wave function according to the light-cone perturbation theory at leading order, e.g., Ref.~\cite{Marquet:2007vb,Chirilli:2012jd}.
The mass counterterm obtained according to Eq.~\eqref{eq:bi_dPmn}, in combination with instantaneous corrections, agrees with the equal time counterterm at the next-to-next-to-leading order, which we show in Appendix~\ref{app:inst}.

Knowing the mass counterterm, we can then diagonalize the Hamiltonian matrix and obtain the remaining eigenstates and eigenvalues.
For a given set of $\{ g, N_\perp, K, \tilde m_q \}$, the mass counterterm $\delta \tilde{m} $, the eigenstates, and eigenvalues are invariant upon changing other parameters, such as $m_q$ and $L_\perp$.
However, not all parameters can yield a physically meaningful result.
The dressed quark LFWF is infrared (IR) divergent as can be seen from Eq.~\eqref{eq:q_LFWF_qg_anal}.
We want the mass to be large enough to regulate this IR divergence,
otherwise, the $\ket{qg}$ components will be unnaturally large because the momentum resolution in the IR $1/d_p$ will not be enough to accurately represent the physics at small $\Delta_m$.
On the other hand, the quark mass should not be larger than the largest momentum in the basis, i.e., $m_q< N_\perp d_p=\Lambda_{UV}$, otherwise the quark can barely be dressed by the $\ket{qg}$ components.
In summary, a reasonable basis should satisfy
\begin{align}
  1 < \tilde m_q < N_\perp\;.
\end{align}

In Fig.~\ref{fig:MC_Z2_mq}, we plot the mass counterterm and $Z_2$ as functions of the physical quark mass $m_q$, given $N_\perp=8$, $K=8.5$, and $L_\perp=10~\GeV^{-1}$. 
In the continuum, the mass counterterm depends on the IR and UV cutoffs. In our discrete basis space, it depends on the basis parameters $K$ and $N_\perp$, as shown in Fig.~\ref{fig:MC_Z2_Nperp_K}. The results of $\delta m$ and $Z_2$ in the continuum are obtained by replacing the summations of $z$ and $\vec \Delta_m$ by integrals, according to Eqs.~\eqref{eq:dm_1loop} and ~\eqref{eq:q_LFWF_qg_anal_cont}, with the integral boundaries matching with the cutoffs in the discrete space.

\subsubsection{Mass spectrum and wavefunctions}
\label{subsec:spectrum}
Having found the mass counterterm and the ground state wavefunction analytically according to Eqs.~\eqref{eq:bi_dPmn} and~\eqref{eq:q_LFWF_qg_anal}, we can solve the full spectrum and the excited states by diagonalizing the renormalized Hamiltonian. We will also refer to the eigenstates as coherent states. 
The LFWF solved by diagonalizing the Hamiltonian is free up to an overall phase. We choose the convention that the quark wavefunction [i.e., $a>0$ in Eq.~\eqref{eq:phi_lambda}] is positive in each dressed state. 

\begin{figure*}[htp!]
  \subfigure[Decomposition in the CM transverse momentum space \label{fig:spectrum_a}]{
 \includegraphics[width=0.48\textwidth]{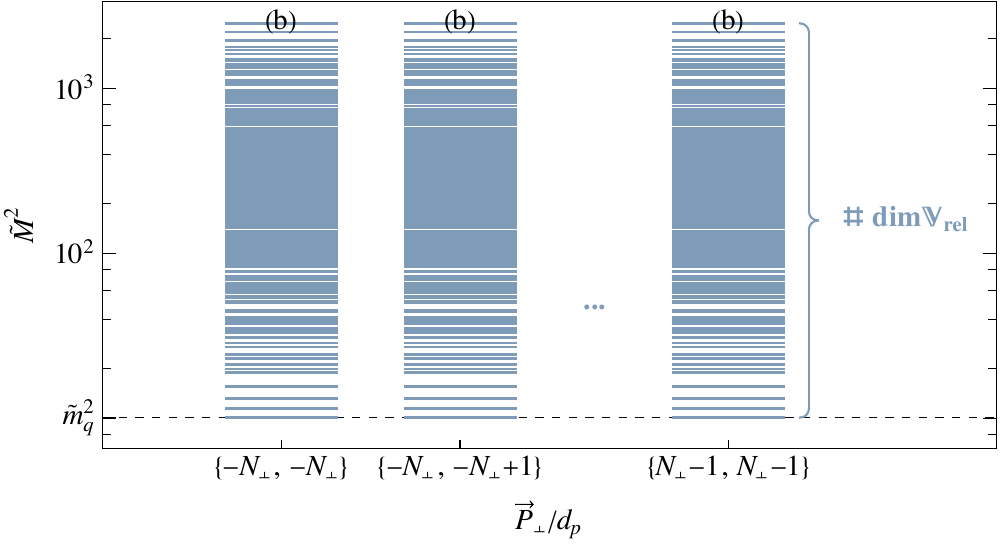}
  }
  \subfigure[Decomposition in color space \label{fig:spectrum_b}]{
  \includegraphics[width=0.48\textwidth]{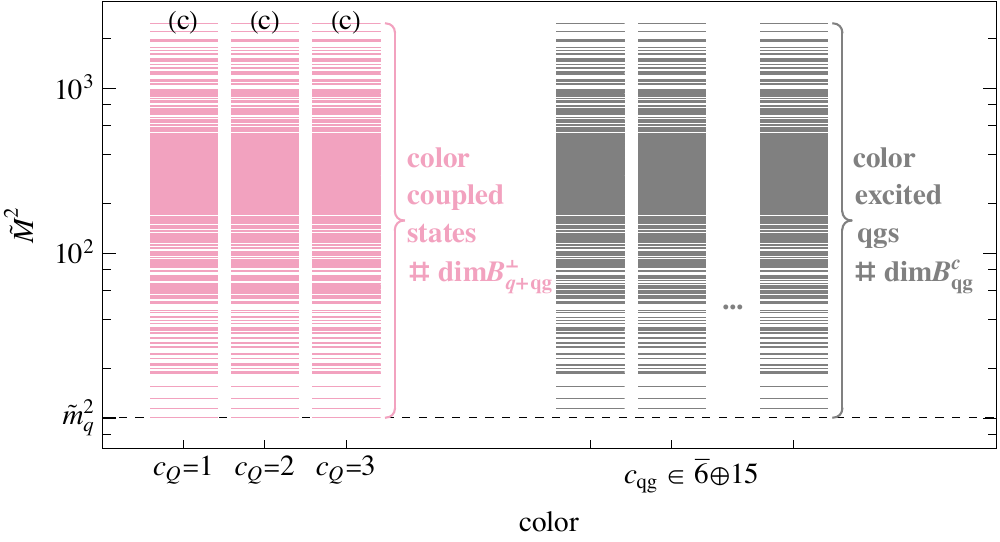}
  }
  \subfigure[Decomposition in  helicity-associated space \label{fig:spectrum_c}]{
  \includegraphics[width=0.48\textwidth]{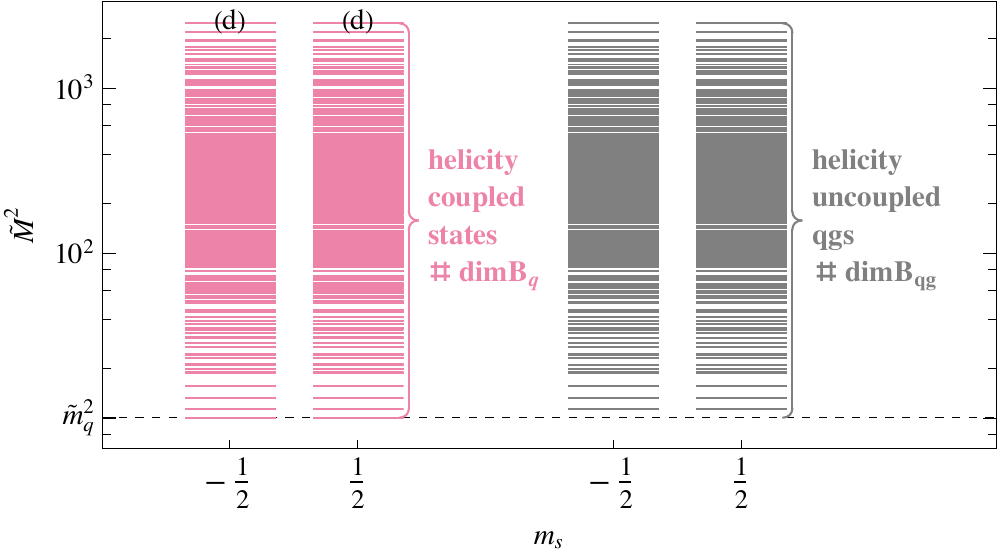}
  }
  \subfigure[Dressed and uncoupled states \label{fig:spectrum_d}]{
  \includegraphics[width=0.48\textwidth]{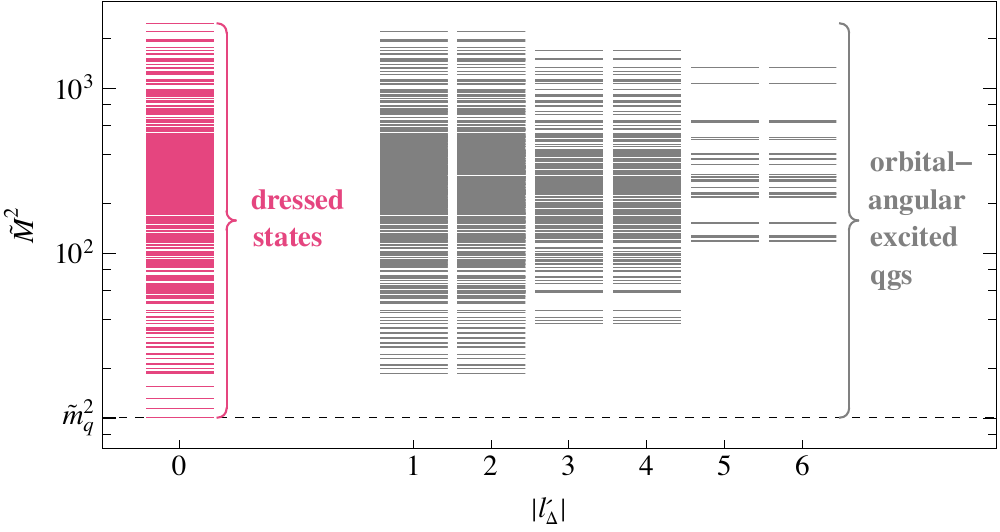}
  }
  \caption{
  Mass spectrum of the coherent $\ket{q}+\ket{qg}$ states solved from diagonalizing the QCD Hamiltonian in the basis space. From panel (a) to (d), the spectrum is decomposed sequentially in a hierarchy of symmetries as introduced in the text. In panel (a), the label ``(b)'' on each column indicates that the contains the whole spectrum presented in (b), and likewise for other labels.
  Basis parameters: $N_\perp=\floor*{K}=8$, $\tilde m_q=3.18$.
  }
  \label{fig:spectrum}
\end{figure*}

To exemplify, we present an obtained mass spectrum of the QCD eigenstates in the $\ket{q}+\ket{qg}$ space in Fig.~\ref{fig:spectrum}. We renormalize the Hamiltonian to match an on-shell physical quark, i.e., taking $\lambda=0$ in Eq.~\eqref{eq:bi_dPmn}.
The calculation is done in the basis of $N_\perp=8$, $K=8.5$, and $\tilde m_q=3.18$; in the case of $L_\perp=10 \GeV^{-1}$, the quark mass is $m_q=1 \GeV$, as in Fig.~\ref{fig:MC_Z2_mq}. The obtained mass counterterm is $\delta \tilde m=3.90124$.
From panels \ref{fig:spectrum_a} to \ref{fig:spectrum_d}, the spectrum is decomposed sequentially in a hierarchy of symmetries as introduced in Sec.~\ref{sec:symmetry}. 
At each CM transverse momentum sections in panel \ref{fig:spectrum_a}, the invariant mass spectrum is the same with that shown in panel \ref{fig:spectrum_b}. Within \ref{fig:spectrum_b}, the mass spectrum in each one of the color triplet subspace is the same with that shown in panel \ref{fig:spectrum_c}, and the states in the $\bar 6\oplus 15$ subspace are pure quark-gluon states. In \ref{fig:spectrum_c}, the left two columns contain the helicity coupled states, i.e., eigenstates of the $B_q$ block, which is further decomposed in panel \ref{fig:spectrum_d}, and the right two columns contain the pure quark-gluon states that can not couple to the quark due to helicity-associated reason explained in Sec.~\ref{sec:symmetry}. Finally in \ref{fig:spectrum_d}, we partition the remaining states into the dressed states (as in the leftmost column) and orbital-angular excited quark-gluon states. 
The ground dressed state, which is also the ground state of the full spectrum, represents the on-shell physical quark. This state predominantly consists of a single-quark component, dressed by quark-gluon contributions, making it a dressed quark state.
The excited dressed states, in a similarly way of thinking, are dressed quark-gluon states, a dominant single-quark state (i.e., an $l_\Delta'=0$ state) dressed by a single-quark component. We have grouped the orbital-angular excited quark-gluon states according to their $|l_\Delta'|$ values in the right columns.

\begin{figure*}[htp!]
  \centering 
  \subfigure[ The dressed quark \label{fig:LFWF_dressed_q}]{
   \includegraphics[width=\textwidth]{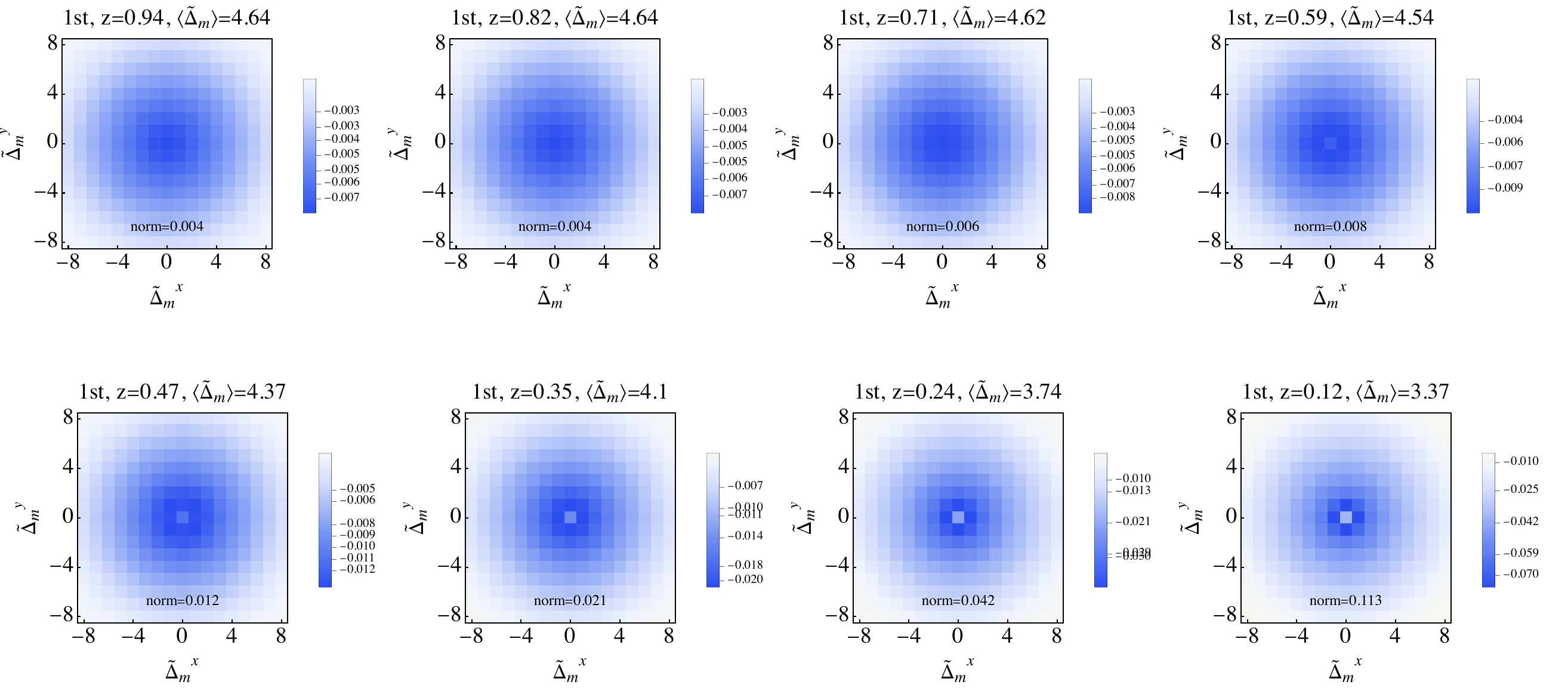} 
   }
   \subfigure[ A dressed quark-gluon state \label{fig:LFWF_dressed_qg}]{
    \includegraphics[width=\textwidth]{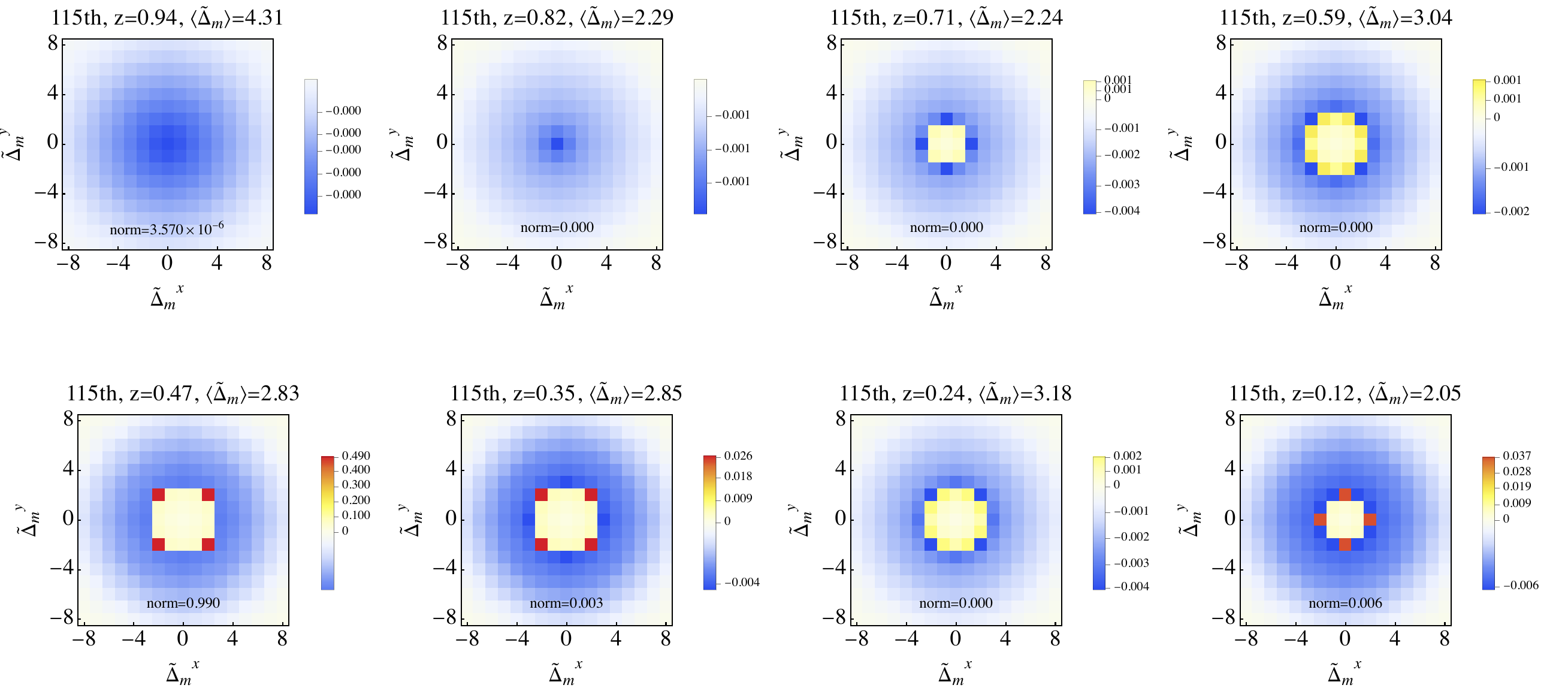} 
    }
  \caption{ The LFWFs of the QCD eigenstates plotted in the $\{z,\vec \Delta_m\}$ space, (a) the dressed quark state and (b) a dressed quark-gluon state. Basis parameters are the same as those in Fig.~\ref{fig:spectrum}. The rank of the state as indicated in the plot is the rank among the eigenstates in the increasing order of the invariant mass, as in Fig.~\ref{fig:spectrum_d}.
  }
  \label{fig:LFWFs_dressed}
\end{figure*}

\begin{figure*}[htp!]
  \centering 
  \subfigure[ A set of $d=4$ degenerate pure quark-gluon states \label{fig:LFWF_pure_d4}]{
   \includegraphics[width=\textwidth]{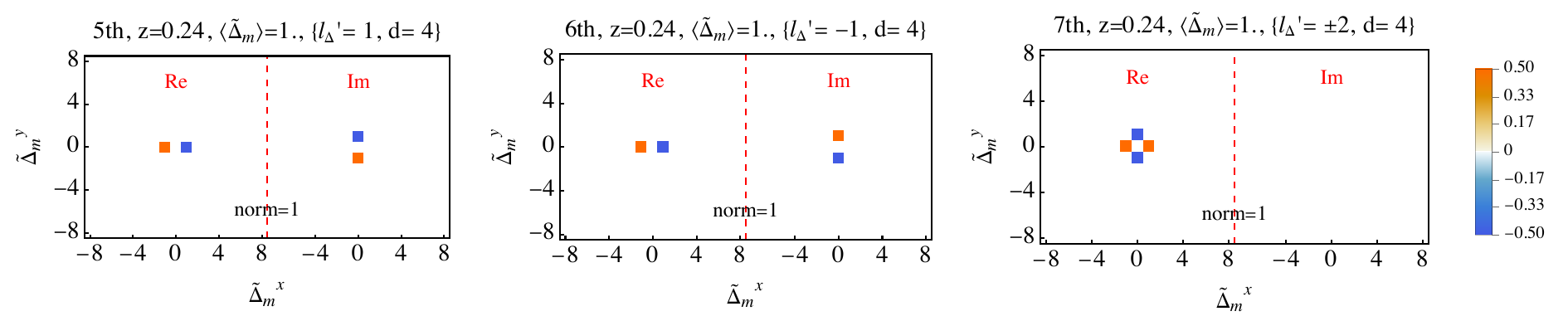} 
   }
   \begin{minipage}{\textwidth}
    \flushleft
    \vspace{10pt}
    \includegraphics[width=.93\textwidth]{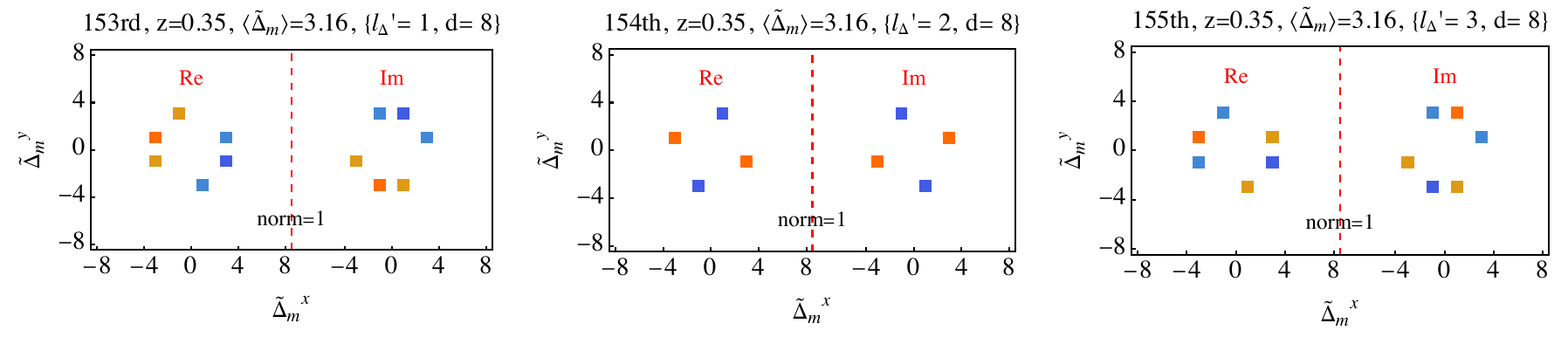} 
    \includegraphics[width=.93\textwidth]{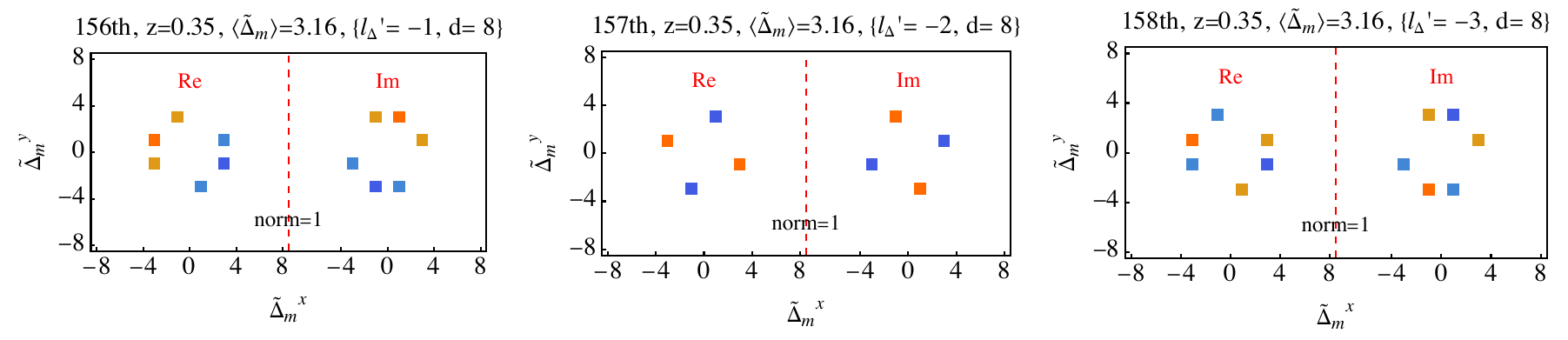} 
  \end{minipage}
   \subfigure[ A set of $d=8$ degenerate pure quark-gluon states  \label{fig:LFWF_pure_d8}]{
    \includegraphics[width=.39\textwidth]{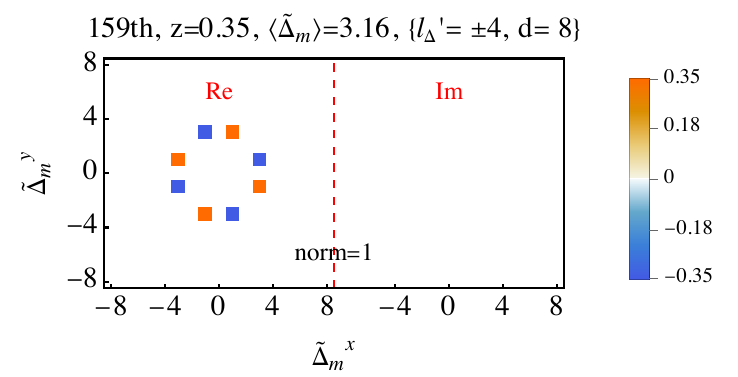} 
    }
  \caption{ 
    The LFWFs of selective QCD eigenstates that are pure quark-gluon states, plotted in the $\{z,\vec \Delta_m\}$ space. Basis parameters are the same as those in Fig.~\ref{fig:spectrum}. The rank of the state as indicated in the plot is the rank among the eigenstates in the increasing order of the invariant mass, as in Fig.~\ref{fig:spectrum_d}. The states in (a) are degenerate with $d=4$, and those in (b) are degenerate with $d=8$.
  }
  \label{fig:LFWFs_pure}
\end{figure*}

To have an intuitive interpretation of the eigenstates, we present selected LFWFs in the $\{z,\vec \Delta_m\}$ space in Figs.~\ref{fig:LFWFs_dressed} and ~\ref{fig:LFWFs_pure}. The states are calculated in the CHD basis, so the color and helicity dependence is implicit. 
The ground state, as shown in Fig.~\ref{fig:LFWF_dressed_q}, is the dressed quark state. The dominant $\ket{q}$ component is ``dressed" by quark-gluon states with different $z$s, and the smaller the $z$, the more it contributes. In other words, the quark is more dressed by the softer gluons. The quark-gluon component at each $z$, is azimuthal symmetric in $\vec \Delta_m$, since they are the $l_\Delta'=0$ states. 
The dressed quark-gluon states, as the one shown in Fig.~\ref{fig:LFWF_dressed_qg}, have a larger quark-gluon component than the bare quark component. Compared to the ground state, its quark-gluon components are radial excited, as can be seen from the peaked ``ring'' pattern in the plots.
The pure quark-gluon states that are orbital-angular excited in $l_\Delta'$ are plotted in Fig.~\ref{fig:LFWFs_pure}. In contrast to the dressed states, those states are single-valued in $z$ and $|\vec \Delta_m|$, and are angularly excited in $\vec \Delta_m$.

\subsubsection{Off shell quarks}
\label{subsec:offshell}
By requiring the ground state invariant mass $M_0^2=m_q^2$ in the mass renormalization, we obtained the on shell quark, e.g., the ground state in Fig.~\ref{fig:spectrum_d} with its LFWF presented Fig.~\ref{fig:LFWF_dressed_q}.

\begin{figure*}[t!]
  \centering 
  \subfigure[
    The invariant mass squared
    \label{fig:M0_dm}
  ]{ \includegraphics[width=0.44\textwidth]{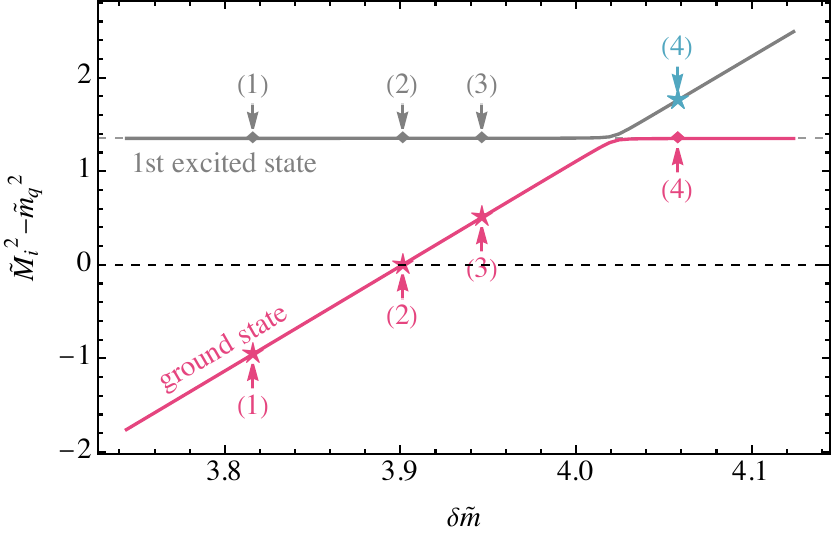} }\qquad
  \subfigure[
    The single-quark probability 
    \label{fig:Z2_dm}
  ]{ \includegraphics[width=0.44\textwidth]{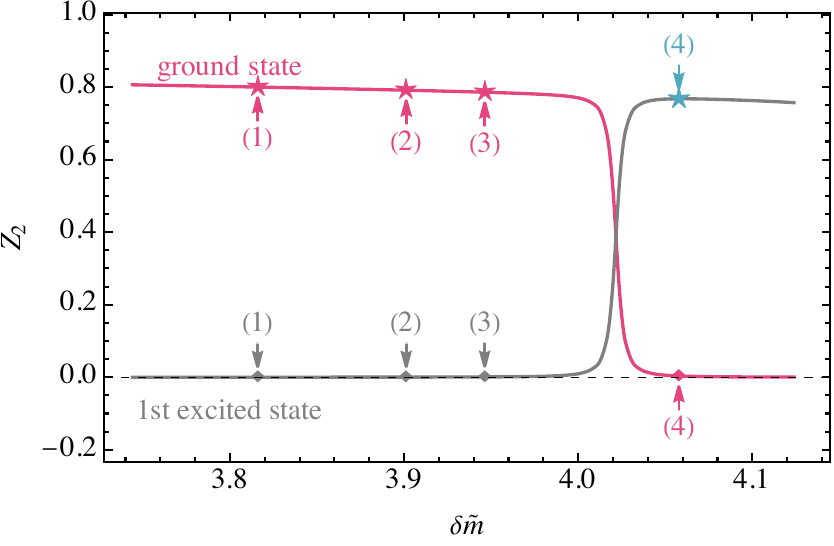} }
  \subfigure[
    The low-lying dressed states
    \label{fig:spectrum_dm}
  ]{ \includegraphics[width=0.44\textwidth]{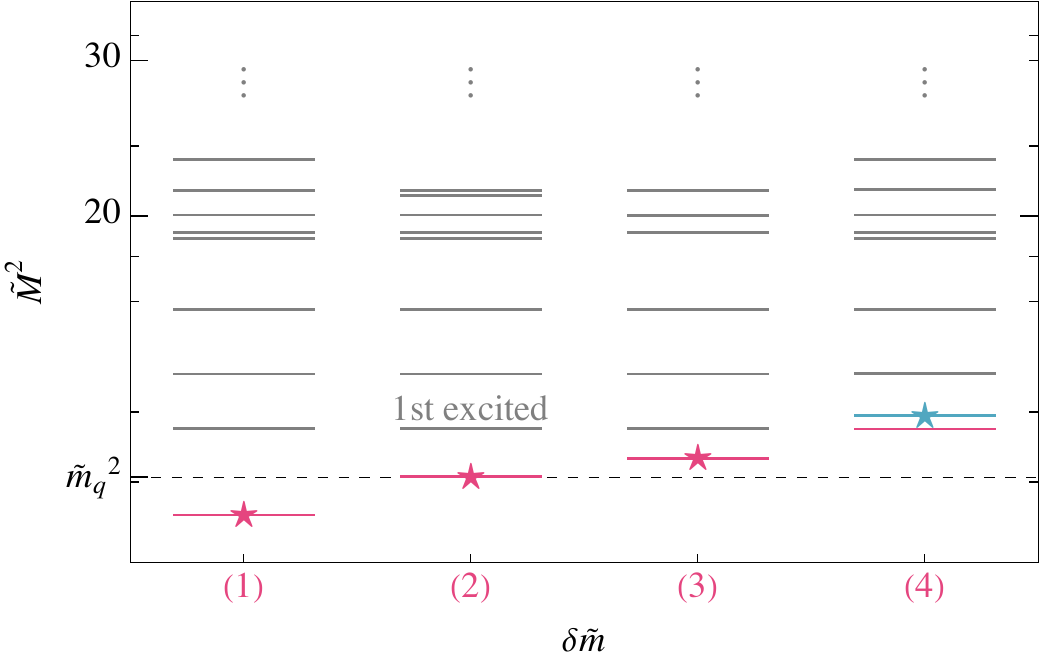} }
  \caption{Panel \ref{fig:M0_dm} shows the dependence of the ground state (the 1st excited state) invariant mass square on the mass counterterm $ \delta \tilde m$, in the solid red (gray) line. The lower dashed line indicates the physical quark mass, and the upper dashed line indicates the first excited state when the ground state is an on-shell quark.
   Panel \ref{fig:Z2_dm} shows the dependence of the ground state (the 1st excited state) single-quark probability, i.e., $Z_2$,  on the mass counterterm $ \delta \tilde m$.
   Panel \ref{fig:spectrum_dm} shows mass spectrum of the low lying states at the four different cases of $ \delta \tilde m$ as marked in panel \ref{fig:M0_dm} with the ``$(i)$"s. The ground states are plotted in red. The states labeled with a star are the intended states for a given $M_{\text{target}}^2$.
   Basis parameters are the same as those in Fig.~\ref{fig:spectrum}. 
  }
  \label{fig:spectrum_MC}
\end{figure*}

Nevertheless, we can also obtain states with an invariant mass different from the physical quark, which we will call off-shell quarks,  by choosing a different mass counterterm. 
Note that the spectrum also has the excited dressed states (i.e., the dressed quark-gluon states), e.g., those in Fig.~\ref{fig:spectrum_d} and the LFWF presented in Fig.~\ref{fig:LFWF_dressed_qg} that have a higher invariant mass. 
However, we do not refer to those states as time-like quarks, even though they share the quark's quantum numbers and have a larger invariant mass $M^2>m_q^2$, because their dominant component is a quark-gluon state, which corresponds to their asymptotic state in scattering. 
The procedure to obtain an off-shell quark state is similar to that for the on-shell quark. Instead of matching the state's eigenvalue to $m_q^2$, we match it to the intended virtuality $Q^2$. 
More generally, if we choose the mass counterterm to produce a state with a given $M_{\text{target}}^2$, one can find the associated mass counterterm according to Eq.~\eqref{eq:bi_dPmn}, and the corresponding eigenstates. 
We show the  dependence on the mass counterterm $ \delta \tilde m$ of the invariant mass square  of the ground state $\tilde M_0^2$ and of the first excited state $\tilde M_1^2$  in Fig.~\ref{fig:spectrum_MC}. 
Both the values of $\tilde M_0^2$ and $\tilde M_1^2$ increase as $ \delta \tilde m$ increases. The former saturates at 
the value of the original first excited state (shown as the top dashed line in the figure), and subsequently it is the value of $\tilde M_1^2$ that grows with increasing  $ \delta \tilde m$.
We note that the two eigenvalues do not cross, featuring the phenomenon of avoided crossing.
In the regime that $M_{\text{target}}^2< m_q^2$, the ground state is a spacelike quark, i.e., $M_0^2 = M_{\text{target}}^2$. We take the point marked as ``$(1)$" as an example and plot the mass spectrum of the low lying states.   
At $M_{\text{target}}^2=m_q^2$, the ground state is the on-shell quark, marked as ``$(2)$" in the figure.
In the regime where $M_{\text{target}}^2>m_q^2$, the target state corresponds to a time-like quark, though it is not necessarily the ground state.
When $M_{\text{target}}^2$ is between the original ground and excited states, the target (mostly quark) state is still the ground state, for example, the ``$(3)$" in the figure. 
However, once $\tilde M_{\text{target}}^2$ exceeds the invariant mass of the original first excited state, the ``target'' state that is predominantly a quark state becomes an excited state. 
An example of this is the blue-starred state in case ``$(4)$" in the figure. 
The original lowest excited state, which is predominantly a $qg$ state, becomes the new ground state, as seen in case ``$(4)$".
One can see from Fig.~\ref{fig:Z2_dm} that when this transition happens the state with the highest single-quark probability $Z_2$ becomes an excited state, and the lowest of the $qg$ states becomes the new ground state. 

Overall, when $M_{\text{target}}^2$ increases, the associated eigenstate, which is mostly a bare quark state, moves upward through the original spectrum, slightly changing the other dressed states. As the only part of the  Hamiltonian that is changed is the mass counterterm, the eigenstates that are pure quark-gluon states always stay the same.

\begin{figure}[t!]
  \centering 
    \includegraphics[width=0.48\textwidth]{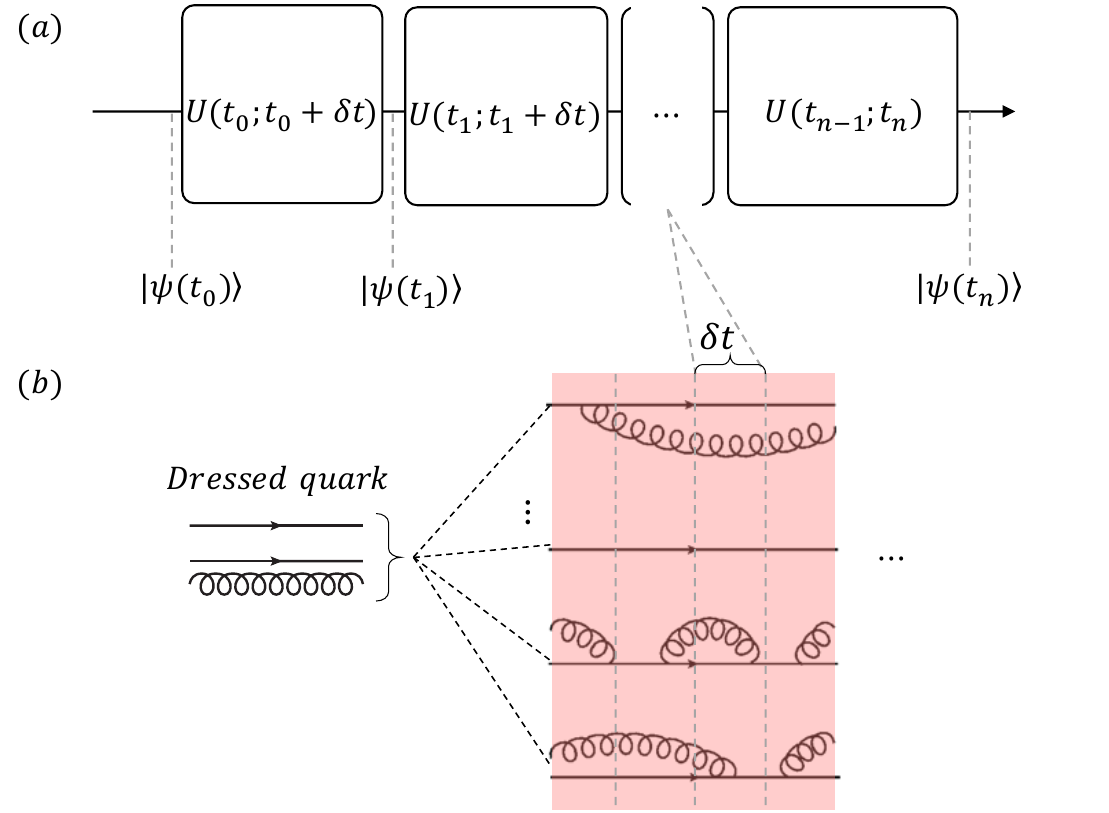} 
  \caption{
  Schematic representation of the time-dependent non-perturbative Hamiltonian approach: (a) a “step-by-step” treatment, (b) quark is a superposition of different quantum states ($\ket{q}$, $\ket{qg}$) traversing the medium (shown in red band).
  }
  \label{fig:setup_evolution}
\end{figure}

\subsection{Time evolution}\label{sec:time_evolution}

In the light-front Hamiltonian formalism, the time evolution of a quantum state obeys the equation of motion,
\begin{align}
  i\frac{\partial}{\partial x^+}\ket{\psi;x^+}=\frac{1}{2}P^-(x^+)\ket{\psi;x^+}\;.
\end{align}
The solution describes the state of the investigated system at any given light-front time $x^+$,
\begin{align}\label{eq:ShrodingerEqSol}
 \ket{\psi;x^+}=\mathcal{T}_+\exp\left[-\frac{i}{2}\int_0^{x^+}\diff z^+ P^-(z^+)\right]\ket{\psi;0}\;,
\end{align}
where $\mathcal{T}_+$ denotes light-front time ordering. 
In our developed computational framework~\cite{Li:2021zaw, Li:2023jeh}, tBLFQ, the time evolution equation is solved by sequentially applying the evolution operators for many small time steps to the state vector in the basis space. 
The method is illustrated in Fig.~\ref{fig:setup_evolution}. 

In the discrete momentum basis representation, which is also the computational basis space, one can write out an arbitrary state in the $ \ket{q}+\ket{qg}$ Fock space schematically as 
\begin{align}\label{eq:state_in_beta}
    \ket{\psi;x^+} = \sum_\beta \psi_\beta(x^+)\ket{q (\beta)}+\sum_\alpha \psi_\alpha(x^+) \ket{qg (\alpha )}\;,
\end{align}
where $\beta=\{p^+, \vec p_\perp, c, \lambda\}$ and $\alpha=\{p_q^+, \vec p_{\perp,q}, c_q, \lambda_q;p^+_g, \vec p_{\perp,g},$  $  c_g, \lambda_g\}$ are the quantum numbers of $\ket{q}$ and $\ket{qg}$ basis states, respectively. 
The basis coefficients $\psi_\beta(x^+)$s and $\psi_\alpha(x^+) $s, which represent the wavefunction, encode the information of the state and allow us to extract the relevant physical observables.

The acquired knowledge of the eigenstates of the QCD Hamiltonian in vacuum, as discussed in Sec.~\ref{sec:phys_q}, provides a new perspective on understanding the jet state.
Giving the wavefunction of each eigenstate, $ \phi_l$, in the momentum basis space
\begin{align}\label{eq:eigenstate_beta}
  \ket{\phi_l} = \sum_\beta c_{l,\beta}\ket{q (\beta)}+\sum_\alpha  c_{l,\alpha}\ket{qg (\alpha )}\;,
\end{align}
we can write the same quantum state in Eq.~\eqref{eq:state_in_beta} as an expansion in the eigenstate basis as well, 
\begin{align}\label{eq:state_in_coherent}
  \begin{split}
    \ket{\psi;x^+}=&\sum_{l} \tilde\psi_l(x^+) \ket{\phi_l}\\
    =&\sum_{d}^{\text{dressed q}} \tilde\psi_d(x^+) \ket{\phi_d}
    +\sum_{e}^{ \text{excited states}} \tilde\psi_e(x^+) \ket{\phi_e}
    \;.
  \end{split}
\end{align}
Here, $l$ contains the corresponding quantum numbers for each eigenstate $\phi_l$, and we partition the states into on-shell dressed quark states $\phi_d$s, and all other excited states $\phi_e$s.

By analyzing the evolved state wavefunction, we can examine the details of the scattering and transition processes of a given initial state.
Consider the initial state as an on-shell physical quark characterized by the quantum numbers $\{P^+, \vec P_{\perp,Q}, c_Q, h_Q\}$. From the perspective of the eigenstate spectrum, this represents the ground state with momentum $\{P^+, \vec P_{\perp,Q}\}$, assigned color charge $c_Q$ and helicity $h_Q$.
Having the evolved state expressed in the form of Eq.~\eqref{eq:state_in_coherent}, we can categorize the scattering and transition mechanisms as follows.
\begin{itemize}
  \item \emph{Elastic scattering}: the transition to the ground states of all $ \vec P_{\perp}$ space, i.e., all the $\phi_d$ states. Those transitions are elastic because the internal structure of the initial state, the $\{z,\vec \Delta_m\}$-dependence, remains the same. The elastic scattering can be further classified according to the component of the evolved state, not mutually exclusively, 
\begin{itemize}
  \item CM momentum excitation: $ \vec P_{\perp}\neq \vec P_{\perp,Q}$ states;
  \item color rotation: a different color state than $c_Q$;
  \item helicity flip: a different helicity state than $h_Q$.
\end{itemize}
  \item  \emph{Inelastic scattering/gluon radiation}: the transition to all other coherent states, i.e., all the $\phi_e$ states. Those transitions are inelastic because the internal structure of the initial state changes. As those states are either the dressed quark-gluon states or the pure quark-gluon states, the transition to them can be interpreted as gluon radiation. Like the elastic scattering, the inelastic scattering can also be further classified into
  \begin{itemize}
    \item CM momentum excitation: $ \vec P_{\perp}\neq \vec P_{\perp,Q}$ states;
    \item relative momentum excitation: the $\vec P_{\perp,Q}$ states;
    \item color excitation: the color excited quark-gluons;
    \item helicity flip: a different helicity than $h_Q$.
  \end{itemize}
\end{itemize}

\section{Results and discussion}\label{sec:results}
We consider four different initial states, the bare quark, the on-shell dressed quark, a timelike quark, and a spacelike quark. By running in-medium jet simulations, we study gluon radiation, cross section, momentum broadening, and invariant mass distribution in these different scenarios.

In the simulations, unless specified otherwise, we take $N_\perp=\floor*{K}=8$, $g=1$, $L_\perp=50 ~\GeV^{-1}$, and $ m_q=0.2 ~\GeV$, so that the UV cutoff is $\Lambda_{UV}=0.5~ \GeV$. 
The longitudinal momentum is given by $P^+= 2\pi/L\times K=5.34~\GeV$ with $ L=10 ~\GeV^{-1}$.
The total evolution time is $L_\eta = 50 ~\GeV^{-1}$, and we examine a range of $g^2\tilde\mu = 0\sim 0.06 ~\GeV^{3/2}$, giving the saturation scale around $Q_s^2=0\sim 0.04 ~\GeV^2$. 
The IR regulator in the medium field is $m_g=0.08 ~\GeV$.
For each $g^2\tilde\mu$, we sample 10 different configurations of the medium source charge. In evaluating physical quantities, we take the average over different configurations at the observable level, and use the standard deviation to quantify the uncertainty.

The timelike(spacelike) quark wavefunction is obtained by taking $\lambda=\tilde M^2- \tilde m_q^2 =(-) 3$ in Eq.~\eqref{eq:q_LFWF_qg_anal}, which yields a different mass counterterm than the on-shell case. In the evolution Hamiltonian $P^-(x^+)$, the vacuum term $P^-_{QCD} $ is always mass-normalized according to the on-shell dressed quark.

\subsection{Gluon radiation}
The interpretation of gluon radiation is closely tied to the configuration of the jet, and the corresponding process. We describe two scenarios, as illustrated in Fig.~\ref{fig:setup}. 

In the first scenario, the initial jet state is a bare quark, and we used the probability of the  $\ket{qg}$ states developing during the time evolution, $P_{\ket{qg}}$, to quantify gluon radiation~\cite{Li:2023jeh}. Using the notation from Eq.~\eqref{eq:state_in_beta}, we have
\begin{align}
  \begin{split}
  & P_{\ket{q}}\equiv \sum_\beta |\psi_\beta|^2 \;,
  \qquad
  P_{\ket{qg}}\equiv \sum_\alpha |\psi_\alpha|^2\;.
\end{split}
\end{align}
The state is normalized such that $P_{\ket{q}} + P_{\ket{qg}}=1$.
Then the medium-induced gluon emission can be evaluated by~\cite{Li:2023jeh}
\begin{align}\label{eq:Delta_Pqg}
  \delta P_{\ket{qg}}(Q_s^2) = P_{\ket{qg}}(Q_s^2) -P_{\ket{qg}}(0) \;.
\end{align}
Here, $ P_{\ket{qg}}(0)$ is the value obtained in the vacuum for the same amount of evolution time as in the medium.

In the second scenario, the initial jet is a dressed quark. There are already $\ket{qg}$ components at the starting point, so the amount of the gluon radiation can no longer be simply quantified by $P_{\ket{qg}}$. 
Rather, we define the probability of finding the jet state in the dressed quark states and that in the other excited states, using the notation from Eq.~\eqref{eq:state_in_coherent}, 
\begin{align}
  \begin{split}
  & P_{\text{dressed q}}\equiv \sum_{d}^{\text{dressed q}} |\tilde\psi_d |^2\;,
 \\
 & P_{ \text{excited}}\equiv \sum_{e}^{ \text{excited states}}|\tilde\psi_e|^2
  \;.
\end{split}
\end{align}
The state is normalized such that $P_{\text{dressed q}} + P_{ \text{excited}}=1$.
As mentioned above, the transition to the dressed quark states at another momentum does not change the internal structure of the dressed quark state, and we interpret such a transition as elastic scattering. 
On the contrary, the transition to all other excited states adds an asymptotic, dressed gluon to the state, and therefore its probability $P_{ \text{excited}}$ can be used to quantify  gluon emission.
Then the medium-induced gluon emission can be evaluated by
\begin{align}\label{eq:Delta_Pexcited}
  \delta P_{ \text{excited}}(Q_s^2) = P_{ \text{excited}}(Q_s^2) -P_{ \text{excited}}(0) \;.
\end{align}
Here, $ P_{ \text{excited}}(0)$ is the value of the initial state, which is also the value in the vacuum.

\begin{figure*}[htp!]
    \centering 
      \vspace{-1cm}
    \subfigure[
      Bare quark initial state
      \label{fig:q_Pg}
    ]{ \includegraphics[width=0.44\textwidth]{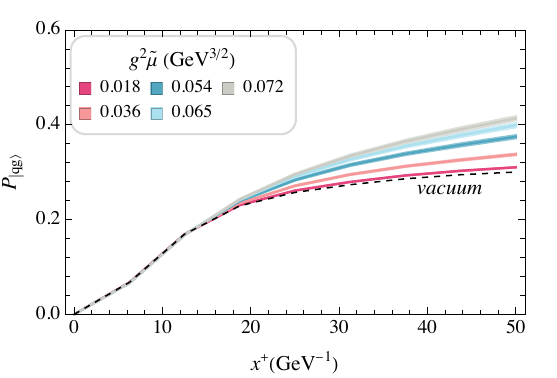}
    \includegraphics[width=0.44\textwidth]{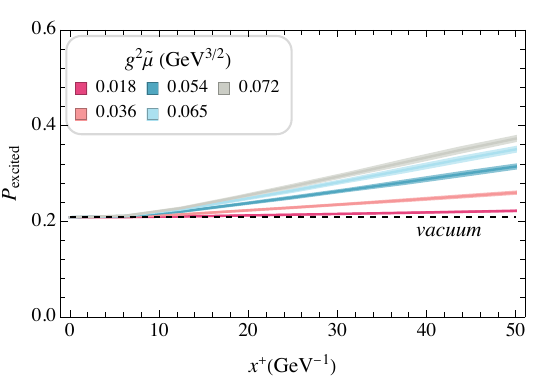} 
    }
    \vspace{-.2cm}
    \subfigure[
        Dressed/on-shell quark initial state
      \label{fig:qdress_Pg}
    ]{ \includegraphics[width=0.44\textwidth]{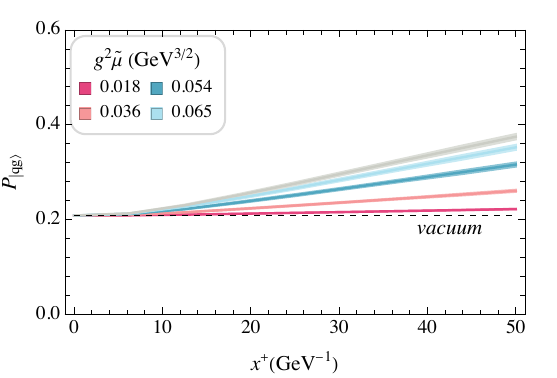}
    \includegraphics[width=0.44\textwidth]{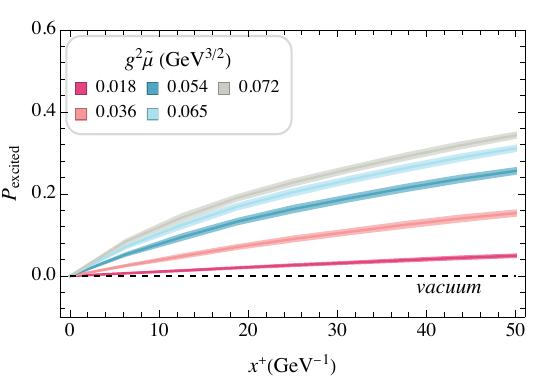}  }
        \vspace{-.2cm}
    \subfigure[
       Timelike quark initial state
      \label{fig:qtimelike_Pg}
    ]{ \includegraphics[width=0.44\textwidth]{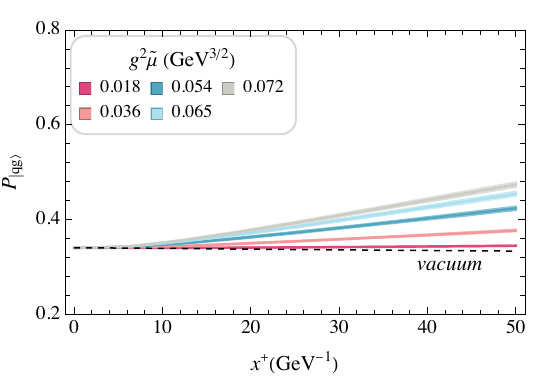}
    \includegraphics[width=0.44\textwidth]{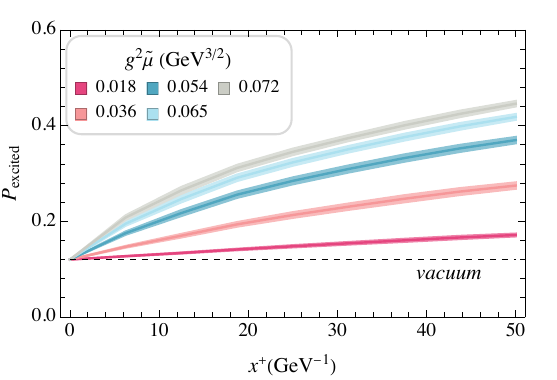}  }
        \vspace{-.2cm}
    \subfigure[
       Spacelike quark initial state
      \label{fig:qspacelike_Pg}
    ]{ \includegraphics[width=0.44\textwidth]{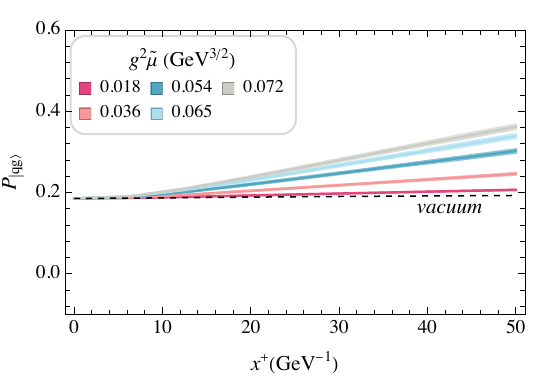}
    \includegraphics[width=0.44\textwidth]{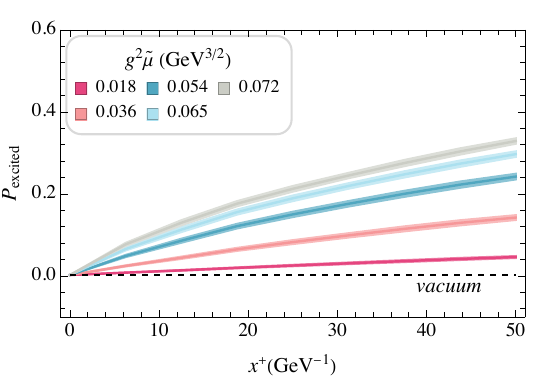}  }
    \caption{
    Evolution of the gluon radiation for (a) bare quark initial state, (b) dressed quark initial state, (c) timelike quark initial state,  and (d) spacelike quark initial state.
    The left panels plots $P_{\ket{qg}}$ and the right panels $P_{\text{excited}}$.
    The black dashed line represents the behavior in the vacuum.
    }
    \label{fig:gluon_radiation}
  \end{figure*}

Using both interpretations, $P_{\ket{qg}}$ and $P_{ \text{excited}}$, we calculate and compare the gluon radiation for different jet initial states, the bare quark, the dressed/on-shell quark, and the off-shell dressed quarks. The results are shown in Fig.~\ref{fig:gluon_radiation}. In all four cases, using either interpretation, we see that the stronger the medium, the more gluons  are emitted at the final state, indicating that the medium promotes gluon emission.

In the case of the bare quark initial state, as shown in Fig.~\ref{fig:q_Pg}, the probability $P_{\ket{qg}}$ initially experiences a rapid increase in vacuum, then after around $x^+=15~\GeV^{-1}$, the rate of increase slows down. Notably, in the early stage, the dominant activity is the formation of the $\ket{qg}$ component, and the presence of medium does not have a significant influence. Such process has also been examined in our previous work without implementing the mass renormalization~\cite{Li:2023jeh}. In contrast, the probability $P_{ \text{excited}}$ is constant over time in the vacuum, which is a consequence of the corresponding operator commuting with the QCD Hamiltonian. This provides an alternative explanation: while the bare quark emits gluons in the vacuum, the probabilities of all the eigenstates of the Hamiltonian stay constant.
Similar to $P_{\ket{qg}}$, there is an early time stage during which the medium does not have a significant effect. However, the transition point occurs much earlier, around $x^+=6~\GeV^{-1}$. After this stage, the increase follows a linear pattern.

In the case of the dressed/on-shell quark initial state, as shown in Fig.~\ref{fig:qdress_Pg}, both probabilities $P_{\ket{qg}}$ and $P_{ \text{excited}}$ are constant over time in the vacuum, which is a consequence of the initial state being the eigenstate of the evolution Hamiltonian. Note that the value of $P_{\ket{qg}}$ in vacuum here is the same with the value of $P_{ \text{excited}}$ in vacuum for the bare quark initial state, both being $1-Z_2$. 
The behavior of $P_{\ket{qg}}$ in presence of the medium closely resembles, though not equivalent to, that of $P_{ \text{excited}}$ for bare quark initial state. Making a quantitative comparison between the two, the values at the final time is different by $0.3\%$ in average. 
Unlike $P_{\ket{qg}}$, the probability $P_{ \text{excited}}$ here starts increasing at the beginning of the in-medium evolution and maintains a constant rate of growth, except in the case of the largest $g^2 \tilde{\mu}$, where the behavior suggests some form of saturation.
In the scenarios of off-shell dressed quark initial state, as shown in Figs.~\ref{fig:qtimelike_Pg} and ~\ref{fig:qspacelike_Pg}, the overall pattern of the gluon emission is very similar to that in the on-shell quark case, with small differences in quantity. 
The selected off-shell states are closely aligned with the on-shell quark in terms of their wavefunctions. The calculated overlaps are
$|\braket{\psi_\text{onshell q}|\psi_\text{timelike q}}|^2=0.880$ and $|\braket{\psi_\text{onshell q}|\psi_\text{spacelike q}}|^2=0.998$.

\begin{figure*}[htp!]
    \centering 
      \includegraphics[width=0.44\textwidth]{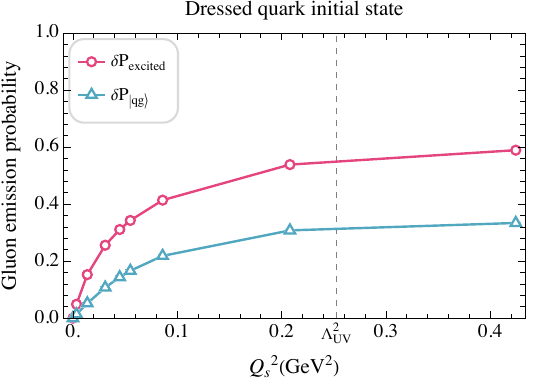}
      \includegraphics[width=0.44\textwidth]{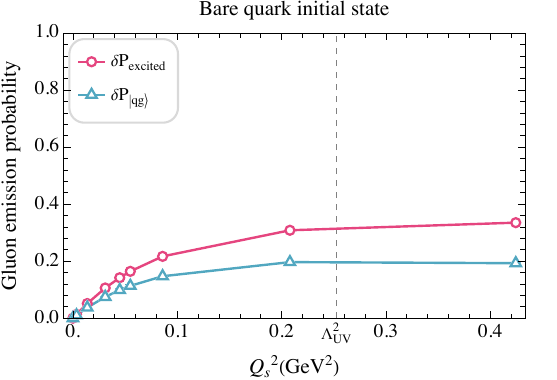} \\
      ~\\
      \includegraphics[width=0.44\textwidth]{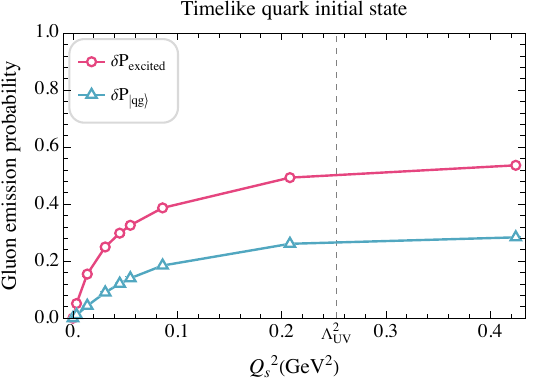} 
    \includegraphics[width=0.44\textwidth]{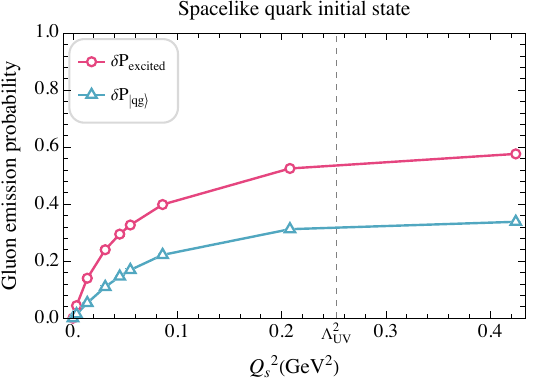} 
    \caption{
    Medium-induced gluon radiation at different $Q_s^2$ for bare quark , dressed/on-shell quark, timelike quark,  and spacelike quark initial states.
In each case, $\delta P_{\ket{qg}}$ and $\delta P_{\text{excited}}$, as defined in Eqs.~\eqref{eq:Delta_Pqg} and ~\eqref{eq:Delta_Pexcited}, are plotted as a function of the saturation scale. 
    }
    \label{fig:gluon_radiation_Qs}
  \end{figure*}

  To further quantify the medium-induced gluon emission, we plot the dependence of $ \delta P_{\ket{qg}}$ and $\delta P_{ \text{excited}}$ at different $Q_s^2$ for the four different initial states in Fig.~\ref{fig:gluon_radiation_Qs}.
  In each plot, the UV cutoff of the basis space is marked by the vertical dashed line, and the points with $Q_s^2 > \Lambda_{UV}^2 $ receive sizable lattice effects and are therefore less trustworthy. 
  In all four cases, $\delta P_{ \text{excited}}$ is sizably larger than $ \delta P_{\ket{qg}}$. 
  Out of these two the quantity $\delta P_{ \text{excited}}$ is the relevant measure when considering medium-induced gluon emission going out of a jet. 

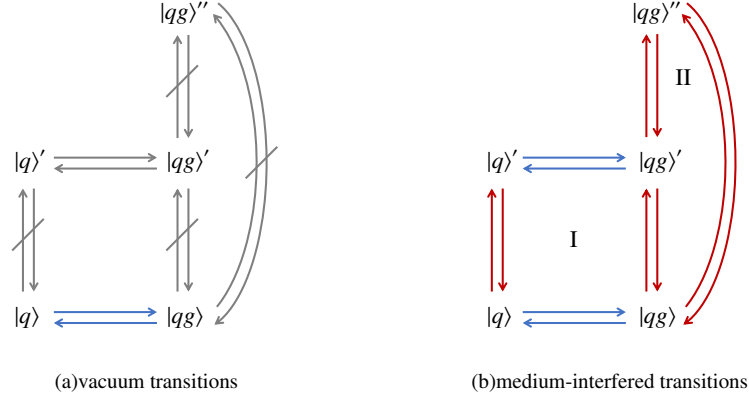
\begin{figure*}[htp!]
    \centering 
    \subfigure[vacuum transitions\label{fig:interference_V}]{
      \begin{minipage}[c]{.3\textwidth}
        \centering
     \begin{overpic}[height=4.4cm]{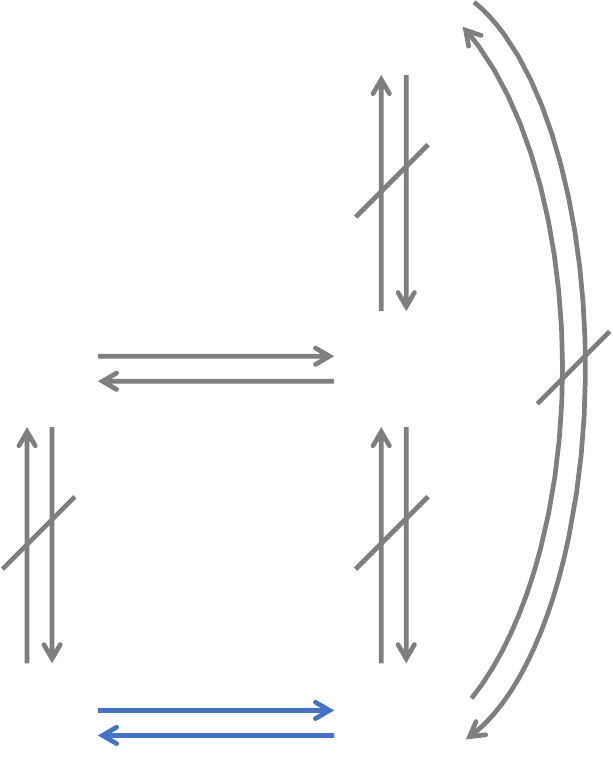} 
      \put (1,3) {$\ket{q}$}
      \put (47,3) {$\ket{qg}$}
      \put (1,50) {$\ket{q}'$}
      \put (47,50) {$\ket{qg}'$}
      \put (45,95) {$\ket{qg}''$}
    \end{overpic}
    \vspace{0.5cm}
    \end{minipage}
    }  
    \hspace{0.5cm}
    \subfigure[medium-interfered transitions \label{fig:interference_A}]{
      \begin{minipage}[c]{.3\textwidth}
        \centering
     \begin{overpic}[height=4.4cm]{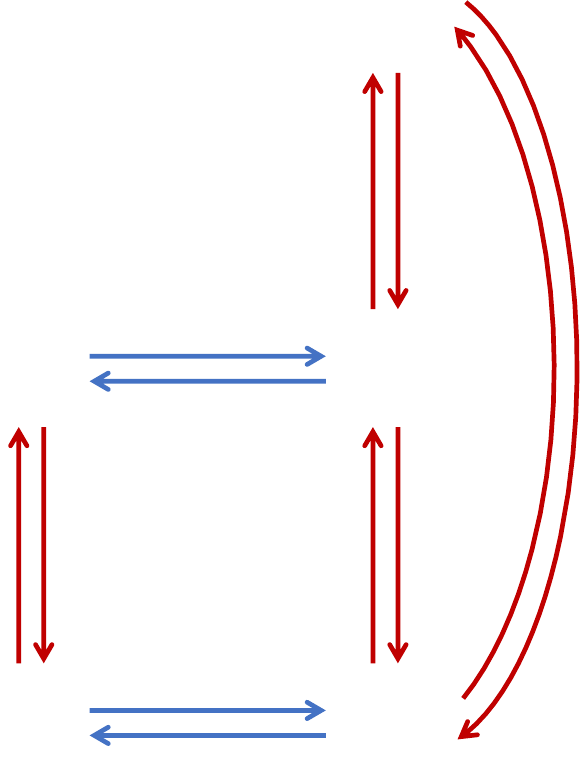} 
      \put (1,3) {$\ket{q}$}
      \put (47,3) {$\ket{qg}$}
      \put (26,26) {I}
      \put (1,50) {$\ket{q}'$}
      \put (47,50) {$\ket{qg}'$}
      \put (45,95) {$\ket{qg}''$}
      \put (58,75) {II}
    \end{overpic}
    \vspace{0.5cm}
    \end{minipage}
    }  
    \caption{
    Schematic illustration of the transitions between different Fock states (a) in the vacuum and (b) in the presence of the medium, with the two processes of I. \emph{phase mixing} and II. \emph{induced excitation} labeled.
    }
    \label{fig:interference}
  \end{figure*}

  The interference between the momentum transfer from the medium and the gluon emission/absorption process renders a non-trivial medium-induced gluon emission pattern. 
  There are two underlying mechanisms of the interference, \emph{phase mixing} and \emph{induced excitation}, as illustrated in Fig.~\ref{fig:interference}.
  In the vacuum, as demonstrated in Fig.~\ref{fig:interference_V}, only the transition between the initial quark (denoted as $\ket{q}$ in the figure) and those quark-gluon states of the same quantum number (denoted as $\ket{qg}$) can happen. The transition is bidirectional and given a sufficiently large number of different quark-gluon states, the transition will eventually reach an equilibrium such that the probability $P_{\ket{qg}}$ stabilizes. 
  In the presence of the medium, as demonstrated in Fig.~\ref{fig:interference_A}, the transitions to the quark states with a different momentum or color (denoted as $\ket{q}'$) are activated, likewise for the quark-gluon states. 
  Note that in theory, there are multiple different sets of single quark state with quark-gluon states that can couple to it, and for the simplicity of illustration in the figure, we put here just one of these sets, namely, $\{\ket{q}', \ket{qg}'\}$. 
  The medium interaction immediately after the first $q\leftrightarrow qg$ transition mixes different quark-gluon states. Though such mixing does not change $P_{\ket{qg}}$ directly, it changes the relative phase among the different states in $\ket{qg}$ and $\ket{qg}'$ respectively, which we call \emph{phase mixing}.
  Then in the next timestep, the $q\leftrightarrow qg$ transition happens within each set of the quark and its coupled quark-gluon states, $\{\ket{q}, \ket{qg}\}$ and $\{\ket{q}', \ket{qg}'\}$ respectively.
  Because now the states in each set have a different relative phase compared to that in the vacuum, the resulting $P_{\ket{qg}}$ can vary, becoming either larger or smaller depending on whether the interference is constructive or destructive.
  In the meantime, the medium also enables the transition to the quark-gluon states that do not couple to any single quark state (denoted as $\ket{qg}''$), an \emph{induced excitation}. 
  As the reachable phase space of of the quark-gluon states is enlarged by having the excited states that do not go back to single quark states directly, the total quark-gluon probability, $P_{\ket{qg}}$, can become larger than that in the vacuum. 

\begin{figure*}[htp!]
    \centering 
    \subfigure[
      Bare quark initial state
      \label{fig:UVg_bare}
    ]{ 
        \includegraphics[width=0.44\textwidth]{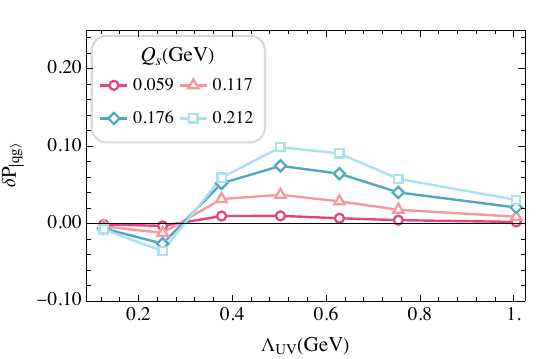} 
    \includegraphics[width=0.44\textwidth]{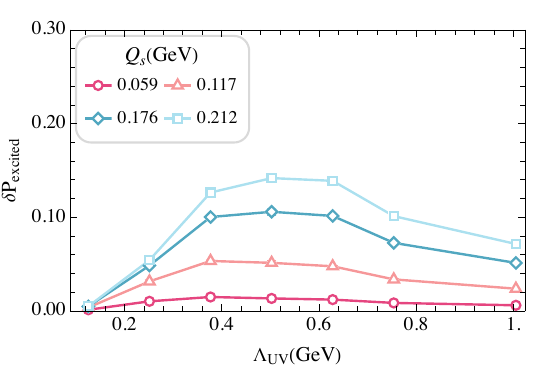} 
    }
    \subfigure[
      Dressed/on-shell quark initial state
      \label{fig:UVg_dressed}
    ]{ 
\includegraphics[width=0.44\textwidth]{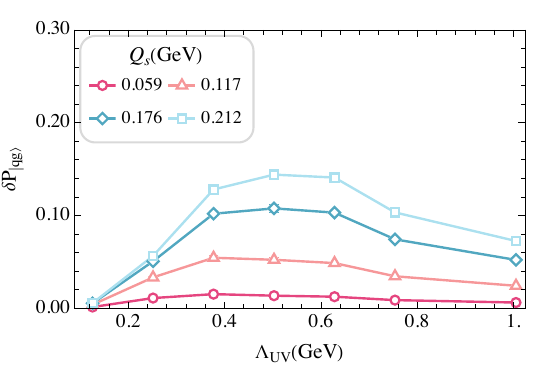} 
\includegraphics[width=0.44\textwidth]{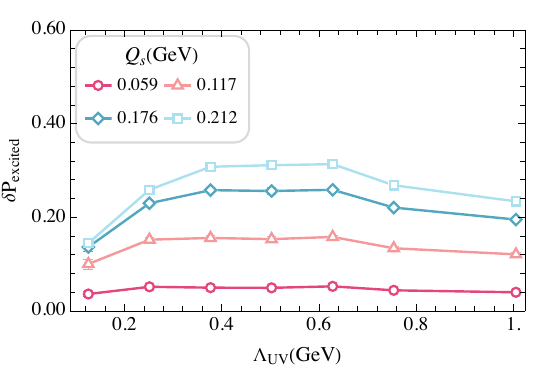} 
    }
    \caption{
    Dependence of gluon radiation on the UV cutoff $\Lambda_{UV}$, for (a) bare quark initial state and (b) dressed/on-shell quark initial state.
    The left panels plots $\delta P_{\ket{qg}}$ and the right panels plots $\delta P_{\text{excited}}$. 
    }
    \label{fig:gluon_radiation_UV}
  \end{figure*}

  To examine the dependence of gluon radiation on the UV cutoff $\Lambda_{UV}$, we fix $L_\perp=50 \GeV^{-1}$ and take $N_\perp=2,4,6,8,10,12,16$ for the simulations. The results are shown in Fig.~\ref{fig:gluon_radiation_UV}. We see that in all four cases, the quantity is not monotonic in $\Lambda_{UV}$. 
  Although increasing $\Lambda_{UV}$ expands the phase space available for gluon emission, it simultaneously increases the phase space for momentum broadening, which has a logarithmic dependence on $\Lambda_{UV}$. 

The bare quark initial state, which is not an eigenstate of the vacuum Hamiltonian, has a coherence time that is $\Lambda_{UV}$ dependent \cite{Li:2021zaw}, 
  \begin{align}
    \tau_{\text{coh}} \approx \frac{2 P^+}{\Lambda_{UV}^2}\;.
  \end{align}
  This means that if we consider the medium length $L_\eta$ as fixed, the final state at a smaller $\Lambda_{UV}$ is at a relatively earlier stage in the sense of coherence. 
  Before the different emergent quark-gluon states fully decohere, the \emph{phase mixing} is more likely to destruct the intrinsic pattern of the relative phase, as the medium randomizes the states. Later on, when the states decohere, their intrinsic pattern is less regular, so the effect from \emph{phase mixing} is less profound, whereas the effect of \emph{induced excitation} becomes dominant.
  Therefore, the medium-induced suppression of the gluon emission is more likely to be observed at the early stage, or effectively at smaller $\Lambda_{UV}$. 
  We see this in the left panel of Fig.~\ref{fig:UVg_bare}, where the $ \delta P_{\ket{qg}}$s are negative at the two lower $\Lambda_{UV}$ values.
 A similar suppression at small energy scale, or at short evolution time, has also been studied in the multiple soft scattering approximation in Refs.~\cite{Wiedemann:2000tf,Salgado:2003gb}.
In the right panel of Fig.~\ref{fig:UVg_bare}, the values of $ \delta P_{\text{excited}}$ are all positive. Different from $ \delta P_{\ket{qg}}$ on the left, now the counting of the emitted gluon is in terms of the excited eigenstates. As the medium induces the transition from the ground state to the excited states, the latter having larger quark-gluon components, the gluon emission is enhanced.

For the dressed quark initial state, which is an eigenstate with $\vec P_\perp=\vec 0_\perp$, its coherence length is given by its light-front energy $P^-$
\begin{align}
  \tau_{\text{coh}} = \frac{2 }{P^-}= \frac{2 P^+ }{m_q^2}\;,
\end{align}
not depending on $\Lambda_{UV}$.
But note that a larger $\Lambda_{UV}$ cutoff allows higher invariant mass  excited states.
In the left panel of Fig.~\ref{fig:UVg_dressed}, the behavior of $ \delta P_{\ket{qg}}$ of the dressed quark initial state closely resembles that of the bare quark's $ \delta P_{\text{excited}}$.
In the right panel of Fig.~\ref{fig:UVg_dressed}, $ \delta P_{\text{excited}}$ of the dressed quark initial state is the least sensitive to $\Lambda_{UV}$ among the four quantities being presented. At the largest values of $\Lambda_{UV}$ both medium modifications $ \delta P_{\ket{qg}}$ and $ \delta P_{\text{excited}}$ start to decrease with the UV cutoff, in spite of the additional phase space available for medium induced gluon radiation. This is a phenomenon that we do not completely understand at this point, but plan to investigate further in future work.

\subsection{Cross section}
\begin{figure}[htp!]
    \centering 
     \includegraphics[width=0.44\textwidth]{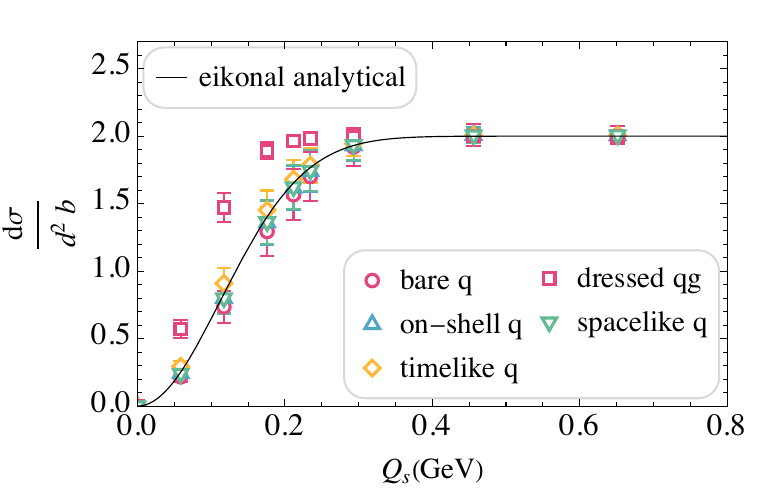}
    \caption{
    The total cross section of different quark initial states.
    }
    \label{fig:crss}
  \end{figure}
The   differential cross section is defined as 
\begin{align}\label{eq:cross}
    \begin{split}
    \frac{\diff\sigma}{\diff^2 b} 
    =&\sum_{\phi_{out}} |M(\phi_{out};\psi_{in})|^2\\
    =&\sum_{\phi_{out}} |\braket{\phi_{out}|S|\psi_{in}}-\braket{\phi_{out}|\psi_{in}}|^2\\
    =&\sum_{\phi_{out}} |\braket{\phi_{out}|\psi_{out}}-\braket{\phi_{out}|\psi_{in}}|^2\;,
\end{split}
\end{align}
in which $\sum_{\phi_{out}}$ is the summation over the final phase space. 
Here our focus is on the scattering of a quark state by an external field and we have solved the eigenstates of the vacuum QCD Hamiltonian. Thus we consider, in the sense of the interaction picture of quantum mechanics,  the ``free'' Hamiltonian to  be $P^-_{QCD}=P^-_{KE} + V_{qg}$, while the full Hamiltonian is given by $P^-(x^+) = P^-_{QCD} + V_A(x^+)$. 

In a quantum mechanical scattering problem, the incoming state is taken to be an eigenstate of the free Hamiltonian, which in our case is $P^-_{QCD}$, say $\ket{\psi(x^+=0)}=\ket{\psi_{in}}=\ket{\psi_{ \lambda_0}}$ with eigenvalue $P^-_{ \lambda_0}$. The evolved state is a superposition of different $P^-_{QCD}$ eigenstates, in the Schr\"odinger picture, 
\begin{align}
    \ket{\psi(x^+)}=\sum_\lambda c_\lambda(x^+)\ket{\psi_\lambda}\;.
\end{align}
The out state is the evolved scattering state in the interaction picture, 
\begin{align}
    \begin{split}
   \ket{\psi_{out}}= &\ket{\psi(x^+)}_{Int.}=e^{iP^-_{QCD}x^+/2}\sum_\lambda c_\lambda(x^+)\ket{\psi_\lambda}\\
    =&\sum_\lambda e^{iP^-_\lambda x^+/2}c_\lambda(x^+)\ket{\psi_\lambda}
    \;.
\end{split}
\end{align}
The cross section according to Eq.~\eqref{eq:cross}, can be written as
\begin{align}\label{eq:cross_lambda}
    \begin{split}
    \frac{\diff\sigma}{\diff^2 b} 
    =&\sum_{\lambda } |\braket{\psi_\lambda|\psi_{out}}-\braket{\psi_\lambda|\psi_{in}}|^2\\
    =&\sum_{\lambda } \left|e^{iP^-_\lambda x^+/2}c_\lambda(x^+)-c_\lambda(0)\right|^2
    \;.
\end{split}
\end{align}

We show in Fig.~\ref{fig:crss} the results of the total cross section of different initial states, including a bare quark, a dressed quark-gluon state with single quark quantum numbers (i.e., a coupled state) an on-shell dressed quark, a timelike quark, and a spacelike quark.  The eikonal analytical result shown in solid black line is the expectation of a single quark in the eikonal limit~\cite{Li:2020uhl}. One can see that the results of the bare quark, the on-shell and off-shell dressed quarks are all very close to the eikonal expectation, whereas the dressed quark-gluon state has a much larger cross section at the same saturation scale. This is because that the total cross section takes into account all transitions that are different from the initial state, including changes in color, momentum, and Fock states. For the quark dominated states, the gluon component is not large, so the state behaves like a single quark, matching the fundamental representation probe in the analytical eikonal limit result. The dressed quark-gluon state, on the other hand, is dominated by a large $\ket{qg}$ component, which enables a more rapid color rotation and thus a larger cross section.

\subsection{Momentum broadening}
Through the interaction with the background field the quark jet state can transition to momentum modes that are different from the initial state.
The transport coefficient $\hat q$ characterizes this momentum broadening as
\begin{align}\label{eq:dpperp}
  \hat q 
  = \frac{\Delta \braket{p^2_\perp ( x^+)}}{\Delta x^+} 
  \;.
\end{align}
We analyze the momentum broadening in terms of two kinds of momentum, the CM transverse momentum and the quark momentum in the jet.
The CM transverse momentum square as defined in Ref.~\cite{Li:2023jeh} is
\begin{align}\label{eq:pT2_CM}
  \braket{P_{\perp, CM}^2}=P_{\ket{q}}\braket{P_\perp^2}_{\ket{q}}
  +P_{\ket{qg}} \braket{P_\perp^2}_{\ket{qg}}\;,
\end{align}
where $P_{\ket{q}}$ and $P_{\ket{qg}}$ are the probabilities for being in a quark and $qg$ state respectively, and $P_\perp^2$ is the total CM momentum of the state. 
Note that these are different from the transverse momenta of the bare quark and gluon separately, which we also calculate, denoting them as
\begin{subequations}
   \begin{align}\label{eq:psq_full_q}
    \braket{P_{\perp,q}^2 }
    =& P_{\ket{q}}\braket{p_\perp^2}_{\ket{q}}
    +P_{\ket{qg}} \braket{p_{\perp,q}^2}_{\ket{qg}}
    \;,
 \end{align}
 \begin{align}\label{eq:psq_full_g}
    \braket{P_{\perp,g}^2 }
    =P_{\ket{qg}} \braket{p_{\perp,g}^2}_{\ket{qg}}
    \;.
 \end{align}
\end{subequations}

The other measure of momentum broadening uses what we call here the quark jet momentum. This is obtained by decomposing the state of the system in terms of the vacuum QCD eigenstates: dressed quarks and excited states which are physically interpreted as states with a radiated gluon. In terms of these we can try to measure the squared momentum  of the quark within the jet, which we define as consisting of the squared CM momentum of the dressed state, and the squared momentum of the quark within the remaining excited state,
\begin{align} \label{eq:pqjet}
  \braket{P_{\perp,\text{jet} }^2 }
  =  \sum_d |\tilde\psi_d|^2 \braket{P_\perp^2}_d
  +\sum_e |\tilde\psi_e|^2 \braket{P_{\perp,q}^2 }_e
  \;.
 \end{align}
Here, $\braket{P_\perp^2}_d $ is the expectation value of the total momentum $P_\perp^2$ of the dressed quark state, and $\braket{P_{\perp,q}^2 }_e$ is the expectation value of the quark momentum $P_{\perp,q}^2$ in the excited state. 
The latter gets a contribution both from the CM momentum of the $\ket{q}$ components and the quark momentum of the $\ket{qg}$ components so that
\begin{align}
  \braket{P_{\perp,q}^2 }_e
  =
      \sum_\beta|c_{e,\beta}|^2 p_\perp^2(\beta)
      +\sum_\alpha|c_{e,\alpha}|^2 p_{\perp,q}^2(\alpha)
 \;.
\end{align}
 Recall that $\beta$  and $ \alpha $ are the uncoupled basis states in the $\ket{q}$ and the $\ket{qg}$ sector respectively, as defined in Eq.~\eqref{eq:state_in_beta}, and $c$s are the corresponding basis coefficients, as given in Eq.~\eqref{eq:eigenstate_beta}.

\begin{figure*}[htp!]
  \centering 
    \includegraphics[width=0.44\textwidth]{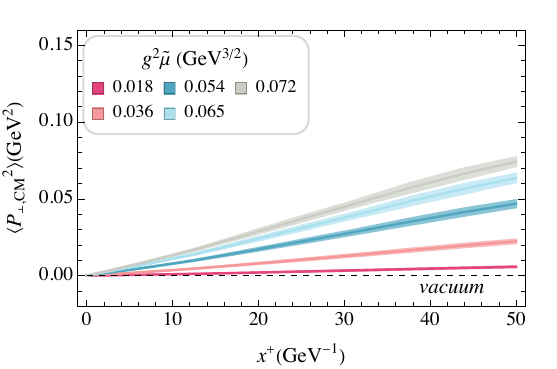}
    \includegraphics[width=0.44\textwidth]{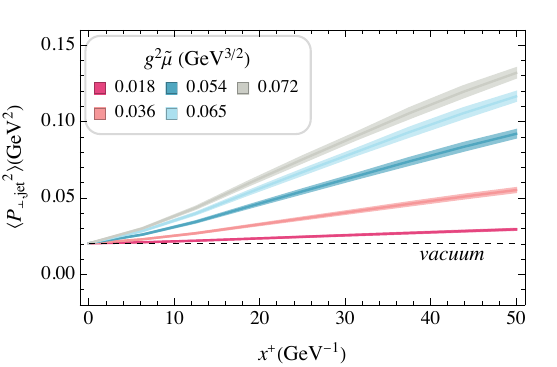}
  \subfigure[
    Bare quark initial state
    \label{fig:pperp_bare}
  ]{
  \includegraphics[width=0.44\textwidth]{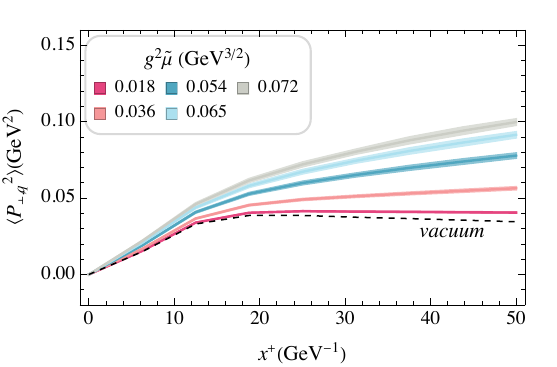}
  \includegraphics[width=0.44\textwidth]{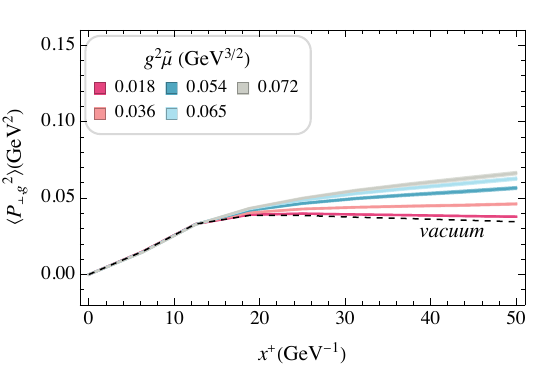}
  }
  \includegraphics[width=0.44\textwidth]{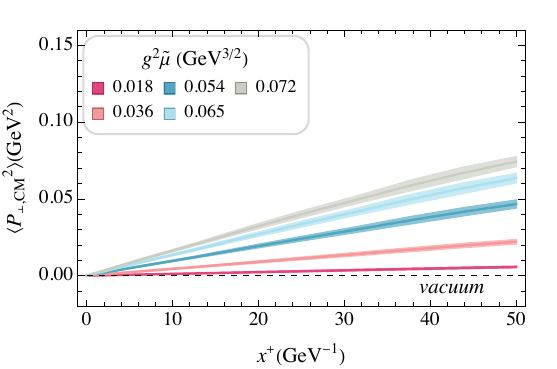}
\includegraphics[width=0.44\textwidth]{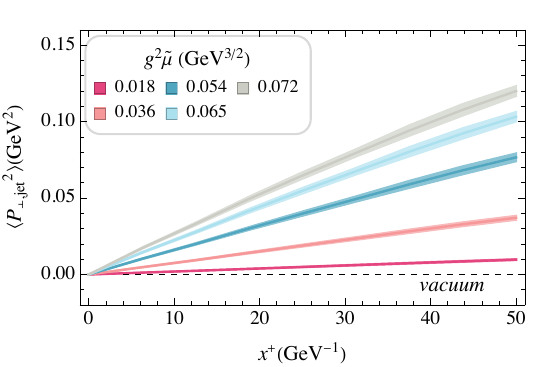}  
  \subfigure[
    Dressed quark initial state
  \label{fig:pperp_dress}
]{  
\includegraphics[width=0.44\textwidth]{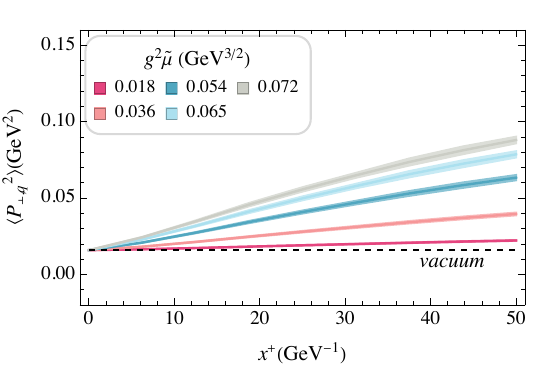}
\includegraphics[width=0.44\textwidth]{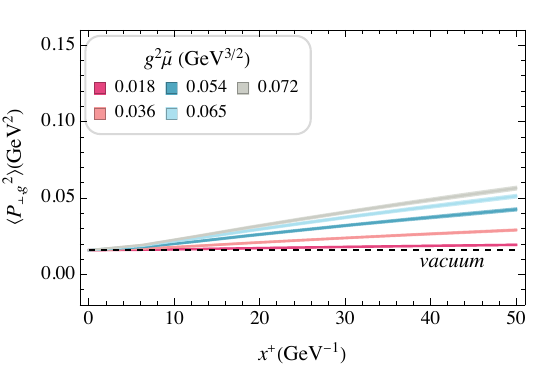}  
}
  \caption{
Momentum broadening of (a) bare quark initial state, and (b) dressed quark initial state.
  In each case, the momentum square $\braket{P_{\perp,\text{CM}}^2}$, $\braket{P_{\perp,q}^2}$, $\braket{P_{\perp,g}^2}$, and $\braket{P_{\perp,\text{jet}}^2}$ are plotted as a function of the evolution time at various charge density.
  }
  \label{fig:pT_broadening}
\end{figure*}

\begin{figure*}[htp!]
  \centering 
    \includegraphics[width=0.44\textwidth]{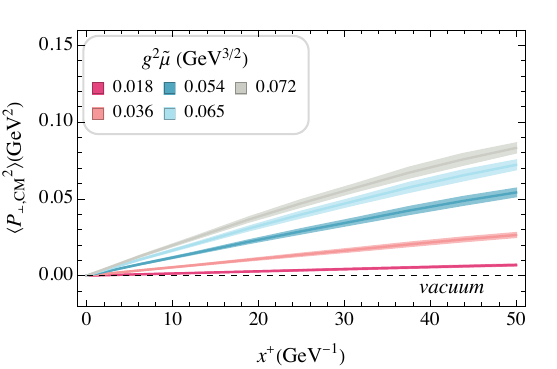}
    \includegraphics[width=0.44\textwidth]{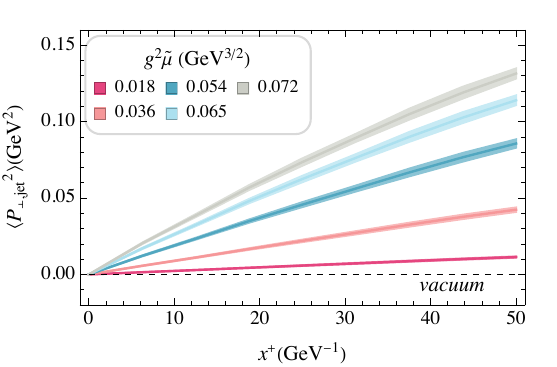}
  \subfigure[
   Timelike quark initial state
    \label{fig:pperp_Time}
  ]{
  \includegraphics[width=0.44\textwidth]{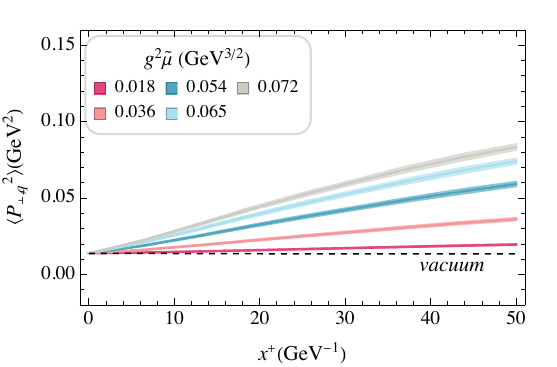}
  \includegraphics[width=0.44\textwidth]{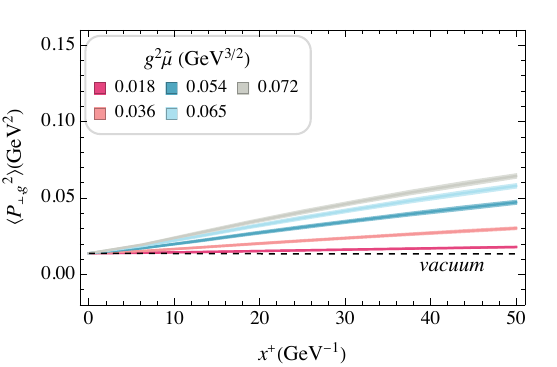}
  }
 \includegraphics[width=0.44\textwidth]{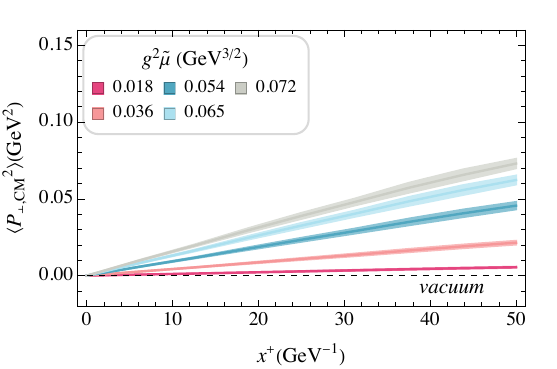}
  \includegraphics[width=0.44\textwidth]{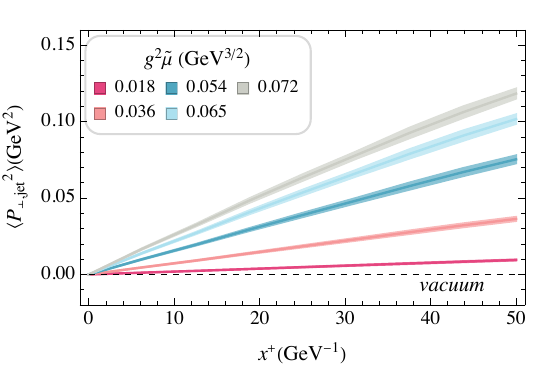} 
  \subfigure[
    Spacelike quark initial state
  \label{fig:pperp_Space}
]{ \includegraphics[width=0.44\textwidth]{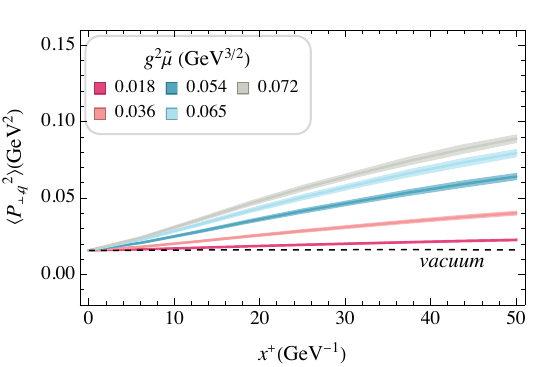}
\includegraphics[width=0.44\textwidth]{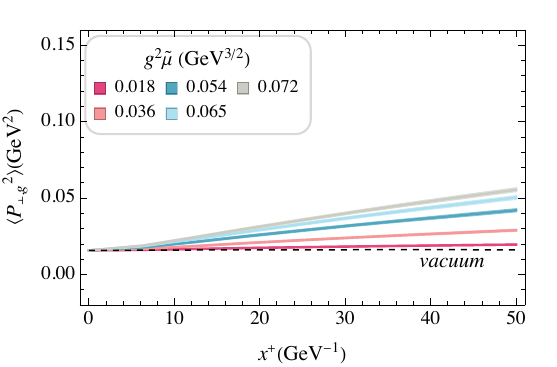}  }
  \caption{
Momentum broadening of (a) timelike quark initial state, and (b) spacelike quark initial state.
In each case, the momentum square $\braket{P_{\perp,\text{CM}}^2}$, $\braket{P_{\perp,q}^2}$, $\braket{P_{\perp,g}^2}$, and $\braket{P_{\perp,\text{jet}}^2}$ are plotted as a function of the evolution time at various charge density.
  }
  \label{fig:pT_broadening_continue}
\end{figure*}

We calculate and compare the momentum broadening for different jet initial states, the bare quark, the dressed/on-shell quark, and off-shell dressed quarks. The results are shown in Figs.~\ref{fig:pT_broadening} and \ref{fig:pT_broadening_continue}. In all four cases, we see that the stronger the medium, the more profound the momentum broadening effect, in terms of $\braket{P_{\perp,\text{CM}}^2}$, $\braket{P_{\perp,q}^2}$, $\braket{P_{\perp,g}^2}$, and $\braket{P_{\perp,\text{jet}}^2}$.

In the case of the bare quark initial state, as shown in Fig.~\ref{fig:pperp_bare}, the $\braket{P_{\perp,\text{CM}}^2}$ momentum remains 0 in the vacuum, and grows linearly over time in medium. The $\braket{P_{\perp,\text{jet}}^2}$ momentum is also constant over time in vacuum but the value is finite, since a bare quark includes some excited states which have a nonzero quark relative momentum of the quark with respect to the CM of the $\ket{qg}$ state. It also grows linearly over time in medium.
On the contrary, the $\braket{P_{\perp,q}^2}$ and $\braket{P_{\perp,g}^2}$ momenta change over time in the vacuum, as a result of the $q\to q+g$ splitting process. The pattern of initially rapid increase and then slows down align with that in the evolution of $P_{\ket{qg}}$ in Fig.~\ref{fig:q_Pg}.

For the dressed quark initial state, as shown in Fig.~\ref{fig:pperp_dress}, the behavior of  $\braket{P_{\perp,\text{CM}}^2}$ is similar to that of the bare quark, whereas the $\braket{P_{\perp,\text{jet}}^2}$ momentum is zero in vacuum, unlike the bare quark.
Since the entire dressed quark state is within the `jet', the two quantities are equal in vacuum. They increase approximately linearly over time in medium.
The quark and gluon momenta $\braket{P_{\perp,q}^2}$ and $\braket{P_{\perp,g}^2}$ have finite values initially and in the vacuum, since the dressed quark state contains  $\ket{qg}$ components with different relative quark and gluon momenta. 
Their in-medium behavior also seem to be linear.
In the scenarios of off-shell dressed quark initial state, as shown in Figs.~\ref{fig:pperp_Time} and ~\ref{fig:pperp_Space}, the overall pattern of the momentum broadening is very similar to that in the on-shell quark case, with small differences in quantity, as the states themselves largely overlap.

\begin{figure*}[htp!]
  \centering
  \includegraphics[width=.44\textwidth]{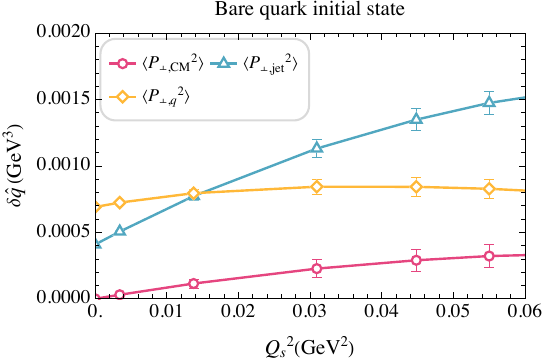}
\includegraphics[width=.44\textwidth]{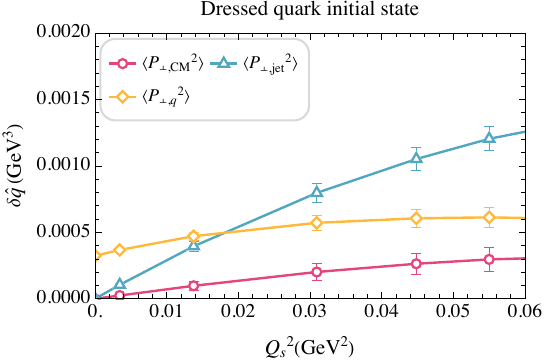}\\
~\\
\includegraphics[width=.44\textwidth]{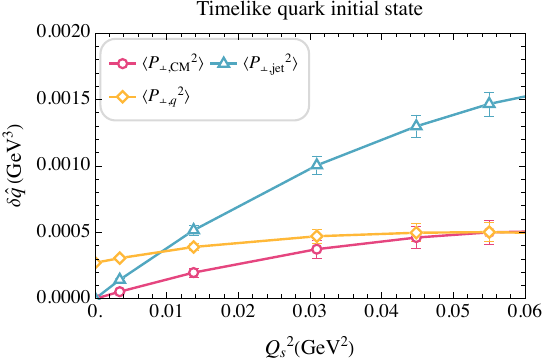}
\includegraphics[width=.44\textwidth]{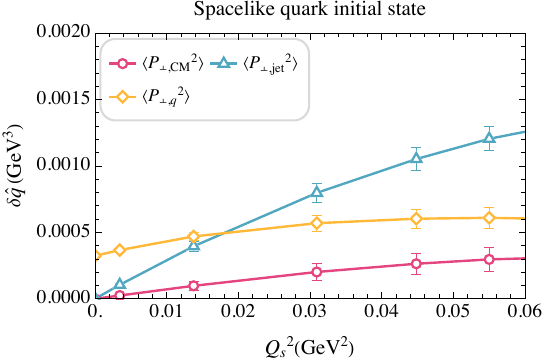}
  \caption{
  The radiative  correction to $\hat q$ as defined in Eq.~\eqref{eq:delta_qhat}.
 }
  \label{fig:delta_qhat}
\end{figure*}

It is also interesting to analyze the radiative correction to the quenching parameter $\hat{q}$. We define $\delta \hat q$ as the difference of the $\hat q$ that is calculated from the dressed quark initial state and that of a single quark with no allowed gluon radiation in the eikonal limit, 
\begin{align}\label{eq:delta_qhat}
  \delta \hat q \equiv \hat q_{\text{dressed q}}-\hat q_{\text{single q}}\;.
\end{align}
We present the results in Fig.~\ref{fig:delta_qhat}, using both the bare quark and dressed quark initial states. For the former, we calculate $\delta \hat q \equiv \hat q_{\text{bare q}}-\hat q_{\text{single q}}$ accordingly. 
In all four cases, the radiative correction in $\braket{P_{\perp,\text{jet}}^2}$ is sizably larger than that in $\braket{P_{\perp,\text{CM}}^2}$ and $\braket{P_{\perp,q}^2}$. 
Note that here $\braket{P_{\perp,\text{CM}}^2}$ is the quantity to indicate the change in the CM motion, and the invariant quantity to quantify the change in the relative motion is the invariant mass square, $\braket{M^2}$, which we will discuss in the next section.

\subsection{Invariant mass distribution}

\begin{figure*}[htp!]
  \subfigure[
      Bare quark initial state
      \label{fig:M2_bare}
    ]{
  \includegraphics[width=0.44\textwidth]{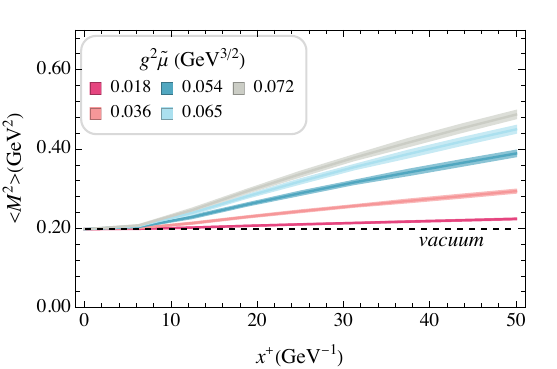}
    }
    \subfigure[
      Dressed/on-shell quark initial state
      \label{fig:M2_dressed}
    ]{
 \includegraphics[width=0.44\textwidth]{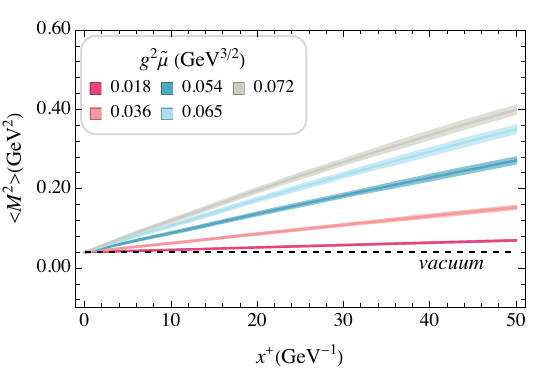}
 }
 \subfigure[
  Timelike quark initial state
   \label{fig:M2_timelike}
 ]{
 \includegraphics[width=0.44\textwidth]{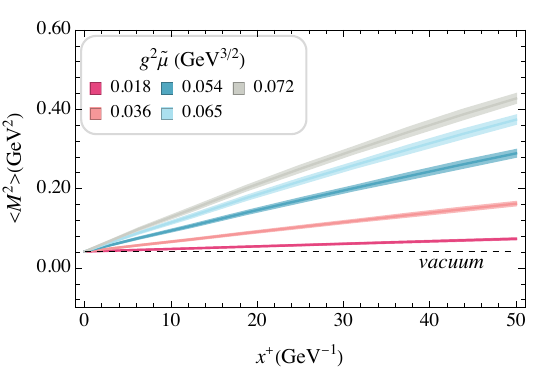}
 }
 \subfigure[
   Spacelike quark initial state
   \label{fig:M2_spacelike}
 ]{
 \includegraphics[width=0.44\textwidth]{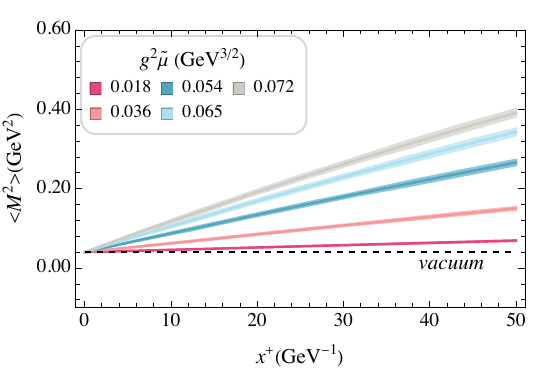}
 }
  \caption{
  The evolution of the invariant mass square in the subspace of (a) bare quark initial state, (b) dressed quark initial state, (c) timelike quark initial state,  and (d) spacelike quark initial state. 
  }
  \label{fig:M2_evolved}
\end{figure*}

We study the evolution of an initially dressed and bare quark states by extracting their invariant mass distribution. In Fig.~\ref{fig:M2_evolved}, we present the expectation value of the invariant mass square $\braket{M^2}$ as a function of the evolution time at various $g^2\tilde\mu$. 
In each of the four cases with a different initial state, the value of $\braket{M^2}$ does not change in the vacuum, as the operator commutes with the evolution Hamiltonian, the same reason we have seen with $ P_{\text{excited}}$ and $ \braket{P_{\perp,CM}^2}$. The invariant mass square of the bare quark is larger than that of the dressed/on-shell quark, indicating that it is not a ground state of the vacuum Hamiltonian. 
Note that the spacelike quark here has $\braket{M^2} > m_q^2$, as the evaluation of the invariant mass square expectation value uses the mass-renormalized Hamiltonian of the on-shell quark, not of the spacelike quark. In other words, the spacelike quark state has been obtained as the ground state for a Hamiltonian with a mass counterterm corresponding to a smaller invariant mass, but then the evolution and the measurement of  $\braket{M^2}$ is done with the mass counterterm corresponding to an on-shell quark.

In the presence of the medium, the bare quark initial state first goes through a slow-changing region, and after around $x^+=6~\GeV$, the growth of $\braket{M^2}$ becomes linear. This behavior aligns with that of $ P_{\text{excited}}$ in Fig.~\ref{fig:q_Pg}, where  the transition to excited states, i.e., larger $\braket{M^2}$ states, becomes significant only after a certain period of time. In comparison, all the three dressed initial states starts to experience the linear growth of $\braket{M^2}$ at the onset of the evolution.

\begin{figure}[t]
  \centering
\includegraphics[width=.44\textwidth]{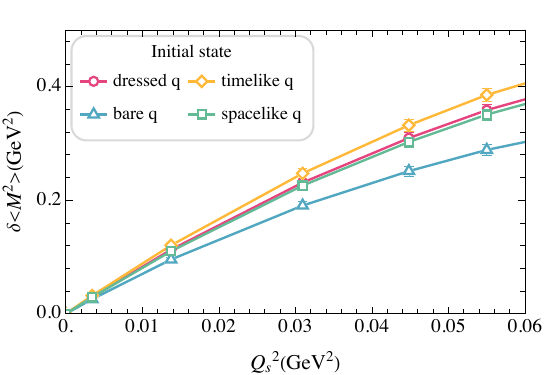}
  \caption{
  Medium-induced invariant mass change at various $Q_s^2$ for different initial states.
 }
  \label{fig:M2_Qs2}
\end{figure}

Then in Fig.~\ref{fig:M2_Qs2}, we present the change of the invariant mass square $M^2$ by the medium at various $Q_s^2$,
\begin{align}
  \delta \braket{M^2}(Q_s^2) = \braket{M^2}(Q_s^2) -\braket{M^2}(0) \;.
\end{align}
Here, $ \braket{M^2}(0)$ is the value of the initial state, i.e., in the vacuum.
The dependence of $ \delta \braket{M^2}$ on $Q_s^2$ is similar to that of $\delta q$ of $\braket{P_{\perp,CM}^2}$ in Fig.~\ref{fig:delta_qhat}, having an increase close to but slower than linear in $Q_s^2$; the timelike quark has the largest values, and the bare quark has the smallest values of the four. In complement to $\braket{P_{\perp,CM}^2}$, which reflects the CM momentum broadening, the change in  $ \braket{M^2}$  indicates the medium induced momentum broadening of the relative momentum between the quark and the gluon.

\begin{figure*}[htp!]
  \subfigure[
    Dressed quark initial state
  ]{
   \begin{minipage}[b]{0.38\textwidth}
    \centering
    \raisebox{0.2\height}{\includegraphics[width=\textwidth]{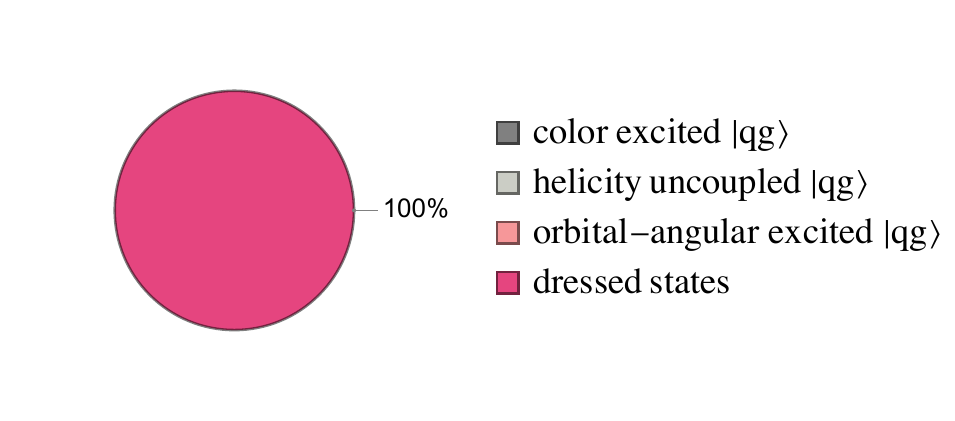}}
  \end{minipage}
  \hspace{0.03\textwidth}
  \begin{minipage}[b]{0.55\textwidth}
    \centering
    \includegraphics[width=\textwidth]{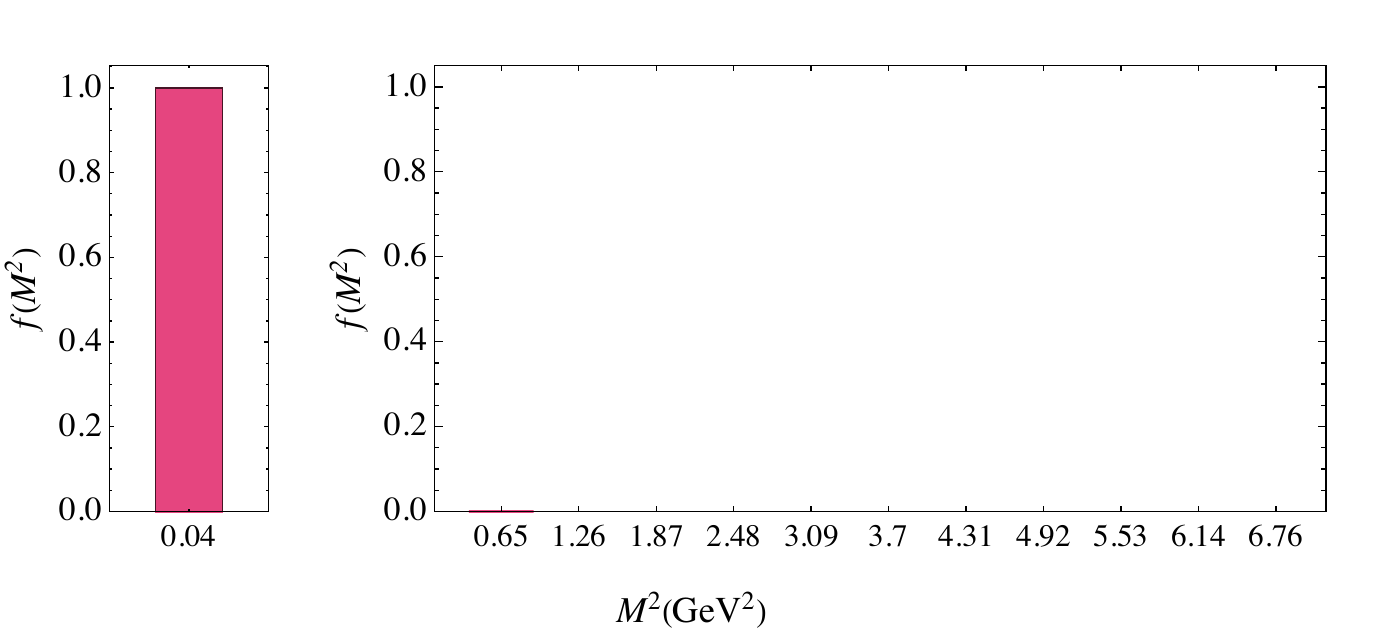}
  \end{minipage}
  }
 \subfigure[
  Evolved dressed quark, $Q_s^2 = 0.013 ~\GeV^2$
]{
 \begin{minipage}[b]{0.38\textwidth}
  \centering
  \raisebox{0.2\height}{\includegraphics[width=\textwidth]{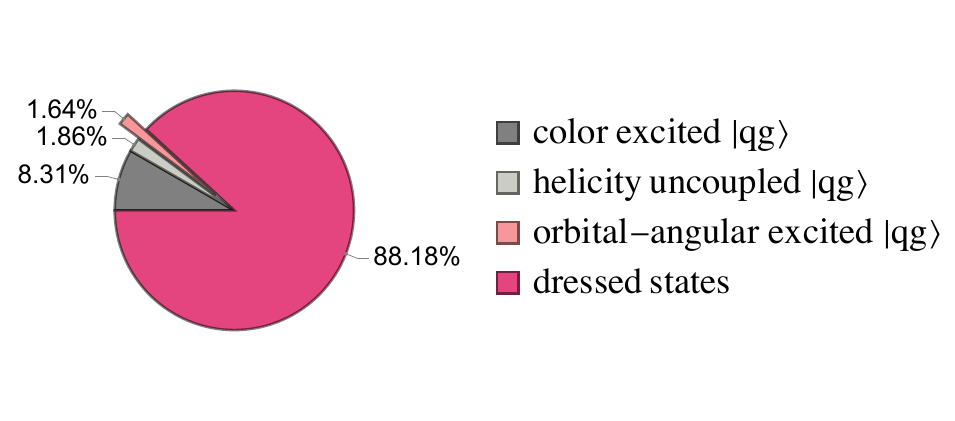}}
\end{minipage}
\hspace{0.03\textwidth}
\begin{minipage}[b]{0.55\textwidth}
  \centering
  \includegraphics[width=\textwidth]{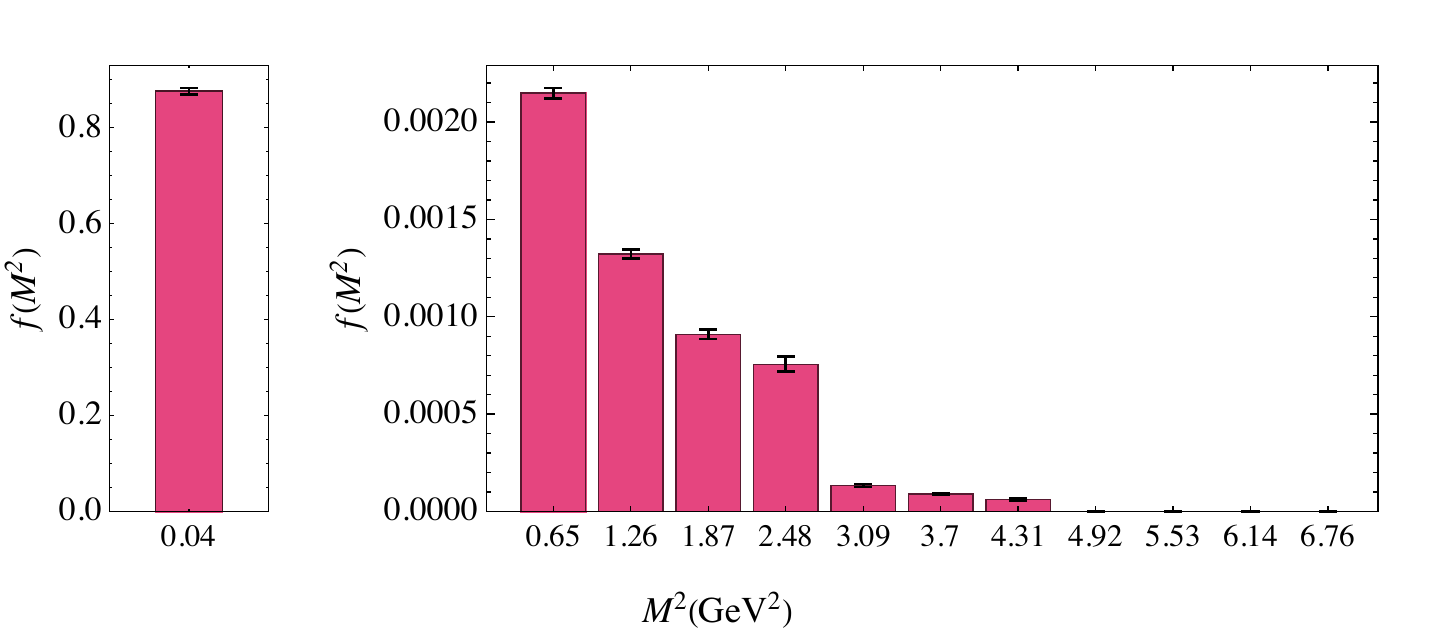}
\end{minipage}
}
 \subfigure[
  Evolved dressed quark, $Q_s^2 = 0.045 ~\GeV^2$
]{
 \begin{minipage}[b]{0.38\textwidth}
  \centering
  \raisebox{0.2\height}{\includegraphics[width=\textwidth]{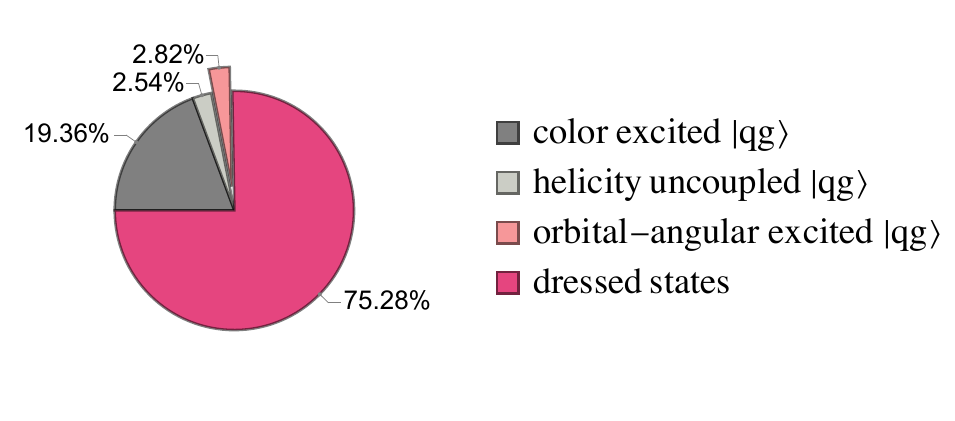}}
\end{minipage}
\hspace{0.03\textwidth}
\begin{minipage}[b]{0.55\textwidth}
  \centering
  \includegraphics[width=\textwidth]{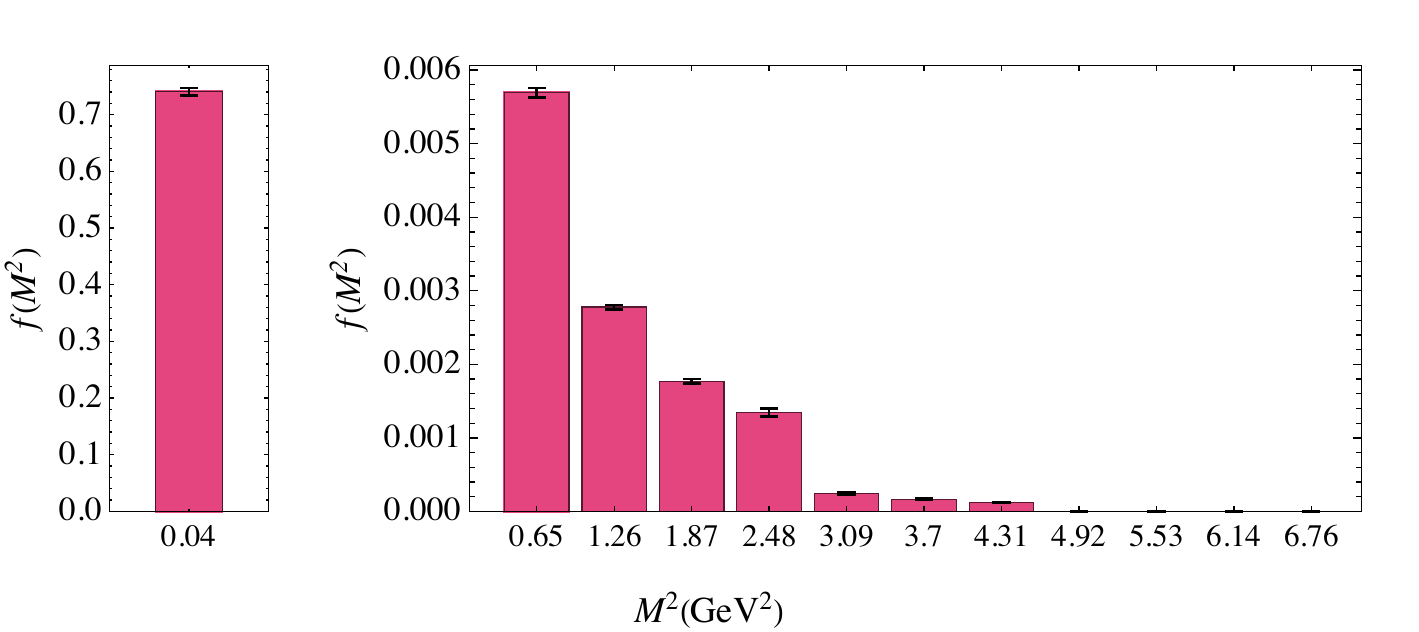}
\end{minipage}
}
  \caption{
  The distribution of the initial and evolved dressed/on-shell quark in the eigenstate space, and the corresponding invariant mass distribution in the dressed states subspace. 
  }
  \label{fig:pie_spectrum_dressed}
\end{figure*}

\begin{figure*}[htp!]
  \subfigure[
    Bare quark initial state
  ]{
   \begin{minipage}[b]{0.38\textwidth}
    \centering
    \raisebox{0.2\height}{\includegraphics[width=\textwidth]{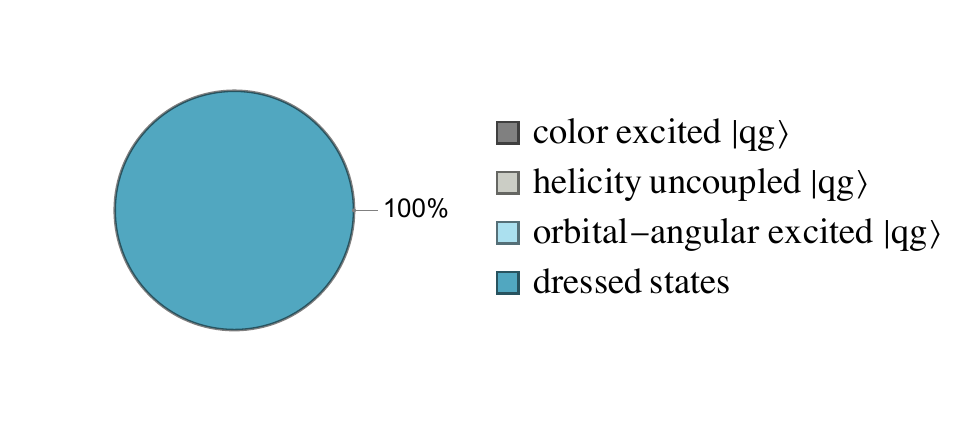}}
  \end{minipage}
  \hspace{0.03\textwidth}
  \begin{minipage}[b]{0.55\textwidth}
    \centering
    \includegraphics[width=\textwidth]{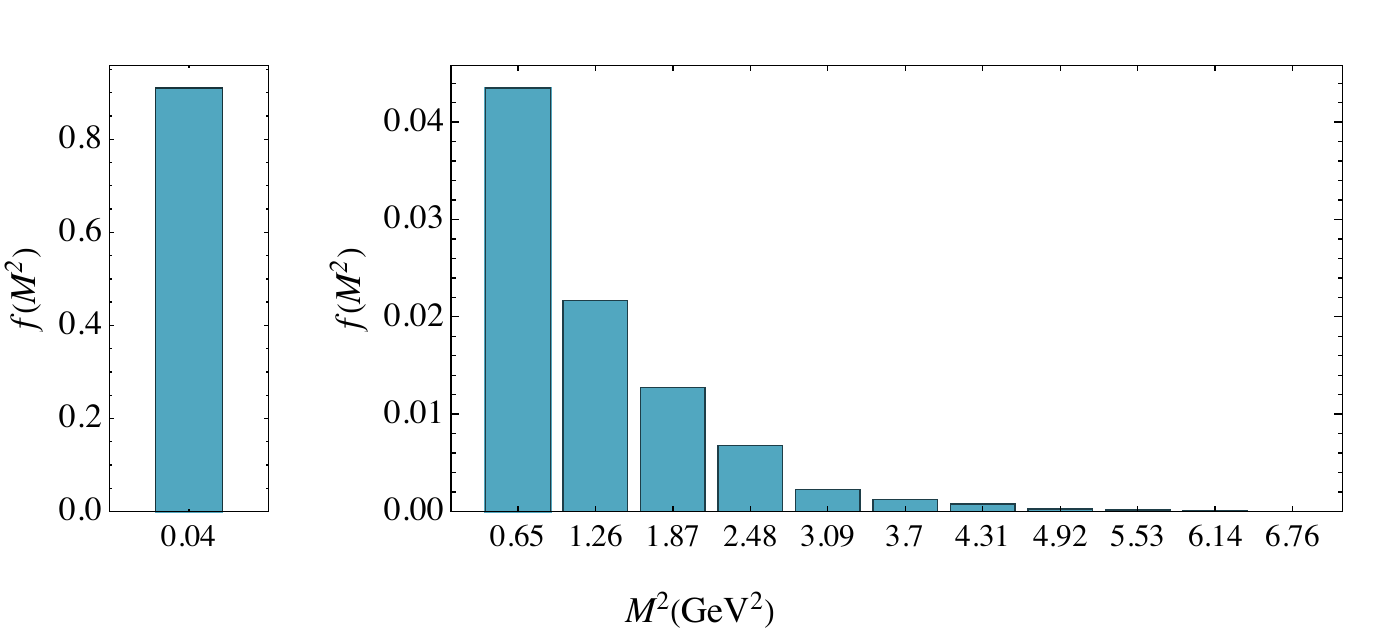}
  \end{minipage}
  }
 \subfigure[
  Evolved bare quark, $Q_s^2 = 0.013 ~\GeV^2$
]{
 \begin{minipage}[b]{0.38\textwidth}
  \centering
  \raisebox{0.2\height}{\includegraphics[width=\textwidth]{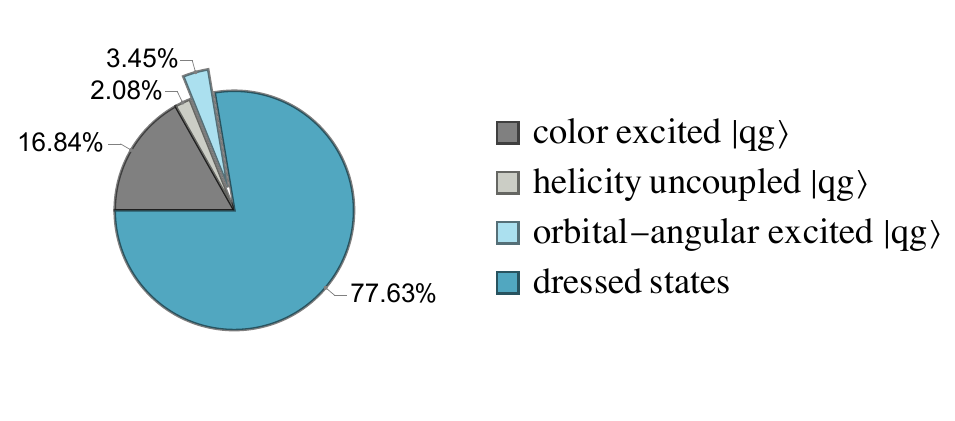}}
\end{minipage}
\hspace{0.03\textwidth}
\begin{minipage}[b]{0.55\textwidth}
  \centering
  \includegraphics[width=\textwidth]{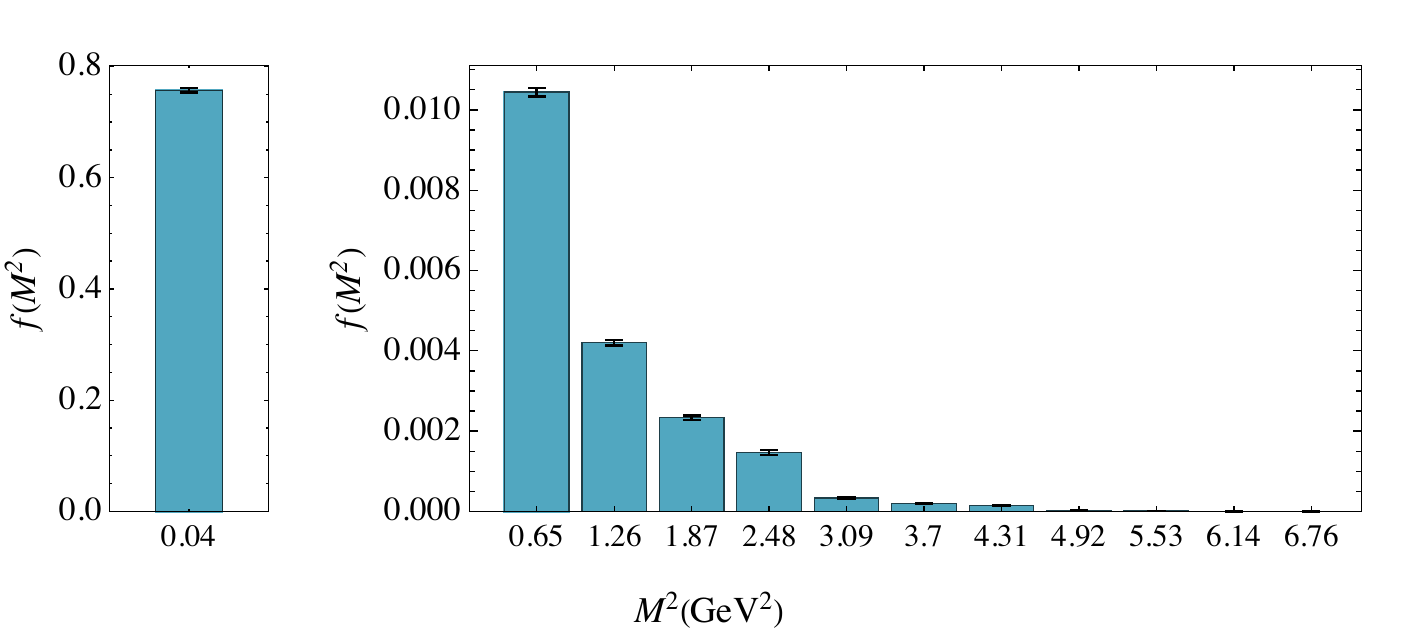}
\end{minipage}
}
 \subfigure[
  Evolved bare quark, $Q_s^2 = 0.045 ~\GeV^2$
]{
 \begin{minipage}[b]{0.38\textwidth}
  \centering
  \raisebox{0.2\height}{\includegraphics[width=\textwidth]{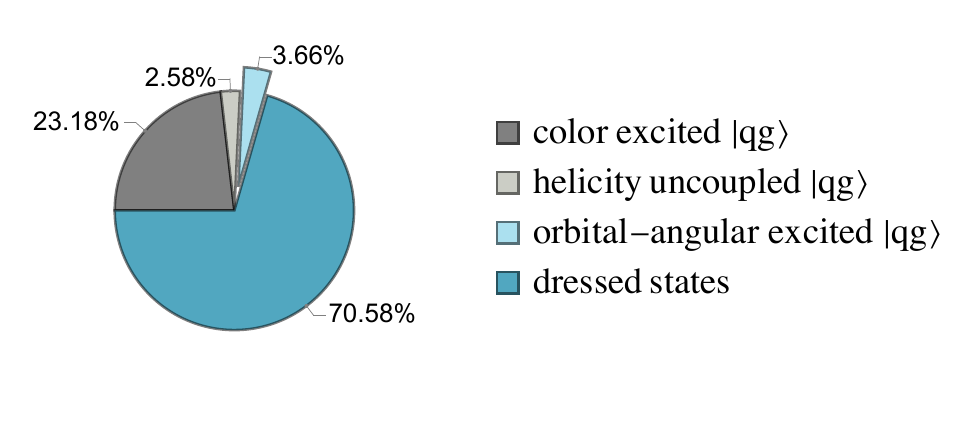}}
\end{minipage}
\hspace{0.03\textwidth}
\begin{minipage}[b]{0.55\textwidth}
  \centering
  \includegraphics[width=\textwidth]{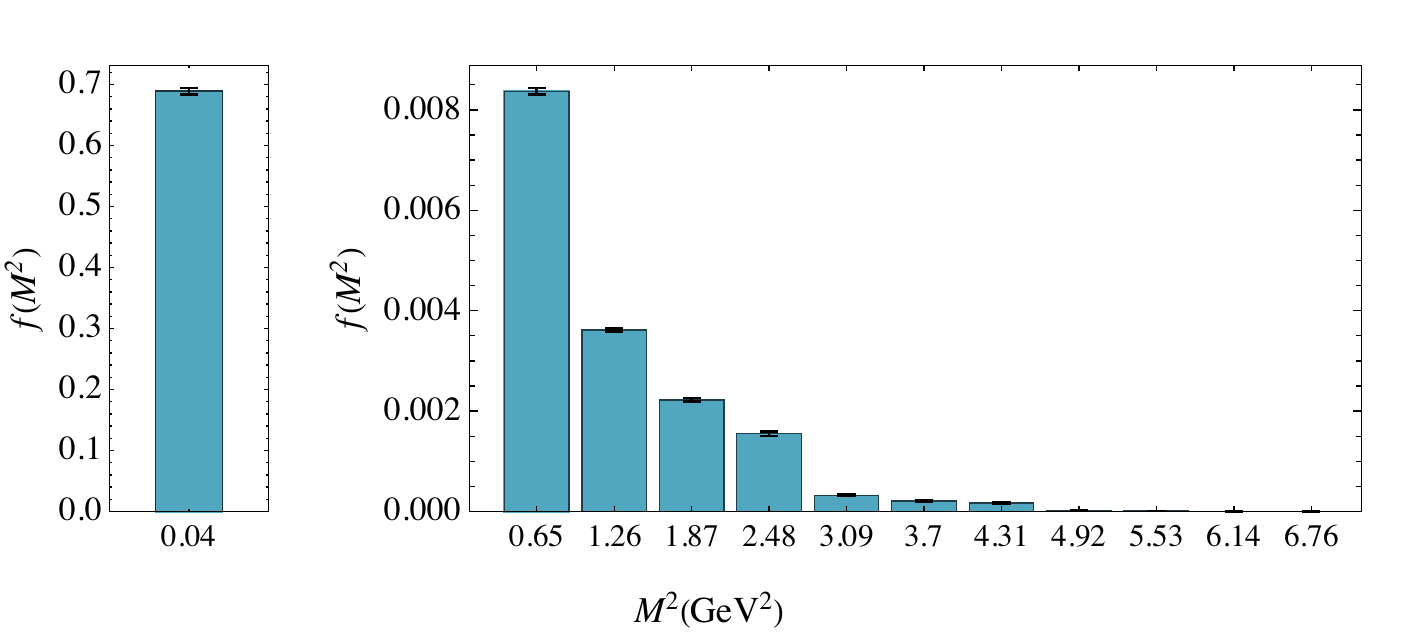}
\end{minipage}
}
  \caption{
  The distribution of the initial and evolved bare quark in the eigenstate space, and the corresponding invariant mass distribution in the dressed states subspace. 
  }
  \label{fig:pie_spectrum_bare}
\end{figure*}

\begin{figure*}[htp!]
  \subfigure[
    Timelike quark initial state
  ]{
   \begin{minipage}[b]{0.38\textwidth}
    \centering
    \raisebox{0.2\height}{\includegraphics[width=\textwidth]{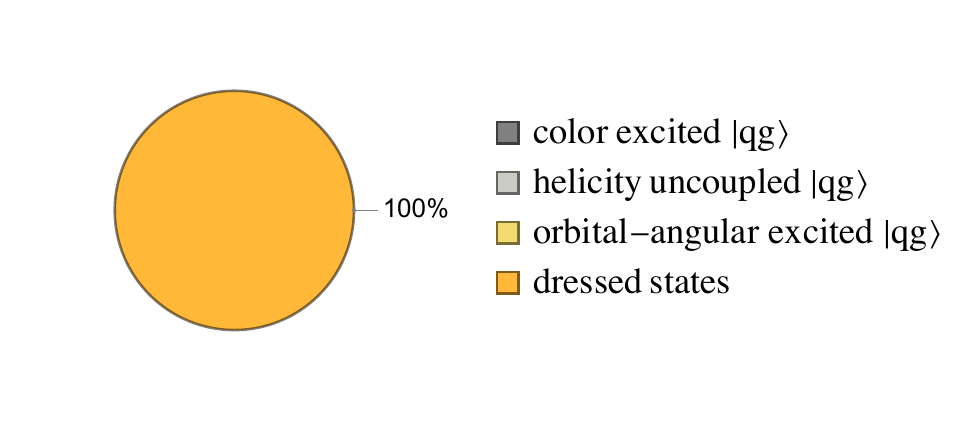}}
  \end{minipage}
  \hspace{0.03\textwidth}
  \begin{minipage}[b]{0.55\textwidth}
    \centering
    \includegraphics[width=\textwidth]{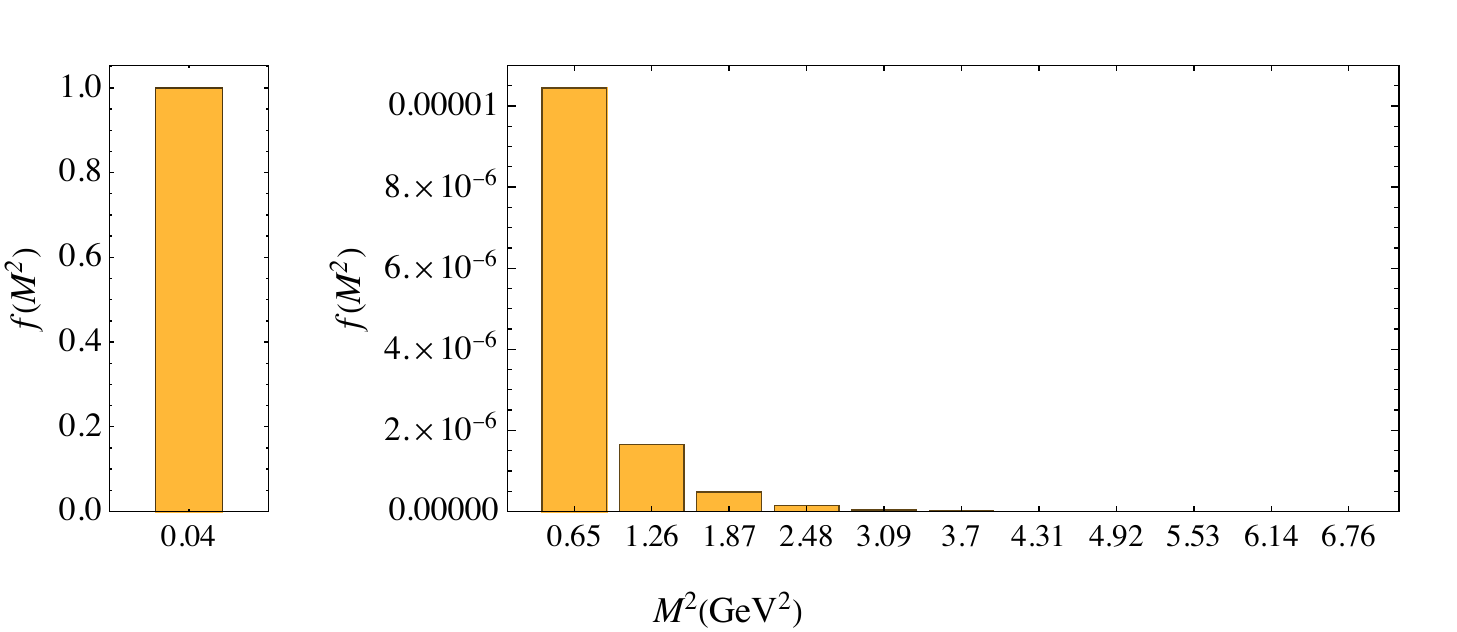}
  \end{minipage}
  }
 \subfigure[
  Evolved timelike quark, $Q_s^2 = 0.013 ~\GeV^2$
]{
 \begin{minipage}[b]{0.38\textwidth}
  \centering
  \raisebox{0.2\height}{\includegraphics[width=\textwidth]{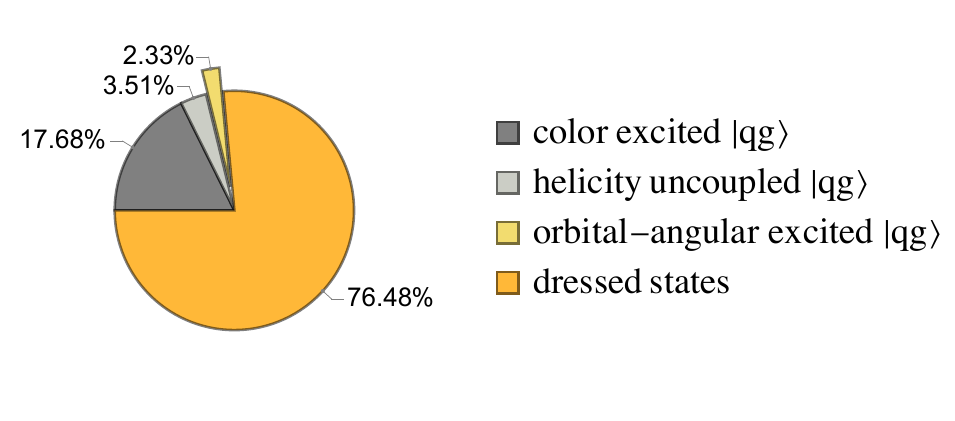}}
\end{minipage}
\hspace{0.03\textwidth}
\begin{minipage}[b]{0.55\textwidth}
  \centering
  \includegraphics[width=\textwidth]{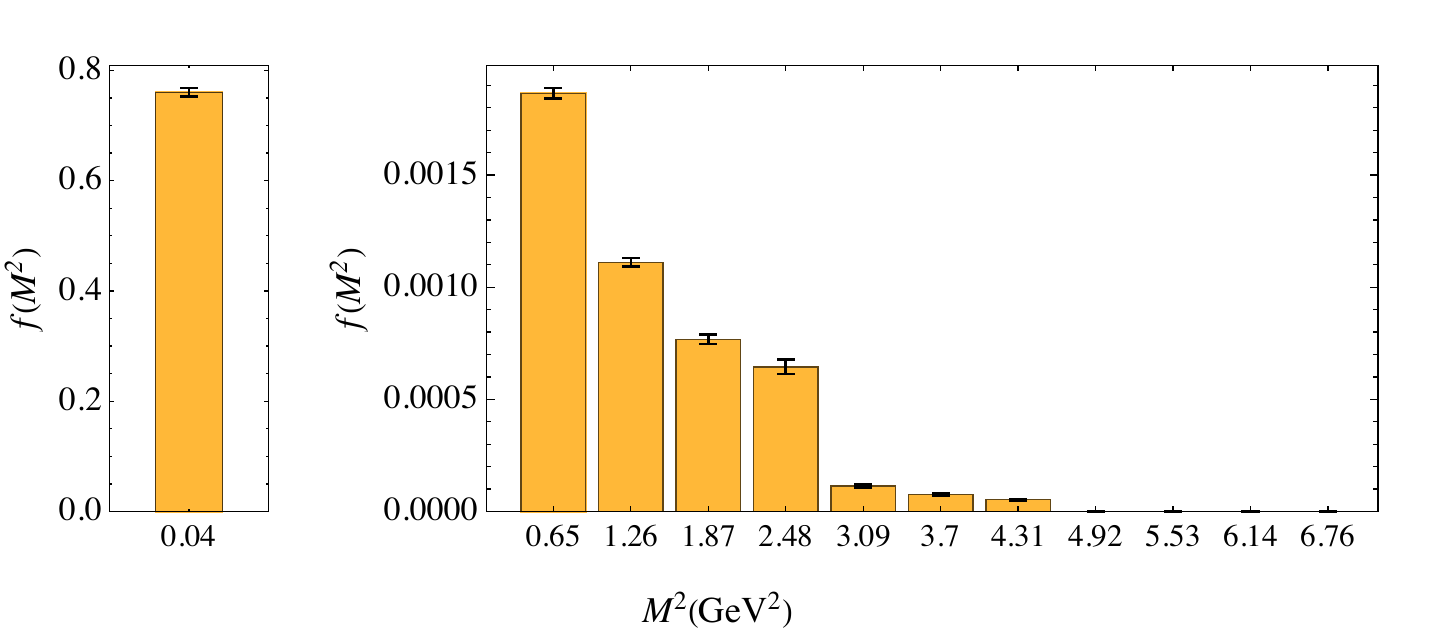}
\end{minipage}
}
 \subfigure[
  Evolved timelike quark, $Q_s^2 = 0.045 ~\GeV^2$
]{
 \begin{minipage}[b]{0.38\textwidth}
  \centering
  \raisebox{0.2\height}{\includegraphics[width=\textwidth]{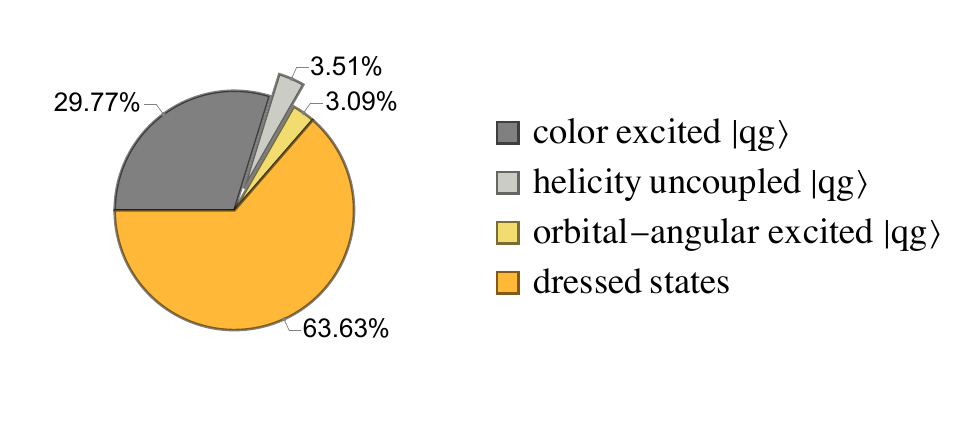}}
\end{minipage}
\hspace{0.03\textwidth}
\begin{minipage}[b]{0.55\textwidth}
  \centering
  \includegraphics[width=\textwidth]{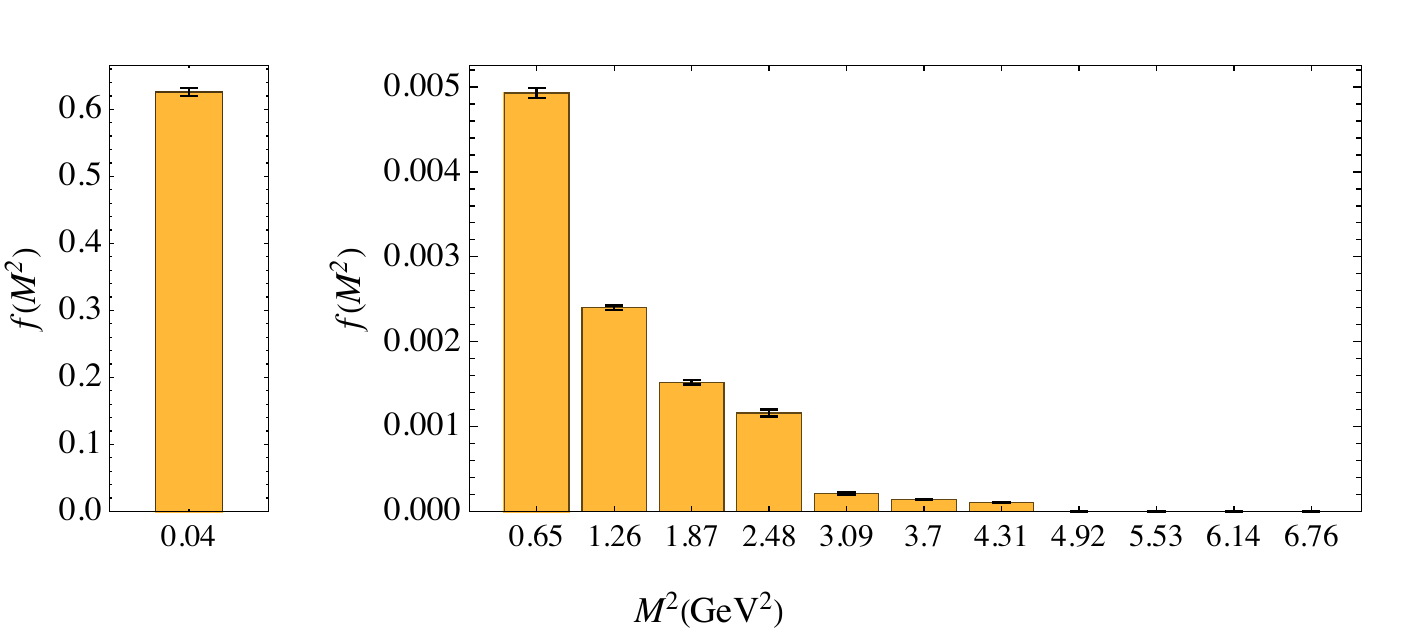}
\end{minipage}
}
  \caption{
  The distribution of the initial and evolved timelike quark in the eigenstate space, and the corresponding invariant mass distribution in the dressed states subspace. 
  }
  \label{fig:pie_spectrum_timelike}
\end{figure*}

\begin{figure*}[htp!] 
  \subfigure[
    Spacelike quark initial state
  ]{
   \begin{minipage}[b]{0.38\textwidth}
    \centering
    \raisebox{0.2\height}{\includegraphics[width=\textwidth]{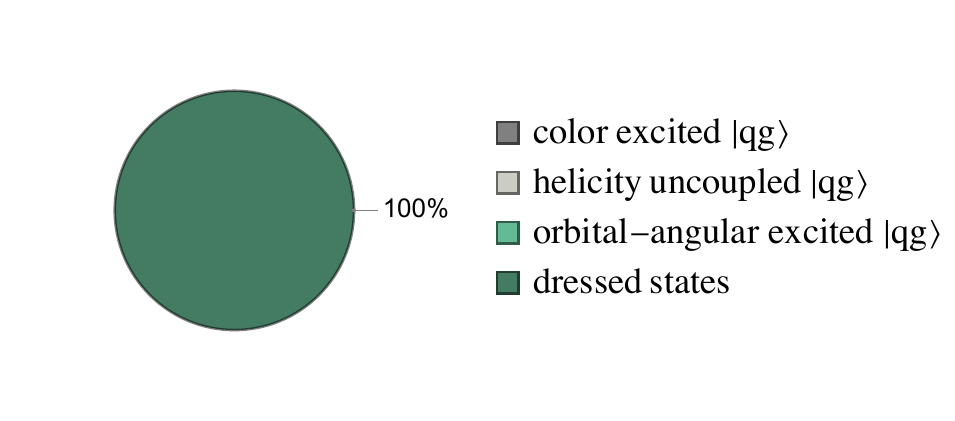}}
  \end{minipage}
  \hspace{0.03\textwidth}
  \begin{minipage}[b]{0.55\textwidth}
    \centering
    \includegraphics[width=\textwidth]{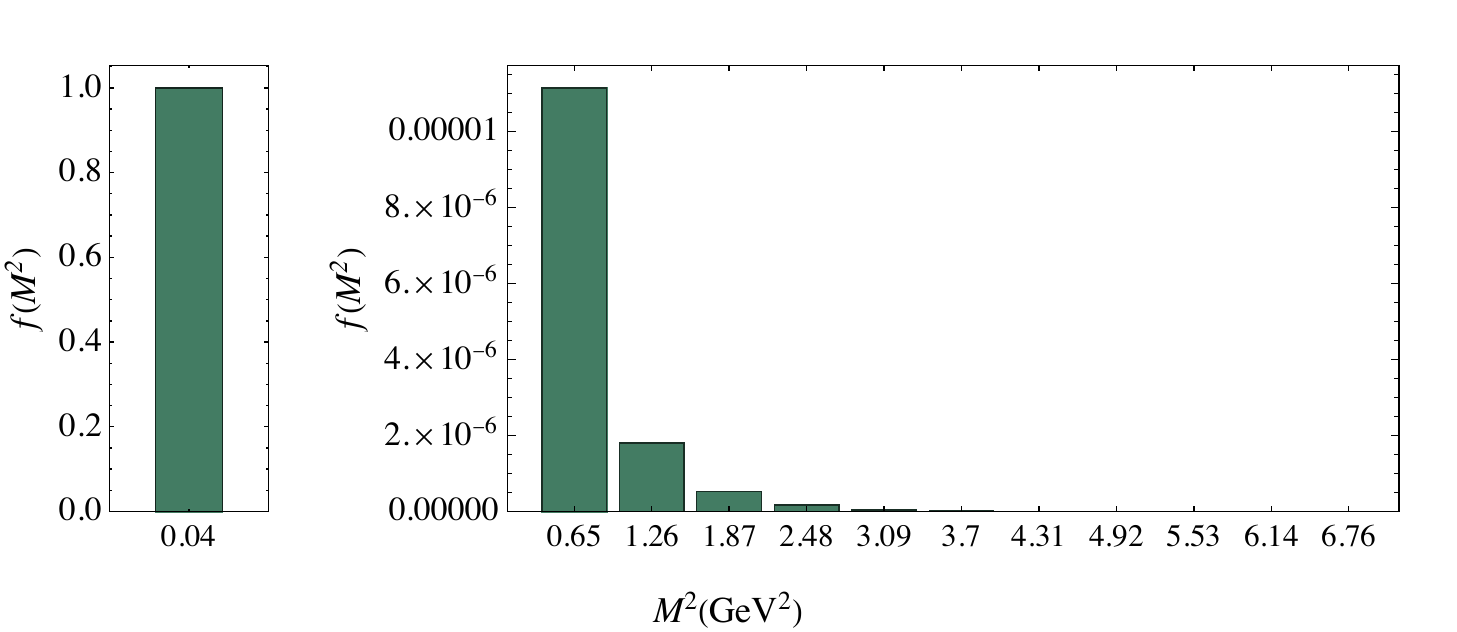}
  \end{minipage}
  }
 \subfigure[
  Evolved spacelike quark, $Q_s^2 = 0.013 ~\GeV^2$
]{
 \begin{minipage}[b]{0.38\textwidth}
  \centering
  \raisebox{0.2\height}{\includegraphics[width=\textwidth]{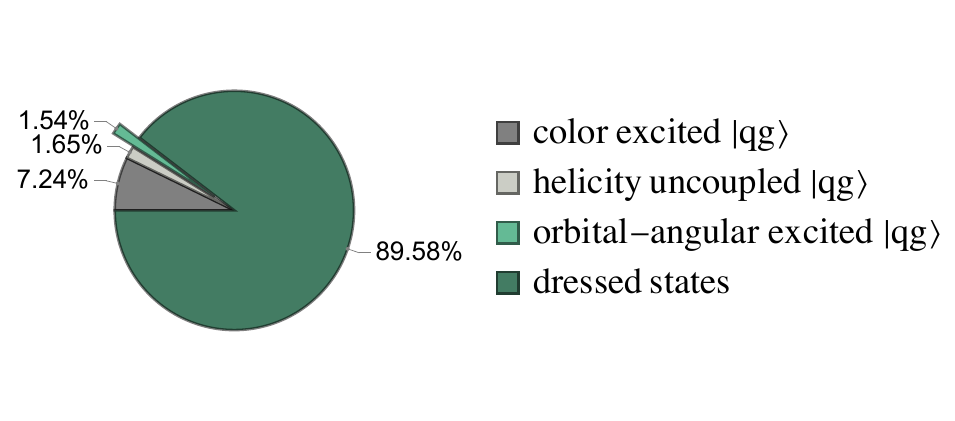}}
\end{minipage}
\hspace{0.03\textwidth}
\begin{minipage}[b]{0.55\textwidth}
  \centering
  \includegraphics[width=\textwidth]{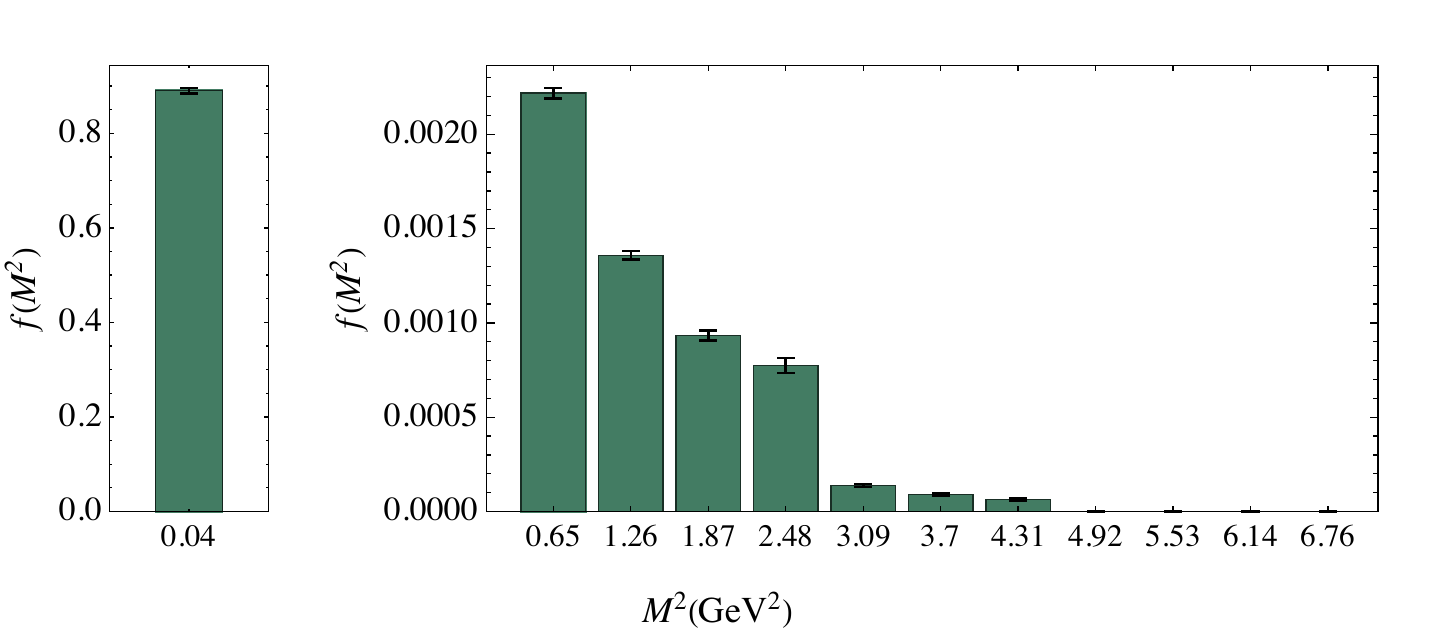}
\end{minipage}
}
 \subfigure[
  Evolved spacelike quark, $Q_s^2 = 0.045 ~\GeV^2$
]{
 \begin{minipage}[b]{0.38\textwidth}
  \centering
  \raisebox{0.2\height}{\includegraphics[width=\textwidth]{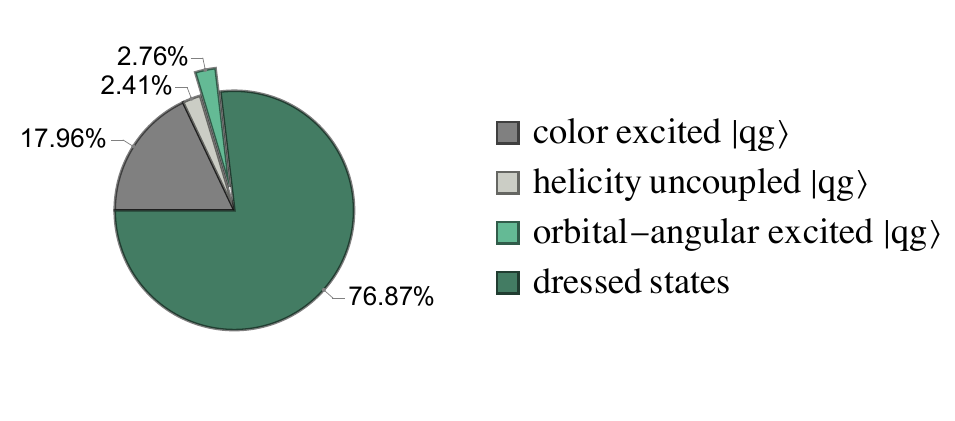}}
\end{minipage}
\hspace{0.03\textwidth}
\begin{minipage}[b]{0.55\textwidth}
  \centering
  \includegraphics[width=\textwidth]{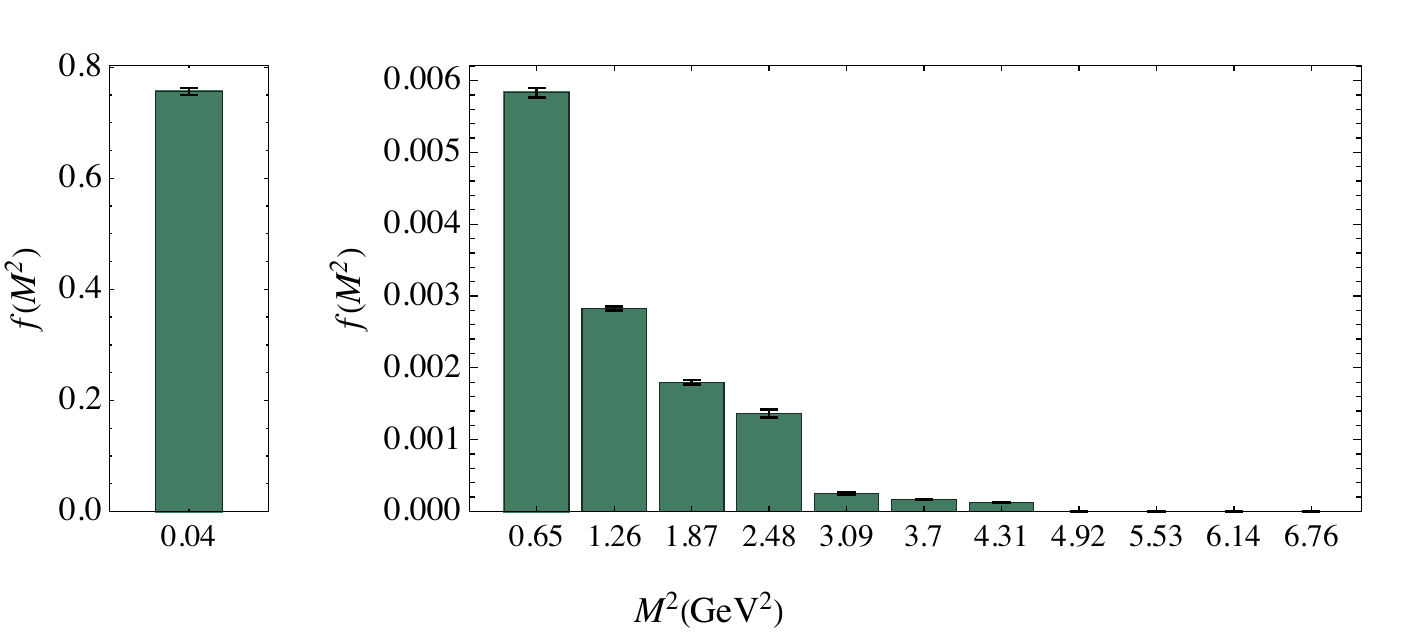}
\end{minipage}
}
  \caption{
  The distribution of the initial and evolved spacelike quark in the eigenstate space, and the corresponding invariant mass distribution in the dressed states subspace. 
  }
  \label{fig:pie_spectrum_spacelike}
\end{figure*}

In Figs.~\ref{fig:pie_spectrum_dressed}, ~\ref{fig:pie_spectrum_bare}~\ref{fig:pie_spectrum_timelike} and ~\ref{fig:pie_spectrum_spacelike}, we present the  distribution of the initial and evolved states in the eigenstate space and the invariant mass distribution in the dressed states subspace, for the four different cases respectively. 
All the initial states are in the dressed state subspace, as they are each assigned with quantum numbers of a physical quark. In each case, the stronger medium results in larger occupation in the excited states, including color excited, helicity uncoupled, orbital-angular excited quark-gluon states, and dressed quark-gluon states. 
The projection onto different partitions of the eigenstate basis offers an opportunity for a deeper analysis of the internal structure of a jet in a colored medium.

\section{Summary and outlook}\label{sec:summary}
In this work, we first presented an extensive study on the formulation of the physical dressed quark states in the light-front Hamiltonian formalism. 
We implement a sector-dependent mass renormalization scheme to the quark mass and solve for the eigenstates via diagonalizing the vacuum QCD Hamiltonian. The resulting wavefunctions and spectra are useful for various studies of jets and high energy scattering processes, such as gluon emission and particle production. We compared the obtained light-front wavefunction with the perturbative result, and examined the basis dependence of the mass counterterm and the wavefunction renormalization coefficient $Z_2$. The developed method is also applicable to gluon jet. We did not implement the renormalization of the coupling constant in this work, since it  only becomes necessary  when including higher Fock sectors.

We then performed numerical simulations of the real-time jet evolution using four different initial states, the bare quark, on-shell dressed quark, timelike quark, and spacelike quark interacting with a colored background field. We studied gluon radiation, cross sections, momentum broadening, and invariant mass distribution in these different scenarios.
The comparison of the two scenarios, with or without gluon-dressing initially, helped reveal the relevance of the gluon component in a dressed asymptotic quark state.
We observed, as shown in Fig.~\ref{fig:gluon_radiation_Qs}, a larger medium-induced gluon emission in the case of the dressed quark initial state compared to that of the bare quark initial state.
The knowledge of the eigenstates also facilitates our understanding of the jet momentum broadening. We quantified this in terms of the total CM momentum of the jet, and in terms of a measure of the quark momentum within the jet.
The results in terms of the jet quark momentum correspond to a larger momentum broadening compared to that using the CM or a single quark transverse momentum.
We extracted the distribution of the initial and the evolved states in the eigenstate space and the invariant mass distribution in the dressed state subspace. 
Such an analysis can be useful for the succeeding process of jet fragmentation. 

With the recent applications of quantum simulation to jet evolution based on tBLFQ ~\cite{ Barata:2022wim,Barata:2023clv,Qian:2024gph}, we also foresee the implementation of quantum simulation to the dressed states in the future. Based on the development in this work, the preparation of the initially dressed state on a quantum computer can be achieved with eigensolvers using the variation methods, and the quantum simulation of the state follows.

\section*{Acknowledgement}
We are grateful to Xabier Feal, Zhongbo Kang, Yang Li, Wenyang Qian, James P. Vary, and Xin-Nian Wang for their valuable discussions. 
ML and CS are supported by the European Research Council under project ERC-2018-ADG-835105 YoctoLHC; by Maria de Maeztu excellence unit grant CEX2023-001318-M and project PID2020-119632GB-I00 funded by MICIU/AEI/10.13039/501100011033; and by ERDF/EU.  TL is supported by the Research Council of Finland, the Centre of Excellence in Quark Matter (projects 346324 and 364191). 
XZ is supported by new faculty startup funding by the Institute of Modern Physics, Chinese Academy of Sciences, by Key Research Program of Frontier Sciences, Chinese Academy of Sciences, Grant No. ZDBS-LY-7020, by the Foundation for Key Talents of Gansu Province, by the Central Funds Guiding the Local Science and Technology Development of Gansu Province, Grant No. 22ZY1QA006, by Gansu International Collaboration and Talents Recruitment Base of Particle Physics (2023-2027), by International Partnership Program of the Chinese Academy of Sciences, Grant No. 016GJHZ2022103FN, by National Natural Science Foundation of China, Grant No. 12375143, by National Key R\&D Program of China, Grant No. 2023YFA1606903 and by the Strategic Priority Research Program of the Chinese Academy of Sciences, Grant No. XDB34000000.

\appendix

\begin{widetext}
  \section{convention}\label{app:convention}
\subsection{Discretization}
  We can consider that the system is contained in a box of finite volume $\Omega=2L{(2L_\perp)}^2$. 
We have introduced two artificial length parameters, $L$ in the longitudinal direction and $L_\perp$ in transverse directions. 
The 2-dimensional transverse space is a periodic lattice extending from $-L_\perp$ to $L_\perp$ for each side. The number of transverse lattice sites is
$2N_\perp$, giving the lattice spacing $a_\perp=L_\perp/N_\perp$.  For any vector in this space, $\vec r_\perp=(r^1,r^2)$,
\[
  r^i= n_i a_\perp (i = 1,2) , \quad n_i=-N_\perp,-N_\perp+1,\ldots,N_\perp-1 .
\]
It follows from the periodic boundary conditions that in the
momentum space, for any vector, $\vec p_\perp=(p^1,p^2)$,
\[
  p^i=k_i d_p (i = 1,2), \quad k_i=-N_\perp,-N_\perp+1,\ldots,N_\perp-1,
\]
where $d_p\equiv \pi/L_\perp$ is the resolution in momentum space. The momentum space extends from $-\pi/a_\perp$ to $\pi/a_\perp$.

The correspondence between the sums and integrals over momenta is
\begin{align}
  \int\frac{\diff^2 \vec p_\perp}{{(2\pi)}^2} \to \frac{1}{{(2L_\perp)}^2}\sum_{k_1, k_2}\;,
  \qquad
  \int \diff^2 \vec r_\perp \to a_\perp^2\sum_{n_1, n_2}\;.
\end{align}
The Dirac delta is converted to the Kronecker delta as follows
\begin{subequations}\label{eq:deltas}
  \begin{align}
    &\int \diff^2 \vec r_\perp e^{-i \vec p_\perp\cdot \vec r_\perp}
    = {(2\pi)}^2\delta^2(\vec p_\perp)
    \to
    \sum_{n_1, n_2}  a_\perp^2 e^{-i (n_1 k_1 + n_2 k_2)\pi/N_\perp} = {(2L_\perp)}^2 \delta_{k_1,0}\delta_{k_2,0}\;,\\
      &\int \diff^2 \vec p_\perp e^{i \vec p_\perp\cdot \vec r_\perp}={(2\pi)}^2\delta^2(\vec r_\perp)
       \to
      \sum_{k_1, k_2} \frac{1}{{(2L_\perp)}^2}  e^{i (n_1 k_1 + n_2 k_2)\pi/N_\perp} =  \frac{1}{a_\perp^2}\delta_{n_1,0}\delta_{n_2,0}\;.
  \end{align}
\end{subequations}
The (inverse-)Fourier transformation becomes
\begin{align}
  \begin{split}
  f(n_1, n_2)= & \frac{1}{ {(2 L_\perp)}^2} \sum_{k_1, k_2} e^{i(n_1 k_1 + n_2 k_2)\pi/N_\perp}\tilde f(k_1, k_2),\\
  \tilde f(k_1, k_2)  = & \sum_{n_1, n_2} a_\perp^2 e^{-i(n_1 k_1 + n_2 k_2)\pi/N_\perp}  f(n_1, n_2)\;.
  \end{split}
\end{align}

In the longitudinal direction, $-L\le x^- \le L$, we impose periodic boundary conditions for bosons and  anti-periodic boundary conditions for fermions such that the longitudinal momentum space is discretized as,
\begin{align}
    &p^+ = 
    \begin{dcases}
        &\frac{2\pi}{L}k^+, ~\text{with}~ k^+ = \frac{1}{2}, \frac{3}{2}, \ldots,\infty 
        ~\text{for fermions}\;, \\
        &\frac{2\pi}{L}k^+, ~\text{with}~ k^+ = 1,2, \ldots,\infty
        ~\text{for bosons}\;.
    \end{dcases}
\end{align}
The unit of $p^+$ is $d_+\equiv 2\pi/L$.

\subsection{Quantization in a discrete space}\label{app:modes}
The mode expansion for field operators on such discrete momentum basis is
\begin{subequations}\label{eq:box_modes}
  \begin{align}
   & \Psi^{\text{Box}}_c(x)=\sum_{\bar{\alpha}} \frac{1}{\sqrt{ p^+ 2L (2L_\perp)^2}}
   [b_{\bar{\alpha},c} u(p,\lambda)
     e^{-ip\cdot x}+d^\dagger_{\bar{\alpha},c} v(p,\lambda) e^{ip\cdot x}]\;, \\
   &A^{\text{Box}}_{\mu,a}(x)=\sum_{\bar{\alpha}} \frac{1}{\sqrt{ p^+ 2L (2L_\perp)^2}}
    [a_{\bar{\alpha},a}\epsilon_\mu(p,\lambda)e^{-ip\cdot x}+a^\dagger_{\bar{\alpha},a}\epsilon_\mu^*(p,\lambda)e^{ip\cdot x}]   \;.   
    \end{align}
\end{subequations}
where $p\cdot x=1/2p^+ x^- =\vec p_\perp\cdot \vec x_\perp$ is the 3-product for the spatial components of $p^\mu$ and $x^\mu$.
Each single particle state is specified by five quantum numbers, $\bar{\alpha}=\{ k^+, k^1, k^2, \lambda \}$ and $c$, where $\lambda$ is the light-front helicity, and $c$ is the color index.
Note that this is the same with the basis number $\beta = \{\bar{\alpha},c\}$ defined in our basis representation.
The creation operators $b^{\dagger}_{\bar{\alpha},c}$, $d^{\dagger}_{\bar{\alpha},c}$ and $a^{\dagger}_{\bar{\alpha},a}$
create quarks, antiquarks and gluons with quantum numbers $\bar{\alpha}$ respectively. 
They obey the following commutation and anti-commutation relations,
\begin{align}
    \begin{split}
 & \{b_{\bar{\alpha},c},b^{\dagger}_{\bar{\alpha}',c'}\}=\{d_{\bar{\alpha},c},d_{\bar{\alpha}',c'}^{\dagger}\}
=\delta_{\bar{\alpha},\bar{\alpha}'}
  \delta_{c,c'}\;,\\
 & [a_{\bar{\alpha},a},a^{\dagger}_{\bar{\alpha}',a'}]=\delta_{\bar{\alpha},\bar{\alpha}'}
  \delta_{a,a'}  \;.  
    \end{split}
\end{align}

\subsection{The continuum limit}\label{app:cons}
Since the calculation is performed in the discrete space, it is important to know how the obtained results are valid in the continuum limit. 
Here, we write out the corresponding field operators in the continuous space,
\begin{subequations}\label{eq:con_modes}
\begin{align}
& \Psi_c(x)=\sum_{\lambda} 
\int\frac{\diff^2 p_\perp \diff p^+}{\sqrt{2 p^+ (2\pi)^3}} 
[b_c(p,\lambda) u(p,\lambda)
 e^{-ip\cdot x}+d^\dagger_c(p,\lambda) v(p,\lambda) e^{ip\cdot x}]\;, \\
&A_{\mu,a}(x)=\sum_{\lambda} 
\int\frac{\diff^2 p_\perp \diff p^+}{\sqrt{2 p^+ (2\pi)^3}} 
[a_a(p,\lambda)\epsilon_\mu(p,\lambda)e^{-ip\cdot x}+a^\dagger_a(p,\lambda)\epsilon_\mu^*(p,\lambda)e^{ip\cdot x}]   \;.   
\end{align}
\end{subequations}
The dimension of the field operaters are the same with those defined in the discrete space as given in Eq.~\eqref{eq:box_modes}, $[\Psi]=1.5$ and $[A]=1$. 
The creation and annihilation operators are dimensionful, and they obey the following commutation and anti-commutation relations,
\begin{align}
  \begin{split}
  & \{b_c(p,\lambda),b^{\dagger}_{c'}(p',\lambda')\}
  =\{d_c(p,\lambda),d_{c'}^{\dagger}(p',\lambda')\}
  =\delta^3(p-p')
  \delta_{\lambda,\lambda'}
  \delta_{c,c'}\;,\\
  & [a_a(p,\lambda),a^{\dagger}_{a'}(p',\lambda')]
  =\delta^3(p-p')
  \delta_{\lambda,\lambda'}
  \delta_{a,a'}  \;.  
  \end{split}
\end{align}
The conversion between the discrete and the continous creation/annihilation operators is,
\begin{align}
\begin{split}
& \{
  b_c(p,\lambda), b^\dagger_c(p,\lambda),  a_c(p,\lambda),  a^\dagger_c(p,\lambda) 
\}
\to 
S_{\text{dis}}
\{
b_{\bar\alpha,c},b^\dagger_{\bar\alpha,c},
a_{\bar\alpha,c}, a^\dagger_{\bar\alpha,c}
\}
  \;, 
\end{split}
\end{align}
with the scaling factor for discretization according to the conversion of delta functions given in Eq.~\eqref{eq:deltas},
\begin{align}\label{eq:S_dis}
S_{\text{dis}}\equiv\sqrt{\frac{1}{d_p^2 d_+}  }= \sqrt{\frac{(2 L_\perp)^2 L}{(2\pi)^3}}\;.
\end{align}
\end{widetext}

\begin{widetext}
  \section{Self-induced inertia}\label{app:inst}
In the $\ket{q}+\ket{qg}$ Fock space, the light-front QCD Hamiltonian, as written in Eq.~\eqref{eq:Pmin_QCD}, should in principle also contain the instantaneous quark/gluon interactions $T_{\text{inst.}}$ in the $\ket{q}$ sector, as illustrated in Figs.~\ref{fig:dm_LFpQCD_inst_f} and \ref{fig:dm_LFpQCD_inst_g}.
The truncation scheme is formulated by Tang, Brodsky, and Pauli~\cite{Tang:1991rc}, that is, the instantaneous parton graph is only retained if the corresponding propagating parton graph contributes in the truncated theory. 
It follows that, in our case, the instantaneous terms appear in the $\ket{q}$ but not the $\ket{qg}$ Fock sector.
In this appendix, we will first show that the contributions from the instantaneous terms can be canceled by the mass counterterm through renormalization, without affecting the invariant mass spectrum or intrinsic structure of the physical quark. 
We will also note that in the light-front perturbation theory, the full mass counter term at second order contains the contribution from both $T_{\text{inst.}}$ and the one-loop term (the latter illustrated in Fig.~\ref{fig:dm_LFpQCD_one_loop}), and agrees with the standard result~\cite{Mustaki:1990im}.

\begin{figure*}[htp!]
  \centering 
  \subfigure[The instantaneous quark term \label{fig:dm_LFpQCD_inst_f}]{
    \includegraphics[width=0.2\textwidth]{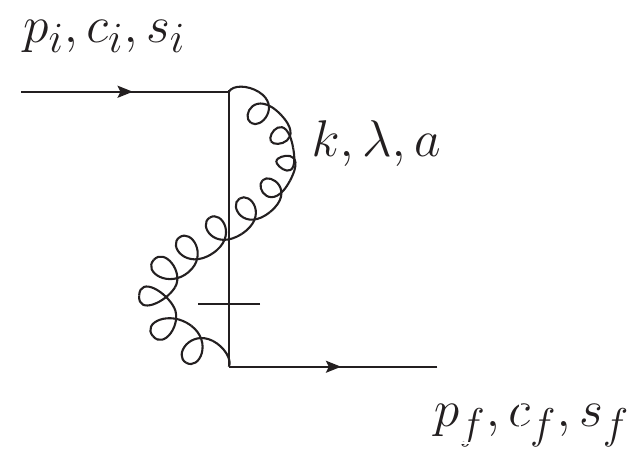} 
  }
  \subfigure[The instantaneous gluon term \label{fig:dm_LFpQCD_inst_g}]{
    \includegraphics[width=0.2\textwidth]{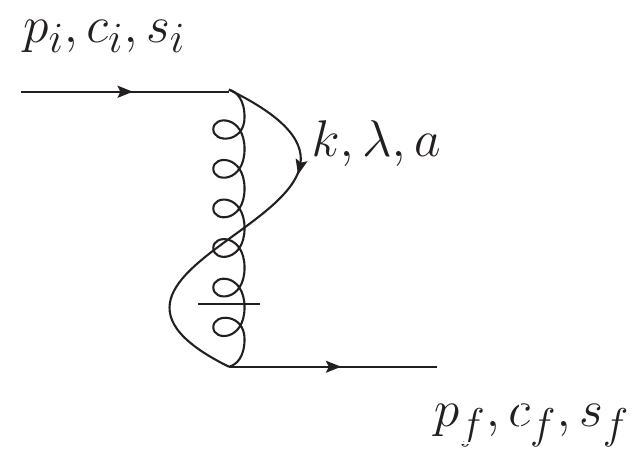} 
    \includegraphics[width=0.2\textwidth]{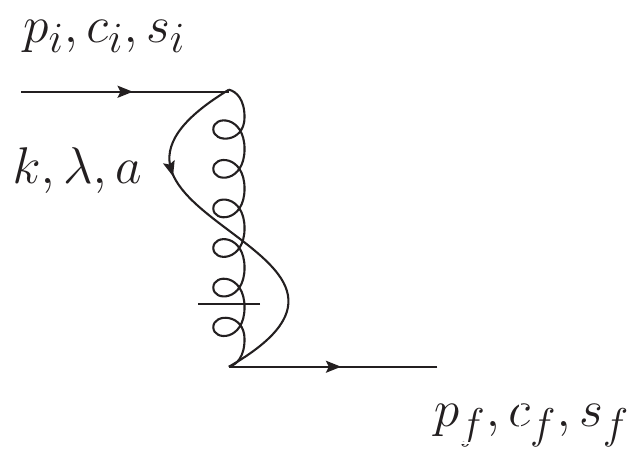} 
  }
    \subfigure[The one-loop term \label{fig:dm_LFpQCD_one_loop}]{
    \includegraphics[width=0.25\textwidth]{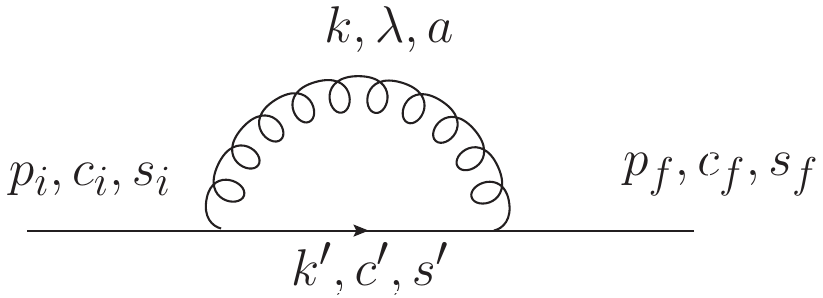} 
  }
  \caption{Diagrams for quark mass counter term at order $\mathcal{O}(g^2)$ in light-front perturbative QCD.}
  \label{fig:dm_LFpQCD}
\end{figure*}

\subsection{Cancellation by mass renormalization}
In contrast to Eq.~\eqref{eq:Pmin_QCD}, let us consider here that the Hamiltonian includes the instantaneous terms according to the aforementioned Fock space truncation scheme~\cite{Tang:1991rc}, 
\begin{align}\label{eq:Pmin_QCD_prime}
  P^-_{QCD, full}=P^-_{QCD}+ T_{\text{inst.}}\;,
\end{align}
in which the instantaneous term only appears in the $\ket{q}$ sector, and is a constant multiplying the identity matrix.
The eigenvalue equation, by mass renormalization with a counterterm $\delta P^-_{full}$, becomes
\begin{align}\label{eq:dPmn_renorm_full}
  (P^-_{QCD}+ T_{\text{inst.}}+\delta P^-_{full})\ket{\phi}=P^-_{\phi}\ket{\phi}\;,
\end{align}
where the ground state eigenvalue $P^-_{\phi,0}$ is equal to the light-front energy of a physical quark. 
In comparison to the procedure without considering $T_{\text{inst.}}$, where
\begin{align}\label{eq:dPmn_renorm}
  (P^-_{QCD}+\delta P^-)\ket{\phi}=P^-_{\phi}\ket{\phi}\;,
\end{align}
with $\delta P^-$ determined by the same $P^-_{\phi,0}$, one can see that
\begin{align}
  \delta P^-_{full} = \delta P^- - T_{\text{inst.}}\;.
\end{align} 
Including the instantaneous term in the Hamiltonian merely shifts the counterterm through mass renormalization, the renormalized Hamiltonian remains the same, $P^-_{QCD, full} +\delta P^-_{full}=P^-_{QCD}+\delta P^-$, therefore so do the eigenvalues and eigenstates.
In other words, the contribution of the instantaneous term in the $\ket{q} $ sector is canceled out by the mass counterterm through renormalization, so we can excluded it in the first place without changing anything upon renormalization.

\subsection{Perturbative mass renormalization}
In the light-front perturbation theory, the full mass counter term at second order in $g$ contains the contribution from both the instantaneous term $T_{\text{inst.}}$ and the one-loop term, as illustrated in Fig.~\ref{fig:dm_LFpQCD}.
We will show that the counterterm from the one-loop diagram agrees with what we obtained through the non-perturbative renormalization, and note that the full perturbative result agrees with the standard result~\cite{Mustaki:1990im}.

In the perturbative renormalization, the energy shift up to the second order can be written as
\begin{align}
 P^-_{QCD,full}\ket{\phi_0}= 
  \left[P^-_{\phi,0} + T_{\text{self energy}} + \mathcal{O}(g^3)\right]
  \ket{\phi_0}\;.
\end{align}
Here, recall that $P^-_{QCD,full}$ is the un-renormalized Hamiltonian, $P^-_{\phi,0}=(m_q^2+\vec P_\perp^2)/P^+$ is the physical quark's light-front energy as well as the unperturbed energy eigenvalue.
The the self-energy amplitude at $g^2$ order, $T_{\text{self energy}}$, is written as
\begin{align}
  \underbrace{T_{\text{self energy}}}_{>0}
      =&  \underbrace{T_{\text{one loop}}}_{<0}
      + \underbrace{T_{\text{inst. f}}}_{>0}
      + \underbrace{T_{\text{inst. g}}}_{>0}
      \;.
  \end{align}
In combination with Eqs.~\eqref{eq:dPmn_renorm_full} and \eqref{eq:dPmn_renorm}, we should make the following comparison between the perturbative and non-perturbative renormalization,
\begin{subequations}
  \begin{align}
    &\delta P^-_{full} \leftrightarrow - T_{\text{self energy}}\;,\\
    &\delta P^- \leftrightarrow - T_{\text{one loop}}\;.
  \end{align}
\end{subequations}
The corresponding mass counter term is then written as
\begin{align}\label{eq:dm_T_inst}
  \delta m =\sqrt{m_q^2-P^+ T_{\#}}- m_q\,,
  \qquad
  T_{\#}=T_{\text{one loop}} \text{ or } T_{\text{self energy}}
  \;.
\end{align}
The one-loop matrix amplitude reads \footnote{
  Note that we are using a different convention in the metric than Refs.~\cite{Mustaki:1990im,Brodsky:1997de}. 
In the $x^+$-ordered perturbation theory, the light-cone energy that appears in the factor for each intermediate state is $p_+$ that is conjugate to $x^+$, so it is written as $p_+=p^-$ in Ref.~\cite{Mustaki:1990im}, and $p_+=1/2 p^-$ here.
Likewise, the $T_{\text{one loop}}$ term is also calculated at the level of $p_+$, so converting it to $p^- ( = 2p_+)$ introduces an overall factor of $2$ here. 

Alternatively, one can, in the first place, think of the expression in Eq.~\eqref{eq:T_oneloop} as an expansion of $p^-$, then the energy denominator would be the $p^- (=2 p_+)$ instead, and each of the two vertex interactions would introduce a factor of $2$ when converting from $p_+$ to $p^-$, making the total difference a factor of $2$.
}
    
    \begin{align}\label{eq:T_oneloop}
      \begin{split}
        \mathcal M_{\text{one loop}}\equiv &
        \braket{p_f,s_f,c_f|T_{\text{one loop}}|p_i,s_i,c_i} \\
        = &2 g^2
        \sum_{c'=1}^{N_c}\sum_{a=1}^{N_c^2-1}
        \sum_{s'=\pm 1/2}\sum_{\lambda=\pm 1}
        \int\frac{\diff^2 k_\perp}{(2\pi)^{3/2}}\diff k^+ \theta(k^+)
        \int\frac{\diff^2 k'_\perp}{(2\pi)^{3/2}}\diff {k'}^+ \theta({k'}^+)\\
        &
        \frac{\bar u(p_f, s_f)\slashed{\epsilon}^*(k,\lambda) u(k', s')}{\sqrt{2 {k'}^+ 2 k^+ 2 p_f^+}} 
        \frac{1}{\frac{1}{2} (p_i^- - k^- -{k'}^-)}
        \frac{\bar u(k', s')\slashed{\epsilon}(k,\lambda) u(p_i, s_i)}{\sqrt{2 {k'}^+ 2 k^+ 2 p_i^+}}
        T^a_{c_i, c'} T^a_{c', c_f}
        \\
        &
        \delta(p_i^+ - k^+ - {k'}^+)
        \delta^{(2)}(\vec p_{i,\perp} - \vec k_\perp - \vec {k'}_\perp)
        \delta(p_f^+ - k^+ - {k'}^+)
        \delta^{(2)}(\vec p_{f,\perp} - \vec k_\perp - \vec {k'}_\perp)\;.
      \end{split}
    \end{align}
The $k'$ integral leads the momentum conservation of the initial and final states, $\delta^{(3)}(\bm p_i-\bm p_f)$. For simplicity, we will use $\bm p=\bm p_i=\bm p_f$ in the following expressions.
The state normalization is 
\begin{align}
  \braket{p_f,s_f,c_f|p_i,s_i,c_i} 
  =\delta_{c_i, c_f}\delta_{s_i, s_f} 
  \delta^{(3)}(\bm p_i-\bm p_f)\;.
\end{align}
  The color part yields
  \begin{align}
    \sum_{c'=1}^{N_c}\sum_{a=1}^{N_c^2-1}T^a_{c_i, c'} T^a_{c', c_f}
      =\frac{N_c^2-1}{2 N_c}\delta_{c_i, c_f}\;.
  \end{align}

  The spinor part becomes
  \begin{align}
    \begin{split}
      \sum_{s'=\pm 1/2}\sum_{\lambda=\pm 1}&
        \bar u(p, s_f)\slashed{\epsilon}^*(k,\lambda) u(k', s')
        \bar u(k', s')\slashed{\epsilon}(k,\lambda) u(p, s_i)\\
       =&4\delta_{s_i, s_f}
  \left[
    \left(
      \frac{2p^+}{k^+}
      +\frac{k^+}{p^+ - k^+}
    \right)(p\cdot k)
    -m_q^2
  \right]
  = 4\delta_{s_i, s_f}
  \frac{1}{2 z^2 (1-z)}
  \left[
    \left(
      2-2 z+z^2
    \right)
      \Delta_m^2
    +z^4 m_q^2
  \right]
       \;.
    \end{split}
  \end{align}
  In the above equation, we recognized $\vec \Delta_m$ and $z= k^+/p^+$, and applied
  \begin{align}
    p\cdot k
    =&\frac{1}{2z} \left[
      (\vec k_\perp - z \vec p_\perp)^2+m_q^2 z^2
    \right]
    =\frac{\Delta_m^2+m_q^2 z^2}{2z} 
    \;.
  \end{align}
  The energy denominator can be written as
  \begin{align}
    p^- - k^- - {k'}^-=-\frac{2}{p^+ - k^+} p\cdot k
    =-\frac{\Delta_m^2+m_q^2 z^2}{z(1-z)p^+}\;,
  \end{align}
  which is $-P^-_{rel,qg}$ as defined in Eq.~\eqref{eq:pmn_qg}.

Let us factorize out the state normalization such that $ T_{\text{one loop}}\delta_{c_i, c_f}\delta_{s_i, s_f} \delta^{(3)}(\bm p_i-\bm p_f)= \mathcal M_{\text{one loop}}$. Then,
\begin{align}\label{eq:T_1loop_res}
  \begin{split}
    T_{\text{one loop}}
    =&-2 g^2
    \frac{N_c^2-1}{2 N_c}
    \frac{1}{2 p^+}
    \int\frac{\diff^2 k_\perp}{(2\pi)^{3}}
    \int_0^{p^+}
    \frac{\diff k^+}{k^+} 
    \left[
      \left(
        \frac{2 p^+}{k^+} + \frac{k^+}{p^+ - k^+}
      \right)
      - \frac{m_q^2}{p\cdot k}
    \right]\\
 =&- 2 g^2
    \frac{N_c^2-1}{2 N_c}
    \frac{1}{2 p^+}
    \int\frac{\diff^2 \Delta_m}{(2\pi)^3}
    \int_0^1 \diff z
  \frac{1}{z^2 (1-z)}
  \frac{[1+(1-z)^2]\Delta_m^2+z^4 m_q^2}{\Delta_m^2+m_q^2 z^2}
   \;.
  \end{split}
\end{align} 
In Sec.~\ref{sec:analytical_solution}, we have obtained the non-perturbative counter term in the discrete basis representation, explicitly according to Eq.~\eqref{eq:bi_dPmn} with $\lambda=0$,
\begin{align}
  \begin{split}
  \delta P^- =\sum_{i=1}^n  \frac{ |V_i|^2}{K_i} 
  =&\sum_{z}\sum_{\vec \Delta_m} 
  \frac{g^2}{(2L_\perp)^2L (P^+)^2 }
  \frac{N_c^2-1}{2 N_c}
  \frac{1}{z^2(1-z)  }
  \frac{[1+(1-z)^2]\Delta_m^2+z^4 m_q^2 }{\Delta_m^2 + z^2 m_q^2}
  \;.
\end{split}
\end{align}
Taking into account the conversion from continuum to discrete space, 
\begin{align}
  \int\frac{\diff^2 k_\perp}{(2\pi)^{3}} \int_0^{p^+}\diff k^+ 
  = \int\frac{\diff^2 \Delta_m}{(2\pi)^{3}}  p^+ \int_0^1 \diff z
  \to
  \sum_{z}\sum_{\vec \Delta_m} \frac{1}{(2L_\perp)^2 L }\;,
\end{align}
and identifying $p^+$ and $P^+$, we have
\begin{align}
  T_{\text{one loop}}=-\delta P^- \;.
\end{align}
This means that the non-perturbative mass counter term in the $\ket{q}+\ket{qg}$ Fock space without the instantaneous interactions agrees exactly with the perturbative mass counter term resulting from the one-loop amplitude.

To compare with the result in Ref.~\cite{Mustaki:1990im}, one can proceed with the first line of Eq.~\eqref{eq:T_1loop_res}.
The first two terms have singularities, and can be regularized by introducing small cutoffs $\alpha$ and $\beta$,
\begin{align}
  \alpha < k^+ < p^+ - \beta\;,
\end{align}
so that the $k^+$ integral of the first two terms in Eq.~\eqref{eq:T_1loop_res} becomes
\begin{align}
  2\left[
    \frac{p^+}{\alpha}-1
  \right]
  +\ln\frac{p^+}{\beta}
\end{align}
Therefore,
\begin{align}
  \begin{split}
    T_{\text{one loop}}
    =& g^2
    \frac{N_c^2-1}{2 N_c}
    \frac{1}{ p^+}
    \int\frac{\diff^2 k_\perp}{(2\pi)^{3}}
    \left[
      \int_0^{p^+} \frac{\diff k^+}{k^+} 
      \frac{m_q^2}{p\cdot k}
      - 2\left[
    \frac{p^+}{\alpha}-1
  \right]
  - \ln\frac{p^+}{\beta}
    \right]
    \;.
  \end{split}
\end{align}
Inserting back into Eq.~\eqref{eq:dm_T_inst}, one can obtain the corresponding mass counter term, which is the $\delta m_a$ in Ref.~\cite{Mustaki:1990im}'s Eq. (3.13).

To find out the continuous limit of the mass counter term in our basis representation, let us carry out the integrals in Eq.~\eqref{eq:T_1loop_res}.
We introduce cutoffs on $z$, which can be related to the basis parameter $K$,
\begin{align}
  \frac{1}{K+0.5}=z_{min} \leq z\leq z_{max}=\frac{K}{K+0.5}\;,
\end{align}
and a transverse momentum cutoff, which also emerges in the finite discrete basis space,
\begin{align}\label{eq:UV_perp}
  p_\perp \leq \Lambda_\perp \approx \sqrt{2}\Lambda_{UV}= \sqrt{2}\frac{\pi}{a_\perp}\;.
\end{align}
We have
\begin{align}\label{eq:dm_1loop}
  \begin{split}
    T_{\text{one loop}}
    =&
    2 g^2
    \frac{N_c^2-1}{2 N_c}
    \frac{1}{2 p^+}
    \frac{1}{ 4 \pi^2}
    \Bigg\{
     \Lambda_\perp^2
      \left(
         \frac{1}{z_{max}} - \frac{1}{z_{min}} 
          +\frac{1}{2}\ln \left[\frac{1-z_{max}}{1-z_{min}} \right]
      \right)
       + 2 m_q \Lambda_\perp \left(
        \arctan\left[\frac{m_q z_{max}}{\Lambda_\perp}\right]
        -\arctan\left[\frac{m_q z_{min}}{\Lambda_\perp}\right]
      \right)\\
      &
      + m_q^2 z_{max} \ln \left[
        1+\frac{\Lambda_\perp^2}{m_q^2 z_{max}^2}
         \right]
      - m_q^2 z_{min} \ln \left[
        1+\frac{\Lambda_\perp^2}{m_q^2 z_{min}^2}
        \right]
    \Bigg\}\;.
  \end{split}
\end{align}
 
The two terms in the first line do not depend on $m_q$, and will partially cancel out with the instantaneous contributions. For completeness, the instantaneous fermion matrix amplitude as in Fig.~\ref{fig:dm_LFpQCD_inst_f} is
\begin{align}
  \begin{split}
    \mathcal M_{\text{inst. f}}\equiv &\braket{p_i,s_i,c_i|T_{\text{inst. f}}|p_f,s_f,c_f} \\
    =&2g^2 \delta^{(3)}(\bm p_i-\bm p_f)
    \sum_{c'=1}^{N_c}\sum_{a=1}^{N_c^2-1}
    \sum_{\lambda=\pm 1}
    \int\frac{\diff^2 k_\perp \diff k^+ \theta(k^+)}{(2\pi)^{3}}
    \frac{
      \bar u(p_f, s_f)\slashed\epsilon^*(k,\lambda) \gamma^+
      \slashed\epsilon(k,\lambda) u(p_i, s_i)
    }{ \sqrt{2 p_f^+} \sqrt{2 k^+} 2(p^+ -  k^+) \sqrt{2 k^+} \sqrt{2 p_i^+} }
 T^a_{c_i, c'} T^a_{c', c_f}
    \;.
  \end{split}
\end{align}
 
The color part works out in the same way as in the one-loop term, and the spinor part becomes
\begin{align}
    \sum_{\lambda=\pm 1}\slashed{\epsilon}^*(k,\lambda)
    \gamma^+\slashed{\epsilon}(k,\lambda)
    =2\gamma^+\;,
\qquad
  \bar u(p_f, s_f)\gamma^+ u(p_i, s_i)
    =2\sqrt{p_i^+ p_f^+}\delta_{s_i, s_f} \;.
\end{align}
After factorizing out the state normalization, we have
    \begin{align}
      \begin{split}
        T_{\text{inst. f}}
    =&2 g^2
    \frac{N_c^2-1}{2 N_c}
    \frac{1}{ 2 p^+}
    \int\frac{\diff^2 k_\perp}{(2\pi)^{3}}
    \int_0^\infty \diff k^+ 
    \left[
      \frac{1}{ k^+ }
      + \frac{1}{ p^+ -  k^+}
    \right]
    = g^2
    \frac{N_c^2-1}{2 N_c}
    \frac{1}{ p^+}
    \int\frac{\diff^2 k_\perp}{(2\pi)^{3}}
    \ln\left[\frac{p^+}{\alpha} \right]
   \;.
  \end{split}
\end{align}
In the second equation, the $k^+$ integral is carried out with the cutoff $ |k^+ - 0|\geq \alpha$. Similarly, the instantaneous gluon self-energy terms as in Fig.~\ref{fig:dm_LFpQCD_inst_g} is
\begin{align}
  \begin{split}
    T_{\text{inst. g}}
    =&g^2
    \frac{N_c^2-1}{2 N_c}
    \int\frac{\diff^2 k_\perp}{(2\pi)^{3}}
    \diff k^+ \theta(k^+)
    \left[
      \frac{1}{(p^+ - k^+)^2}
      -\frac{1}{(p^+ + k^+)^2}
    \right]
    =g^2
    \frac{N_c^2-1}{2 N_c}
    \frac{1}{p^+}
    \int\frac{\diff^2 k_\perp}{(2\pi)^{3}} 
    2\left[\frac{p^+}{\alpha} -1 \right]
   \;.
  \end{split}
\end{align}

The full self-energy correction at the $g^2$ order is therefore
\begin{align}
  \begin{split}
    T_{\text{self energy}}
    =&  T_{\text{one loop}}
    + T_{\text{inst. f}}
    + T_{\text{inst. g}}
    =2 g^2
    \frac{N_c^2-1}{2 N_c}
    \frac{1}{2 p^+}
    \int\frac{\diff^2 k_\perp}{(2\pi)^{3}}
    \bigg\{
      \int_0^{p^+} \frac{\diff k^+}{k^+}\frac{m_q^2}{p\cdot k}
      +\ln\left[\frac{\beta}{\alpha}\right]
    \bigg\}
   \;.
  \end{split}
\end{align}
To recover with the standard covariant result, the momentum cutoffs need to be implemented in a covariant way $P^2<\Lambda^2$, such that $\alpha$ and $\beta$ are $\vec k_\perp$ dependent, then~\cite{Mustaki:1990im}
\begin{align}\label{eq:log_beta_alpha}
  \begin{split}
    \int\frac{\diff^2 k_\perp}{(2\pi)^{3}}\ln\left[\frac{\beta}{\alpha}\right]
    =
    \int\frac{\diff^2 k_\perp}{(2\pi)^{3}}
      \int_0^{p^+} \frac{\diff k^+}{p^+}\frac{m_q^2}{p\cdot k}
   \;.
  \end{split}
\end{align}
Therefore,
\begin{align}\label{eq:T_g2}
  \begin{split}
    T_{\text{self energy}}
    =&g^2
    \frac{N_c^2-1}{2 N_c}
    \frac{1}{p^+}
    \int\frac{\diff^2 k_\perp}{(2\pi)^{3}}
    \int_0^{p^+} \diff k^+ \frac{m_q^2}{p\cdot k}
    \left(
      \frac{1}{k^+}+\frac{1}{p^+}
    \right)
    = g^2
    \frac{N_c^2-1}{2 N_c}
    \frac{m_q^2}{p^+}
    \int\frac{\diff^2 \Delta_m}{(2\pi)^{3}}
    \int_0^{1} \diff z \frac{2(1+z)}{\Delta_m^2 + m_q^2 z^2}\\
    =&g^2 
    \frac{N_c^2-1}{2 N_c} 
    \frac{m_q^2}{p^+}
    \frac{1 }{8 \pi^2}
    \Bigg\{
      4\frac{\Lambda_\perp}{m_q}
      \text{arccot} \left[ \frac{\Lambda_\perp}{ m_q}\right]
      +\left(
        3+\frac{\Lambda_\perp^2}{m_q^2}
      \right)
      \ln \left[1+\frac{m_q^2} {\Lambda_\perp^2}\right]
      +3\ln \left[\frac{\Lambda_\perp^2}{m_q^2 } \right]
    \Bigg\}
   \;.
  \end{split}
\end{align}
 The $z$ integral is finite, and the transverse integral is carried out with the cutoff in Eq.~\eqref{eq:UV_perp}.
 The $\mathcal O (g^2)$ mass counter term follows directly according to Eq.~\eqref{eq:dm_T_inst}.
 The same result as in Ref.~\cite{Mustaki:1990im} is obtained by dropping out the $\delta m^2$ term,
\begin{align}\label{eq:dm_T_inst_expansion}
  \delta m = -\frac{1}{2 m_q}P^+ T_{\text{self energy}}-\frac{1}{2 m_q}\delta m^2
  \approx
  -\frac{p^+}{2 m_q} T_{\text{self energy}}\;.
\end{align}
In the limit of $\Lambda_\perp\to\infty$, the standard result is recovered,
\begin{align}\label{eq:dm_g2}
  \begin{split}
    \lim_{\Lambda_\perp\to\infty}\delta m
    = \lim_{\Lambda_\perp\to\infty}-\frac{N_c^2-1}{2 N_c} 
    \frac{g^2 m_q }{16 \pi^2}
    \bigg\{
      4
      +1
      +3\ln \left[\frac{\Lambda_\perp^2}{m_q^2 } \right]
    \bigg\}
    = \lim_{\Lambda_\perp\to\infty} -\frac{N_c^2-1}{2 N_c} 
    \frac{3 g^2 m_q}{16 \pi^2}
    \ln \left[\frac{\Lambda_\perp^2}{m_q^2 } \right]\;.
  \end{split}
\end{align}
A comparison of the mass counter term defined in Eq.~\eqref{eq:dm_T_inst} with the self-energy corrections according to Eqs.~\eqref{eq:dm_1loop} and \eqref{eq:T_g2} are shown in Fig.~\ref{fig:dm_LFpQCD_res}.
Notably, the sign of the two counter terms are opposite, and in the massless limit of $m_q=0$, the full second order mass counter term vanishes, whereas the one-loop counter term is finite. 
\begin{figure}[htp!]
  \centering 
    \includegraphics[width=0.4\textwidth]{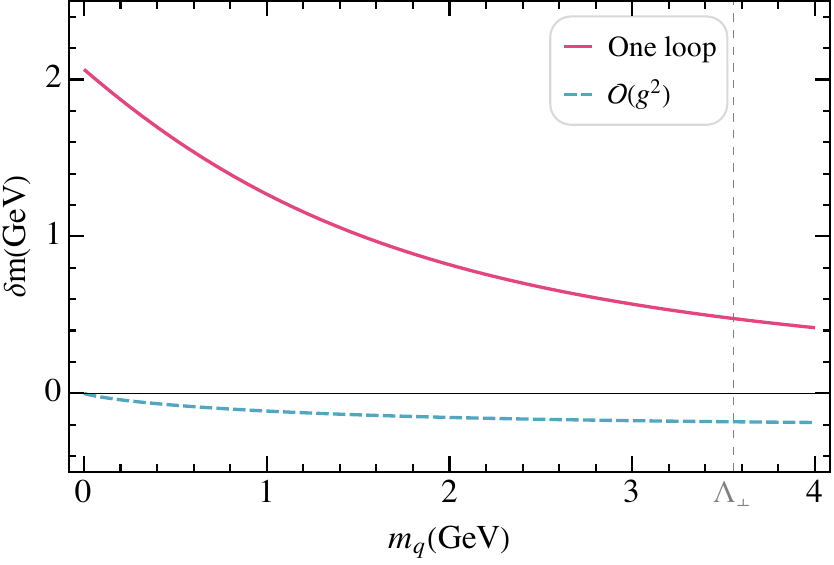} 
    \quad
    \includegraphics[width=0.4\textwidth]{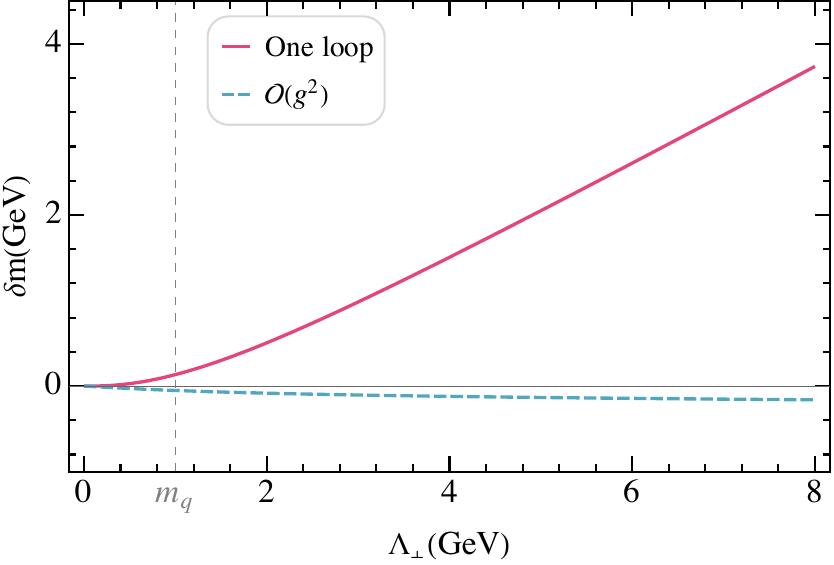} 
  \caption{Mass counter term as a function of (a) the physical quark mass $m_q$, and (b) the transverse momentum cutoff $\Lambda_\perp$. The solid red line contains only the one-loop contribution, and is obtained according to Eqs.~\eqref{eq:dm_1loop} and~\eqref{eq:dm_T_inst}; the blue dashed line contains the full second order self-energy contribution, obtained according to Eqs.~\eqref{eq:dm_1loop} and \eqref{eq:T_g2}. We take $z_{min}=0.1$ and $z_{max}=0.9$ in the plot. }
  \label{fig:dm_LFpQCD_res}
\end{figure}

\section{Perturbative dressed quark}
\subsection{Perturbative light-front wavefunction at leading order}\label{app:LFWF}
The physical quark's LFWF we obtained by diagonalizing the Hamiltonian $P^-_{QCD}$ should be comparable to that calcualted perturbatively.
The derivation of quark LFWF using perturbative expansion in the light-front Hamiltonian formalism can be found in Refs.~\cite{Lepage:1980fj,Harindranath:2001rc,Marquet:2007vb}. 
For a physical quark state with momentum, color $c_Q$, and helicity $\lambda_Q$ specified, its LFWF decomposition in the $\ket{q}+\ket{qg}$ Fock space reads,
\begin{align}\label{eq:pLFWF_cont}
  \begin{split} 
  & \ket{q (p_Q, c_Q, \lambda_Q )}_{phys} =
  \mathcal{Z}
  \bigg\{ \ket{q(p_Q, c_Q, \lambda_Q )}  +
  \sum_{c_q, \lambda_q, c_g, \lambda_g} 
  \int \diff^3 p_q
  \int \diff^3 p_g
  \phi(p_q,c_q,\lambda_q;p_g,c_g, \lambda_g) 
  \ket{qg (p_q, c_q, \lambda_q; p_g, c_g, \lambda_g)}
  \bigg\}  \;,  
  \end{split}
\end{align}
in which we denote the three momentum as $p=(p^+, \vec p_\perp)$. 
The factor $\mathcal{Z}$ is a constant determined by the normalization of the state.
The quark-gluon component of the LFWF is given by
\begin{align}
  \phi(p_q,c_q,\lambda_q;p_g,c_g, \lambda_g) =
  \frac{
    \braket{ qg (p_q, c_q, \lambda_q; p_g, c_g, \lambda_g)|P^-_{QCD}| q(p_Q, c_Q, \lambda_Q )}
    }{
      p^-_Q - p^-_{qg}
    }\;,
\end{align}
which is essentially the unnormalized quark-gluon LFWF that we obtained non-perturbatively in Eq.~\eqref{eq:q_LFWF_qg_anal}. 
Let us write out the expression and take into account the  discretization convention to make an exact comparison.
In the continuum, the matrix element reads,
\begin{align}\label{eq:Vqg_element_con_repeat}
\begin{split}
    &    \braket{ qg (p_q, c_q, \lambda_q; p_g, c_g, \lambda_g)|P^-_{QCD}| q(p_Q, c_Q, \lambda_Q )}=
    \frac{g}{\sqrt{ 2(2\pi)^3 p_q^+ p_g^+ p_Q^+  }}
    T_{c_q,c_Q}^{c_g} 
     \Gamma_{\lambda_q,\lambda_Q}^{\lambda_g}(z, \vec \Delta_m, m_q)
    \delta^{(3)}(p_q-p_Q + p_g)
    \;,
  \end{split}
  \end{align}
  in which, recalling that the relative momenta are defined as $ \vec \Delta_m =-z \vec p_{\perp,q}+(1-z)\vec p_{\perp,g} $ and $z=p^+_g/(p^+_q + p^+_g)$, the spinor part is
  \begin{align}
    \begin{split}
    \bar{u}(p_q,\lambda_q)
    \gamma^\mu 
    u(p_Q,\lambda_Q)
    \epsilon_\mu^*(p_g,\lambda_g) 
    =&
    \frac{\sqrt{2}}{z\sqrt{1-z}}
    \begin{cases}
      &\Delta_m^R \delta_{\lambda_q,\uparrow}\delta_{\lambda_g,\uparrow}
      + (1-z)\Delta_m^L \delta_{\lambda_q,\uparrow}\delta_{\lambda_g,\downarrow}
      -z^2 m_q \delta_{\lambda_q,\downarrow}\delta_{\lambda_g,\uparrow},
      \qquad \lambda_Q=\uparrow\\
      & z^2 m_q\delta_{\lambda_q,\uparrow}\delta_{\lambda_g,\downarrow}
      + (1-z)\Delta_m^R \delta_{\lambda_q,\downarrow}\delta_{\lambda_g,\uparrow}
      + \Delta_m^L \delta_{\lambda_q,\downarrow}\delta_{\lambda_g,\downarrow},
      \qquad \lambda_Q=\downarrow
    \end{cases}
    \\
  \equiv & \frac{\sqrt{2}}{z\sqrt{1-z}} \Gamma_{\lambda_q,\lambda_Q}^{\lambda_g}(z, \vec \Delta_m, m_q)
    \;.
  \end{split}
  \end{align}
  The energy denominator is
\begin{align}\label{eq:DPmin_mq}
  \Delta p^- 
  =p^-_{Q}-p^-_{qg}
  =-\frac{\Delta_m^2 + z^2 m_q^2}{z(1-z)p^+_Q}
  \;,
\end{align}
which is $-P^-_{rel,qg}$ as defined in Eq.~\eqref{eq:pmn_qg}.
  We can therefore write out the quark wavefunction as the following, 
  \begin{align}\label{eq:q_LFWF_qg_anal_con}
    \begin{split} 
  \int \diff^3 p_q
  \int \diff^3 p_g
  \phi(p_q,c_q,\lambda_q;p_g,c_g, \lambda_g)
    & =
    \int_0^1 \diff z 
    \int   \diff^2 \Delta_m
    \underbrace{
    \frac{-1}{\sqrt{(2\pi )^3}}
    \frac{g \sqrt{p^+_Q}}{\Delta_m^2 + z^2 m_q^2}
    \frac{1}{\sqrt{z}} 
    T_{c_q,c_Q}^{c_g} 
    \Gamma_{\lambda_q,\lambda_Q}^{\lambda_g}(z, \vec \Delta_m, m_q)
    }_{\phi_{\text{con.}}}
    \;. 
    \end{split}
  \end{align}  
  In getting the above expression, we first integrate over $p_q$ thus applying the delta function of the momentum conservation. 
  We then make a change of variables $\{\vec p_{\perp,g} \to \vec \Delta_m =\vec p_{\perp,g} -z\vec p_{\perp,Q}, p_g^+ \to z\}$ such that
  \begin{align}
  \int\diff^2 p_{\perp,g} \int_0^{p_Q^+}\diff p_g^+ 
  = \int\diff^2 \Delta_m  p_Q^+ \int_0^1 \diff z
  \;.
\end{align}

In the discrete basis space, Eq.~\eqref{eq:pLFWF_cont} is written as
\begin{align}
  \begin{split} 
  & \ket{q (p_Q, c_Q, \lambda_Q )}_{phys} =
  \mathcal{Z}
  \bigg\{ 
  \ket{\beta_q(p_Q, c_Q, \lambda_Q)} 
  +
  \sum_{\beta_{qg}}
  \phi(\beta_{qg}; \beta_q) 
  \ket{\beta_{qg} (p_q, c_q, \lambda_q; p_g, c_g, \lambda_g)}
  \bigg\}  \;,  
  \end{split}
\end{align}
The Hamiltonian matrix element as of Eq.~\eqref{eq:Vqg_element_con_repeat} is now
\begin{align}\label{eq:Vqg_element_repeat}
  \begin{split}
    \braket{ \beta_{qg}(p_q, c_q, \lambda_q; p_g, c_g, \lambda_g)|P^-_{QCD}| \beta_q(p_Q, c_Q, \lambda_Q )}
    =&
    \frac{g}{\sqrt{ 2L(2L_\perp)^2 p_q^+ p_g^+ p_Q^+}}
  T_{c_q,c_Q}^{c_g}
    \Gamma_{\lambda_q,\lambda_Q}^{\lambda_g}(z, \vec \Delta_m, m_q)
    \delta^3_{p_q - p_Q + p_g}
  \;.
  \end{split}
  \end{align}
It follows that the discrete perturbative quark-gluon wavefunction is written as, 
\begin{align}\label{eq:q_LFWF_qg_anal_dis}
  \begin{split} 
    \sum_{\beta_{qg}}
  \phi(\beta_{qg}; \beta_q) 
  & =
  \sum_{c_q, \lambda_q, c_g, \lambda_g} 
  \sum_{k_g^+} \frac{1}{K}
  \sum_{k_g^x,k_g^y}d_p^2
  \underbrace{
  \frac{- \sqrt{ L(2L_\perp)^2}}{ (2\pi)^3}
  \frac{g\sqrt{ p^+_Q}}{\Delta_m^2 + z^2 m_q^2}
  \frac{1}{\sqrt{ z }}
  \Gamma_{\lambda_q,\lambda_Q}^{\lambda_g}(z, \vec \Delta_m, m_q)
  T_{c_q,c_Q}^{c_g}
  }_{\phi_\text{dis.}}
  \;. 
  \end{split}
\end{align}  
For the convenience of discussion, we define the unit of the summation in the $z$ and $\vec \Delta_m$ space as
\begin{align}
  \delta_{z,\Delta_m}=\frac{1}{K} d_p^2=\frac{(2\pi)^2}{K(2 L_\perp)^2}\;. 
\end{align}
Recall that in the CHD basis, $\Gamma_{\lambda_q,\lambda_Q}^{\lambda_g}(z, \Delta_m, m_q)$ becomes $\sqrt{[1+(1-z)^2]\Delta_m^2 +z^4 m_q^2}$, and $T_{c_q,c_Q}^{c_g} $ becomes $\sqrt{(N_c^2-1)/(2N_c)}$, then the right-hand-side of Eq.~\eqref{eq:q_LFWF_qg_anal_dis} in the summations, $\delta_{z,\Delta_m}\phi_\text{dis.}$, 
are exactly Eq.~\eqref{eq:q_LFWF_qg_anal} at $\lambda=0$, i.e. for the ground state.
This means that in the discrete $\ket{q}+\ket{qg}$ Fock space, the quark LFWF obtained perturbatively is the same with that obtained non-perturbatively, apart from the overall normalization which in the perturbative case is expanded in the coupling, whereas in the full calculation the state is always normalized to unity.

To compare the discrete LFWF $\phi_\text{dis.}$ with the continuous one, $\phi_\text{con.}$, one needs to take into account the discretization scaling factor $S_{\text{dis}}$ as in Eq.~\eqref{eq:S_dis}
Note that the single particle state in the discrete space carries one $S_{\text{dis}}$ factor when compared to that in the continuous space, whereas the two particle-state carries $S_{\text{dis}}^2$. Therefore only one factor $S_{\text{dis}}$ enters the quark-gluon components that is in relative to the quark component, so we have
\begin{align}
  \phi_{\text{con.}} = \frac{\phi_{\text{dis.}}}{S_{\text{dis}}} \;,
\end{align}
which Eqs.~\eqref{eq:q_LFWF_qg_anal_con} and~\eqref{eq:q_LFWF_qg_anal_dis} indeed satisfy. 
Therefore in the $\ket{q}+\ket{qg}$ Fock space, the quark LFWF obtained perturbatively is the same with that obtained non-perturbatively, either in the discrete or in the continuous momentum space.

The normalized LFWF can be written as 
\begin{align}
  \phi_{\text{dis. norm.}}= \mathcal{Z}\delta_{z,\Delta_m}\phi_\text{dis.}\;.
\end{align}
The normalization factor is then $ \mathcal{Z}=\sqrt{Z_2}$ by definition, and $Z_2$ can be interpreted as the probability of finding a constituent quark out of a physical quark~\cite{Zhao:2014xaa}:
\begin{align}
  Z_2=\sum_{\ket{q}} |\braket{q|q_{\text{phys}}}|^2\;,
\end{align}
which is exactly the full solution obtained in the $\ket{q}+\ket{qg}$ space as of Eq.~\eqref{eq:unint_PQ_anal}.
\begin{figure*}[t]
\hspace*{\fill}
  \begin{minipage}[t]{0.35\textwidth}
    \centering
    \vspace{30pt}
    \subfigure[In the CHD basis, for $\lambda_Q=\uparrow,\downarrow$. \label{fig:fz_Dm}]{
    \includegraphics[width=\textwidth]{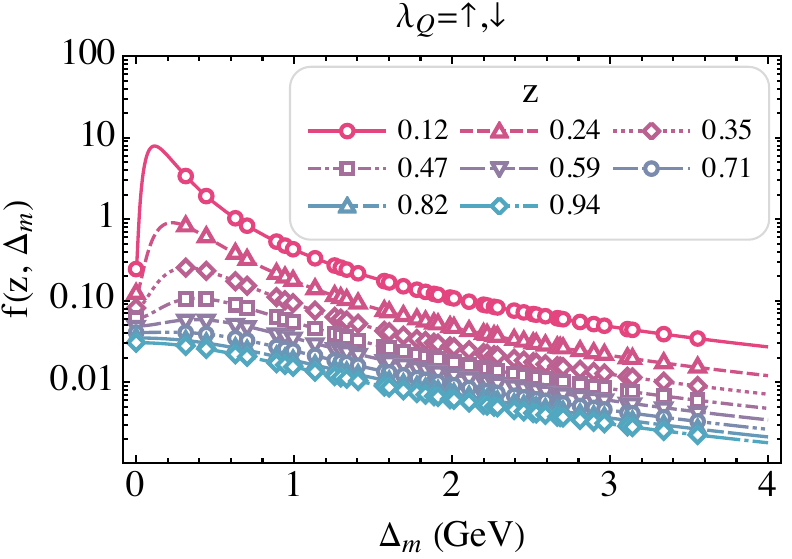} 
    }
  \end{minipage}
  \hfill
  \begin{minipage}[t]{0.6\textwidth}
    \centering
    \vspace{0pt}
    \subfigure[In the single-particle helicity basis, for $\lambda_Q=\uparrow$. \label{fig:fz_Dm_hel}]{
    \includegraphics[width=\textwidth]{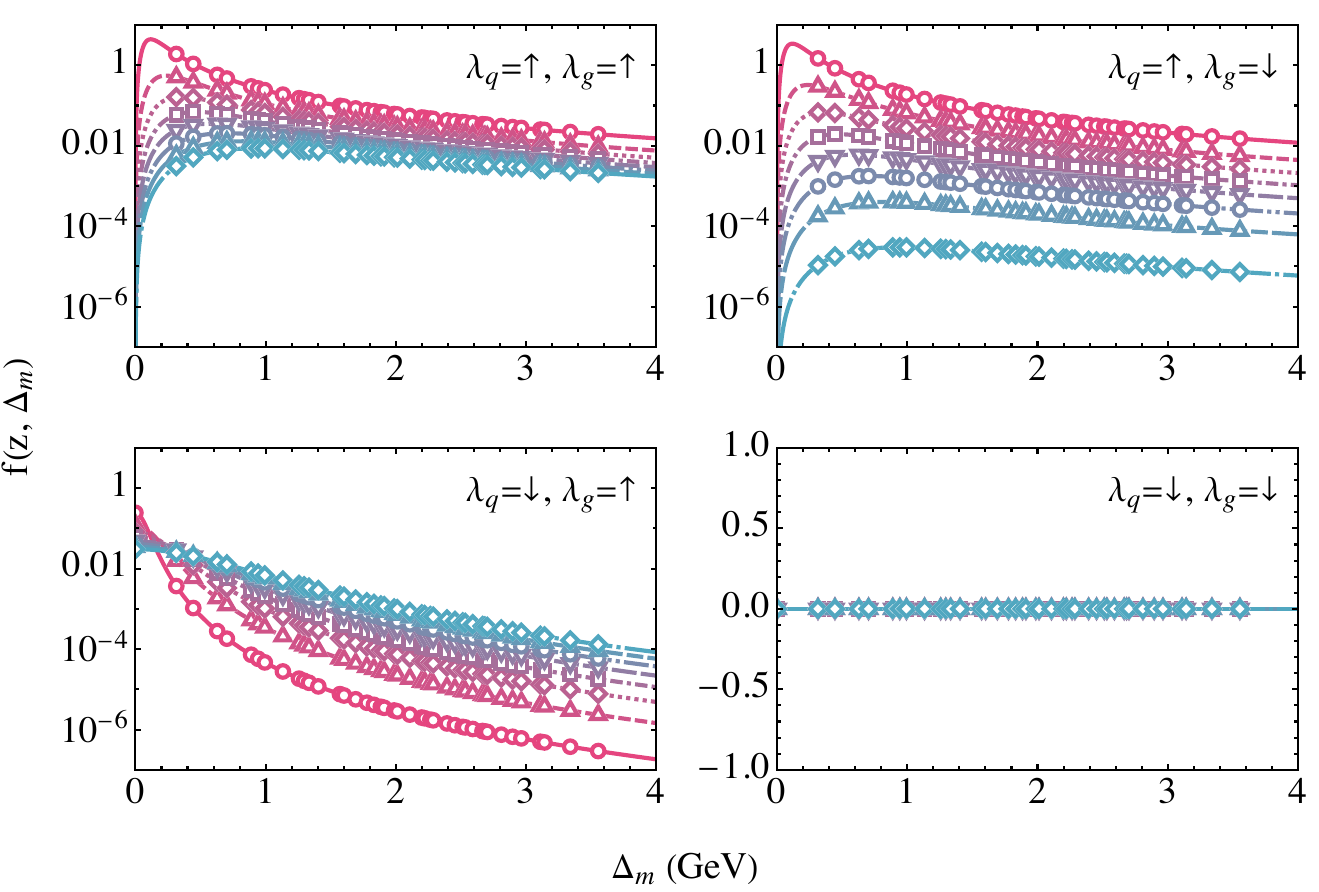} 
    }
  \end{minipage}
\hspace*{\fill}
  \caption{
    Comparison of the dressed quark transverse momentum distribution function between diagonalization (in markers) and perturbative calculations (in lines), at different $z$. 
    The non-perturbative results are rescaled according to Eq.~\eqref{eq:rescaleFN}.
    The two subfigures are results in two different basis as indicated in the captions accordingly.
  }
  \label{fig:LFWF_pQCD}
\end{figure*}

Therefore, the relation between the normalized discrete and the continuous LFWF is
\begin{align}\label{eq:rescaleFN}
 \phi_{\text{con.}} =\frac{1}{ \sqrt{Z_2} \delta_{z,\Delta_m}  S_{\text{dis}}}\phi_{\text{dis. norm.}} \;.
\end{align}
The relation between the discrete and the continuous distribution function $f=|\phi|^2$ follows as
\begin{align}\label{eq:rescalef}
  f_{\text{con.}}(z, \Delta_m)=\frac{1}{Z_2} \frac{K^2 (2L_\perp)^2}{ 2\pi L} f_{\text{dis. norm.}} (z, \Delta_m)\;.
\end{align}
In Fig.~\ref{fig:LFWF_pQCD}, we compare the distribution function obtained by diagonalization (in markers) and by perturbative calculations (in lines), at different $z$; first in the CHD basis, then in the single particle basis so that we can also compare each helicity configuration separately. 
The non-perturbative results are rescaled according to Eq.~\eqref{eq:rescaleFN}.
As expected from the analytical expressions, the non-perturbative and the perturbative distribution functions agree.

\subsection{Renormalization factor $Z_2$}
We have seen in Appendix~\ref{app:LFWF} that the wavefunction and therefore the renormalization factor $Z_2$ agree with that from the leading order perturbation calculation.
Let us write out the continuum limit of $Z_2$ by carrying out integrals instead of summations in Eq.~\eqref{eq:q_LFWF_qg_anal} and thus Eq.~\eqref{eq:unint_PQ_anal},
\begin{align}\label{eq:q_LFWF_qg_anal_cont}
  \begin{split}
    \sum_{i=1}^n \left|\phi_{qg}(z,\vec \Delta_m) \right|^2
  = &
  \sum_{i=1}^n
  g^2 \frac{N_c^2-1}{2 N_c}
  \frac{1}{(2\pi)^{3}K}
  \frac{
    [1+(1-z)^2]\tilde\Delta_m^2 + z^4 \tilde m_q^2 
  }{
z (\tilde\Delta_m^2 +z^2 \tilde m_q^2)^2
  }\\
  \mapsto &
  g^2 \frac{N_c^2-1}{2 N_c}
  \int \frac{\diff^2 \tilde\Delta_m}{(2\pi)^3}
  \int_{z_{min}}^{z_{max}}\diff z 
  \frac{
    [1+(1-z)^2]\tilde\Delta_m^2 + z^4  \tilde m_q^2 
  }{
z (\tilde\Delta_m^2 +z^2 \tilde m_q^2)^2
  }\\
  =&
  g^2 \frac{N_c^2-1}{2 N_c}
  \frac{1}{(2\pi)^2}
 \Bigg\{
  \frac{1}{2}\left(-\Li_2\left[\frac{k^2+\tilde m_q^2 z^2}{\tilde m_q^2 z^2}\right]
 -\ln\left[-\frac{k^2}{\tilde m_q^2 z^2}\right] \ln[k^2+\tilde m_q^2 z^2]\right)
 +\frac{1-2z}{2} \ln[k^2+\tilde m_q^2 z^2]\\
 &
+\frac{(k^2+\tilde m_q^2 z^2) \ln[k^2+\tilde m_q^2 z^2]-k^2}{4 \tilde m_q^2}
 -\frac{k \arctan[\tilde m_q z/k]}{\tilde m_q}+2 \ln[k] \ln[z]
 \Bigg\}
 \Bigg\vert^{k=\tilde \Lambda_\perp}_{k=\tilde\lambda_\perp}
 \Bigg\vert^{z=z_{max}}_{z=z_{min}}
  \;.
\end{split}
\end{align}
Here, the polylogarithm function is defined as $ \Li_n [x]=\sum_{k=1}^\infty x^k/k^n$. The dimensionless transverse momentum cutoffs are defined as $\tilde \Lambda_\perp=\Lambda_\perp/d_p $ and $\tilde \lambda_\perp=\lambda_\perp/d_p $, and to match with the scales on the discrete basis, $ \Lambda_\perp\approx \sqrt{2}\Lambda_{UV}$ and $\lambda_\perp\approx \lambda_{IR}=d_p $.
\end{widetext}

\section{Transformation between single-particle and relative momenta on the discrete and periodic lattice}\label{app:basis}
As introduced in the main text, the transformation between single-particle and relative momenta for a quark-gluon state is given by,
\begin{align}
    \begin{split}
    & \vec P_{\perp}= \vec p_{\perp,q}+\vec p_{\perp,g}\;,\\
    & P^+ = p^+_q+p^+_g\,, \quad z\equiv p^+_g/P^+\;,\\
    & \vec \Delta_m =-z \vec p_{\perp,q}+(1-z)\vec p_{\perp,g}\;,
    \end{split}
\end{align}
and inversely,
\begin{align}
    \begin{split}
        \vec p_{\perp,q}
        =&-\vec \Delta_m+(1-z)\vec P_\perp\;,\qquad
        \vec p_{\perp,g}
        =\vec \Delta_m + z\vec P_\perp\;.
    \end{split}
\end{align}
In the continuous space, one can just go back and forth by following the above equations. However, in our formalism, the transverse momentum space is a finite lattice with periodic boundary conditions, introducing extra complexity to the transformation. One main difficulty is that since $z$ is not an integer, it isn not possible for both the single particle momenta and the relative momentum to be integer multiples of $d_p$ at the same time.
In our preceding work, Ref.~\cite{Li:2021zaw} Appendix C, we introduced a recipe to resolve the ambiguity due to periodicity when calculating the momentum transfer $\vec \Delta_m$, and we will carry on with it in this work.

For simplicity, we discuss the one-dimensional case in the following. The same procedure is applied separately to both the $x$ and the $y$ dimensions in the transverse plane. 
We will also refer to the transverse momentum $p_q $ by its corresponding dimensionless quantum number,$p_q/d_p$, so the accessible quanta on the lattice are in the set
\[
    \Omega_\perp \equiv [-N_\perp, -N_\perp +1, \ldots, N_\perp-1]\;.
\]
In implementing the transformation, we have two major considerations. First, the attainable values of $p_q, p_g, P$, and $\Delta_m$ are all in the set $\Omega_\perp$. In this way, the renormalization and diagonalization procedure stays invariant for different CM momenta, as it should be in the continuum.  
Second, the transformation must be a one-to-one mapping and invertible, so that there is no ambiguity going between the two basis.

We define two operations:
\begin{itemize}
    \item $P.B.[p]$\\
    For a given momentum quanta $p$, the periodic boundary function $P.B.$ brings it to the fundamental period by 
    \begin{align*}
      p' = P.B.[p] = p + i(2 N_\perp) \;,
    \end{align*} 
    with a given $i \in \mathcal{I}$ such that $p'\in\Omega_\perp$.
    \item  $Round[x]$\\
    For a given value $x\in \mathcal{R}$, the operation $Round[x]$ rounds it to the nearest integer.
\end{itemize}

We define the basis transformation from the single-particle momenta to the CM and relative as $ f(\{p_q, p_g\}; z)=\{P, \Delta_m\} $, in which
\begin{align}\label{eq:basis_mapping}
    \begin{split}
        &P = P.B.[p_q + p_g]\;,\\
        &\Delta_m = P.B.[Round[(1-z)P] -p_q ] \;,  
    \end{split}
\end{align}
and the inverse, $f^{-1}(\{P, \Delta_m\} ; z)=\{p_q, p_g\}$, with
\begin{align}\label{eq:basis_mapping_inv}
    \begin{split}
   & p_q = P.B.[Round[(1-z)P]-\Delta_m]\;,\\
   & p_g = P.B.[P - p_q]\;.
\end{split}
\end{align}

We prove in the following that the mapping defined in Eq.~\eqref{eq:basis_mapping} is one-to-one, and the function defined in Eq.~\eqref{eq:basis_mapping_inv} is its inverse.
\begin{enumerate}
    \item Existence:\\
     $\forall \{p_q, p_g\}\in \Omega_\perp$ (i.e., $\forall p_q\in \Omega_\perp$ and $\forall p_g\in \Omega_\perp$), $\exists  \{P, \Delta_m\}\in \Omega_\perp$ according to Eq.~\eqref{eq:basis_mapping}.
  
     \item Uniqueness:\\
      $\forall \{p_q', p_g'\}, \{p_q, p_g\}\in \Omega_\perp $, and $\{p_q', p_g'\}\neq \{p_q, p_g\}$ (i.e., either $p_q'\neq p_q$ or $p_g'\neq p_g$), their respetive relative pairs $\{P', \Delta_m'\}\neq\{P, \Delta_m\}$. Case (i), $P\neq P'$, therefore $\{P', \Delta_m'\}\neq\{P, \Delta_m\}$. Case (ii), $P=P'$, then it must be that $p_q\neq p_q'$ and $p_g\neq p_g'$. Assume $p_q\neq p_q'$ and $p_g= p_g'$, then we have
  \begin{align*}
    P=P' 
    &\Rightarrow P.B.[p_q+p_g]=P.B.[p_q'+p_g']\\
    &\Rightarrow p_q+p_g=p_q'+p_g' + i (2 N_\perp),\quad i=\pm 1\\
    &\Rightarrow p_q=p_q'+ i (2 N_\perp)
  \end{align*}
  which contradicts to $p_q, p_q'\in \Omega_\perp$. In analogy, $p_q= p_q'$ and $p_g\neq p_g'$ can not be true. Therefore, it must be that $p_q\neq p_q'$ and $p_g\neq p_g'$.
  We can proof that $ \Delta_m\neq \Delta_m'$ by contradiction. Assume that $\Delta_m=\Delta_m'$
  \begin{widetext}
  \begin{align*}
    \Delta_m=\Delta_m' 
    &\Rightarrow P.B.[Round[(1-z)P] -p_q ]=P.B.[Round[(1-z)P'] -p_q' ]\\
    &\Rightarrow Round[(1-z)P] -p_q =Round[(1-z)P] -p_q'+ i (2 N_\perp),\quad i=\pm 1\\
    &\Rightarrow p_q=p_q'+ i (2 N_\perp)
  \end{align*}
\end{widetext}
  which contradicts to $p_q, p_q'\in \Omega_\perp$.
  
  \begin{widetext}
 \item Inverse:\\
  $\forall \{p_q, p_g\}\in \Omega_\perp$, $\{p_q, p_g\}=f^{-1}(f(\{p_q, p_g\}; z);z)$, and $\forall \{P, \Delta_m\}\in \Omega_\perp$, $\{P, \Delta_m\}=f(f^{-1}(\{P, \Delta_m\}; z);z)$.
  
  By $f^{-1}(f(\{p_q, p_g\}; z);z)$,
  \begin{align*}
      p_q' 
      = & P.B.[Round[(1-z)P]-\Delta_m]\\
      = & P.B.\left[Round[(1-z)P]-P.B.[Round[(1-z)P] -p_q ]\right]\\
      = & P.B.\left[Round[(1-z)P]-Round[(1-z)P] + p_q + i (2 N_\perp)\right],\quad i=\pm 1\\
      = & p_q\;,\\
    p_g' = &  P.B.[P - p_q]\\
    = &  P.B.[P.B.[p_q+p_g] - p_q]\\
    = &  P.B.[p_q+p_g + i (2 N_\perp)- p_q],\quad i=\pm 1\\
    = & p_g\;.
  \end{align*}
  By $f(f^{-1}(\{P, \Delta_m\}; z);z) $,
  \begin{align*}
    P' = &P.B.[p_q + p_g]\\
    =& P.B.\left[ p_q + P.B.[P - p_q]\right]\\
    =& P.B.\left[ p_q + P - p_q + i (2 N_\perp) \right],\quad i=\pm 1\\
    = & P\;,\\
    \Delta_m' = & P.B.[Round[(1-z)P] -p_q ]\\
    = & P.B.\left[Round[(1-z)P] -P.B.[Round[(1-z)P]-\Delta_m] \right]\\
  = &  P.B.\left[Round[(1-z)P] -Round[(1-z)P]+\Delta_m +i (2 N_\perp)- p_q\right],\quad i=\pm 1\\
  = & \Delta_m\;.
  \end{align*}
\end{widetext}
\end{enumerate}

\bibliography{qA.bib}
\end{document}